\pdfoutput=1
\documentclass[aps,pr,twocolumn,superscriptaddress,preprintnumbers,showpacs,amsmath,amssymb]{revtex4}

\usepackage{graphicx} 
\usepackage{dcolumn}  

\usepackage{epsfig}
\usepackage{epsbox}
\usepackage[T1]{fontenc}
\usepackage{wrapfloat}
\usepackage{subfigure}
\usepackage{here}
\usepackage{array}
\usepackage{amsmath, amssymb, hhline, multir
ow, hangcaption, subfigure}
\usepackage{graphics}
\usepackage{graphicx}
\usepackage{lscape}
\usepackage{fancyhdr}
\usepackage{longtable}
\usepackage{supertabular}
\usepackage{refmerge}
\usepackage{dcolumn}
\usepackage{color}

\def \vec #1{\mbox{{\boldmath $#1$}}}

\def \GeV {{\rm GeV}}

\def \simlt {\stackrel{<}{\sim}}

\begin{document}
\hspace{100mm}
\preprint{\vbox{ \hbox{   }
                \hbox{Belle Preprint 2012-16}
                 \hbox{KEK Preprint 2012-8}
                \hbox{May 2012\ (Rev. Oct. 2012)}
}}
\title
{Measurement of ${\boldmath \gamma \gamma^* \rightarrow \pi^0}$ transition
form factor at Belle}

\begin{abstract}
We report a measurement of the process $\gamma \gamma^* \rightarrow \pi^0$ 
with a 759~fb$^{-1}$ data sample recorded with 
the Belle detector at the KEKB asymmetric-energy $e^+e^-$ collider.
The pion transition form factor, $F(Q^2)$, is measured 
for the kinematical region
$4~\GeV^2 \simlt Q^2 \simlt 40$~GeV$^2$, where $-Q^2$ is 
the invariant mass squared of a virtual photon.
The measured values of $Q^2|F(Q^2)|$ agree well with the
previous measurements below $Q^2 \simeq 9$~GeV$^2$ 
but do not exhibit the rapid growth in the higher $Q^2$ region 
seen in another recent measurement,
which exceeds the asymptotic QCD expectation by as much as 50\%.
\end{abstract}

\normalsize
\affiliation{Budker Institute of Nuclear Physics SB RAS and Novosibirsk State University, Novosibirsk 630090}
\affiliation{Faculty of Mathematics and Physics, Charles University, Prague}
\affiliation{University of Cincinnati, Cincinnati, Ohio 45221}
\affiliation{Department of Physics, Fu Jen Catholic University, Taipei}
\affiliation{Justus-Liebig-Universit\"at Gie\ss{}en, Gie\ss{}en}
\affiliation{Hanyang University, Seoul}
\affiliation{University of Hawaii, Honolulu, Hawaii 96822}
\affiliation{High Energy Accelerator Research Organization (KEK), Tsukuba}
\affiliation{Hiroshima Institute of Technology, Hiroshima}
\affiliation{Indian Institute of Technology Guwahati, Guwahati}
\affiliation{Indian Institute of Technology Madras, Madras}
\affiliation{Institute of High Energy Physics, Chinese Academy of Sciences, Beijing}
\affiliation{Institute of High Energy Physics, Vienna}
\affiliation{Institute of High Energy Physics, Protvino}
\affiliation{Institute for Theoretical and Experimental Physics, Moscow}
\affiliation{J. Stefan Institute, Ljubljana}
\affiliation{Kanagawa University, Yokohama}
\affiliation{Institut f\"ur Experimentelle Kernphysik, Karlsruher Institut f\"ur Technologie, Karlsruhe}
\affiliation{Korea Institute of Science and Technology Information, Daejeon}
\affiliation{Korea University, Seoul}
\affiliation{Kyungpook National University, Taegu}
\affiliation{\'Ecole Polytechnique F\'ed\'erale de Lausanne (EPFL), Lausanne}
\affiliation{Faculty of Mathematics and Physics, University of Ljubljana, Ljubljana}
\affiliation{Luther College, Decorah, Iowa 52101}
\affiliation{University of Maribor, Maribor}
\affiliation{Max-Planck-Institut f\"ur Physik, M\"unchen}
\affiliation{University of Melbourne, School of Physics, Victoria 3010}
\affiliation{Graduate School of Science, Nagoya University, Nagoya}
\affiliation{Kobayashi-Maskawa Institute, Nagoya University, Nagoya}
\affiliation{Nara Women's University, Nara}
\affiliation{National Central University, Chung-li}
\affiliation{National United University, Miao Li}
\affiliation{Department of Physics, National Taiwan University, Taipei}
\affiliation{H. Niewodniczanski Institute of Nuclear Physics, Krakow}
\affiliation{Nippon Dental University, Niigata}
\affiliation{Niigata University, Niigata}
\affiliation{Osaka City University, Osaka}
\affiliation{Pacific Northwest National Laboratory, Richland, Washington 99352}
\affiliation{Panjab University, Chandigarh}
\affiliation{University of Pittsburgh, Pittsburgh, Pennsylvania, 15260}
\affiliation{Research Center for Nuclear Physics, Osaka University, Osaka}
\affiliation{RIKEN BNL Research Center, Upton, New York 11973}
\affiliation{University of Science and Technology of China, Hefei}
\affiliation{Seoul National University, Seoul}
\affiliation{Sungkyunkwan University, Suwon}
\affiliation{School of Physics, University of Sydney, NSW 2006}
\affiliation{Tata Institute of Fundamental Research, Mumbai}
\affiliation{Excellence Cluster Universe, Technische Universit\"at M\"unchen, Garching}
\affiliation{Toho University, Funabashi}
\affiliation{Tohoku Gakuin University, Tagajo}
\affiliation{Tohoku University, Sendai}
\affiliation{Department of Physics, University of Tokyo, Tokyo}
\affiliation{Tokyo Institute of Technology, Tokyo}
\affiliation{Tokyo Metropolitan University, Tokyo}
\affiliation{Tokyo University of Agriculture and Technology, Tokyo}
\affiliation{CNP, Virginia Polytechnic Institute and State University, Blacksburg, Virginia 24061}
\affiliation{Wayne State University, Detroit, Michigan 48202}
\affiliation{Yamagata University, Yamagata}
\affiliation{Yonsei University, Seoul}
  \author{S.~Uehara}\affiliation{High Energy Accelerator Research Organization (KEK), Tsukuba} 
  \author{Y.~Watanabe}\affiliation{Kanagawa University, Yokohama} 
 \author{H.~Nakazawa}\affiliation{National Central University, Chung-li} 
  \author{I.~Adachi}\affiliation{High Energy Accelerator Research Organization (KEK), Tsukuba} 
  \author{H.~Aihara}\affiliation{Department of Physics, University of Tokyo, Tokyo} 
  \author{D.~M.~Asner}\affiliation{Pacific Northwest National Laboratory, Richland, Washington 99352} 
  \author{T.~Aushev}\affiliation{Institute for Theoretical and Experimental Physics, Moscow} 
  \author{A.~M.~Bakich}\affiliation{School of Physics, University of Sydney, NSW 2006} 
  \author{K.~Belous}\affiliation{Institute of High Energy Physics, Protvino} 
  \author{V.~Bhardwaj}\affiliation{Nara Women's University, Nara} 
% ...
\collaboration{The Belle Collaboration}

{\renewcommand{\thefootnote}{\fnsymbol{footnote}}

\setcounter{footnote}{0}
\setcounter{figure}{0}

\normalsize

\maketitle
\normalsize

\section{Introduction}
\label{sec:intro}
Recently, the BaBar collaboration reported a measurement of the 
photon-pseudoscalar meson ($P$) transition
form factors (TFF) from the process $\gamma \gamma^* \rightarrow P$
in the reaction $e^+ e^- \rightarrow (e) e P$, where
$P=\pi^0$~\cite{babar1} (Fig.~\ref{fig:diag}),
$\eta$ or $\eta '$~\cite{babar2}, and
($e$) and $e$  denote an untagged and tagged electron/positron, 
respectively.
The momentum transfer 
$Q^2$ (where $-Q^2$ is the squared invariant mass of a space-like photon)
explored by BaBar extends up to $\sim 40$~GeV$^2$
compared to the previous measurements by CELLO for 
$Q^2 < 2.5$~GeV$^2$~\cite{cello} 
and by CLEO for $Q^2 < 9$~GeV$^2$~\cite{cleo}.
While the $Q^2$ dependence of TFFs measured by BaBar
for the  $\eta$ and $\eta '$
agrees fairly well with theoretical expectations, that for the $\pi^0$ shows 
rapid growth for $Q^2 >10$~GeV$^2$.

\begin{figure}
\centering
\includegraphics[width=4cm]{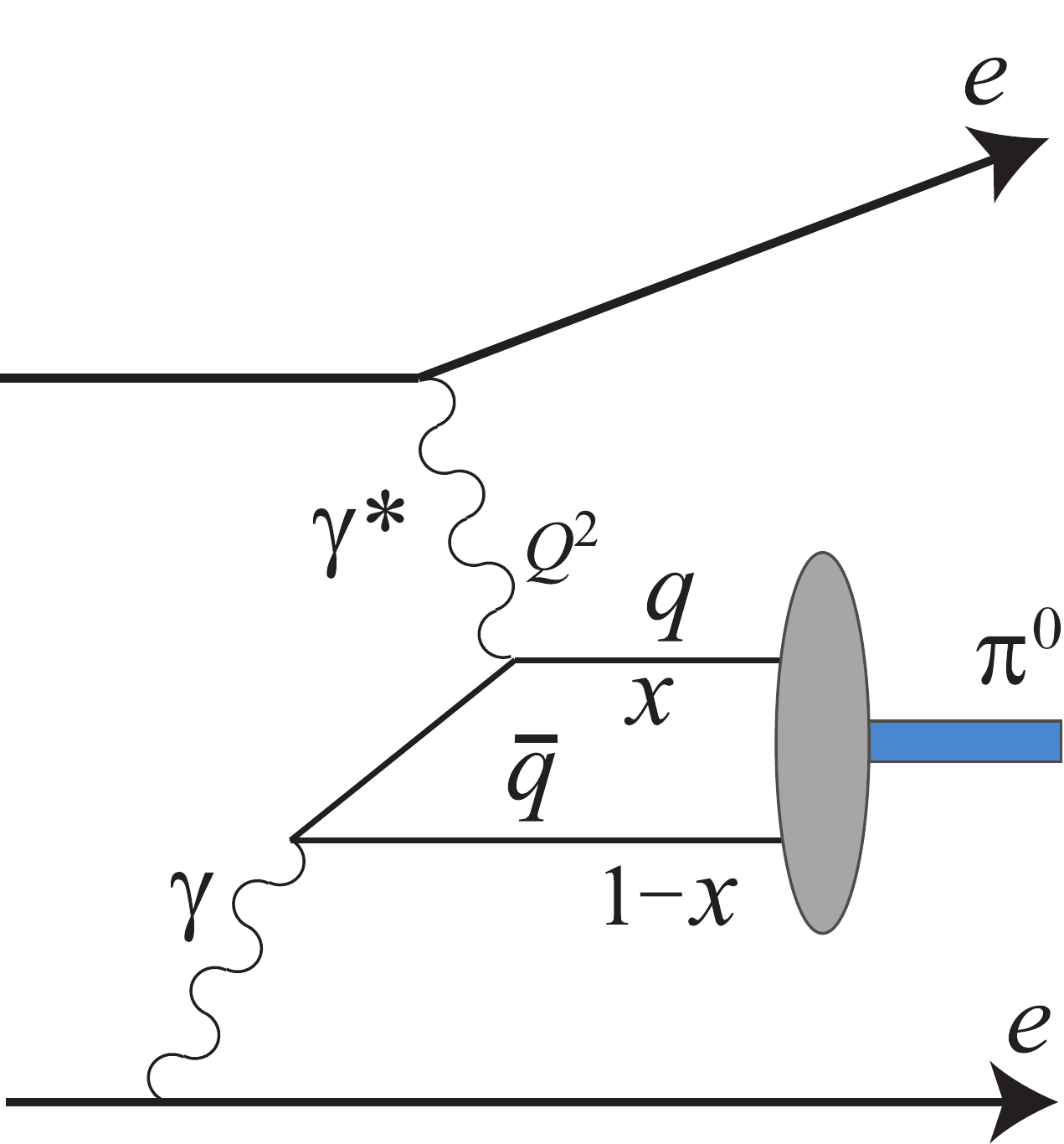}
\centering
\caption{
A Feynman diagram 
for $\gamma\gamma^* \to q \bar{q} \to \pi^0$
in $e^+e^-$ collisions
}
\label{fig:diag}
\end{figure}

Lepage and Brodsky pointed out the importance of
TFF measurements more than 30 years ago and proposed tests of
perturbative QCD (pQCD) in various exclusive processes~\cite{LB}.
We focus our discussion on the TFF for the pion; that for 
the $\eta$ or $\eta '$ is quite similar.
To leading order in 
the QCD coupling constant,
$\alpha_s (Q^2)$, the TFF
can be written in the factorized form~\cite{agaev}
\begin{eqnarray}
F(Q^2) &=& \frac{\sqrt{2} f_{\pi}}{3}
\int_0^1 dx T_H (x, Q^2, \mu, \alpha_s (\mu)) \nonumber \\
& & \times \phi_{\pi}(x, \mu)  ,
\label{eqn:leadingf}
\end{eqnarray}
where $f_{\pi} \simeq 0.131$~GeV is the pion decay constant,
$x$ is the fraction of momentum carried by a quark ($q$) or 
antiquark ($\bar{q}$) in the parent pion,
$T_H (x, Q^2, \mu, \alpha_s (\mu))$ is the hard-scattering amplitude
for $\gamma \gamma^* \rightarrow q \bar{q} $ 
(Fig.~\ref{fig:diag}),
$\mu$ is the renormalization scale, and
$\phi_{\pi}(x, \mu)$ is the leading-twist pion distribution amplitude 
(DA) at $x$ and $\mu$.
Note that the pion DA is a universal function and, once determined,
describes leading terms in all hard exclusive processes involving 
a pion.
The most prominent examples are the pion electromagnetic form factor and
the weak semileptonic decay rate,
$B^{\pm} \rightarrow \pi^0 \ell^{\pm} \nu_{\ell}$~\cite{agaev}.

At sufficiently high $Q^2$, Eq.~(\ref{eqn:leadingf}) can be 
written as~\cite{agaev}
\begin{equation}
F(Q^2) = \frac{\sqrt{2} f_{\pi}}{3}
\int_0^1 dx \frac{\phi_{\pi}(x)}{x Q^2} + {\cal O}\left( 1/Q^4 \right) .
\label{eqn:highq2f}
\end{equation}
Because the asymptotic pion DA is $\phi_{\pi}^{\rm asy} = 6x(1-x)$, 
it follows that $Q^2 F(Q^2)$ at $Q^2 \rightarrow \infty$
is exactly normalized as~\cite{LB}
\begin{equation}
Q^2 F(Q^2) = \sqrt{2} f_{\pi} \simeq 0.185~{\rm GeV} .
\label{eqn:asymptf}
\end{equation}
The measurements reported by CELLO~\cite{cello}
and CLEO~\cite{cleo} support this prediction;
the measured values seem to approach this asymptotic value from below.
The recent measurement by BaBar~\cite{babar1} has changed this situation
drastically.
The measured $Q^2|F(Q^2)|$ values seem to increase with $Q^2$, exceeding
the QCD asymptotic value by as much as 50\%; this has created excitement
and renewed interest in this quantity.

Many theoretical constructs have been made to explain
this $Q^2$ dependence.
Early attempts pointed out that the BaBar result requires
a nearly flat pion DA ({\it e.g.}, \cite{rady, poly}), 
which implies that the pion behaves as a point-like particle.
The conclusions of the theoretical analyses can be divided into two groups: 
those that can explain the BaBar result~({\it e.g.}, \cite{agaev, pham}) 
and those that cannot ({\it e.g.}, \cite{MS, roberts}).

This theoretical controversy seems to have been settled by a recent analysis of
Brodsky, Cao and de T\'{e}ramond~\cite{bcd}, who investigated TFFs of
the $\pi^0$, $\eta$ and $\eta '$ using four typical models of the meson DA.
They concluded that the rapid growth of the $\pi^0$ TFF 
with $Q^2$ at BaBar,
together with the gentle $Q^2$-dependence of
the $\eta$ and $\eta'$ TFFs also measured at BaBar 
are difficult to explain within the present framework of QCD.
They also
pointed out drawbacks of earlier theoretical attempts that explained
the $Q^2$ dependence of the $\pi^0$ TFF at BaBar
and stated that the BaBar $\pi^0$ result, if confirmed, would 
imply new physics beyond standard QCD.

In this Article, we report an independent measurement of the
$\pi^0$ TFF in the $Q^2$ region of interest.

\section{Experimental apparatus and Event Selection}
\label{sec:exper}
We briefly describe the Belle 
detector and then give a description of the event selection.
Finally, we present some event distributions to illustrate
the selection criteria.

\subsection{The Belle detector and data sample}
\label{sub:belle}
We use a 759~fb$^{-1}$ data sample recorded with
the Belle detector at the KEKB asymmetric-energy $e^+e^-$ collider~\cite{kekb}.
We combine data samples 
collected at several beam energies: at
the  $\Upsilon(4S)$ resonance
$(\sqrt{s} = 10.58~{\rm GeV})$ and 
60~MeV below it (637~fb$^{-1}$ in total), 
at the $\Upsilon(3S)$ resonance $(\sqrt{s} = 10.36~\GeV$, 3.2~fb$^{-1}$) and 
near the $\Upsilon(5S)$ resonance $(\sqrt{s} = 10.88~\GeV$, 119~fb$^{-1}$).
When combining the data, the slight dependence of the two-photon cross sections
on beam energy is properly taken into
account.

This analysis is performed in the ``single-tag'' mode, where  
either the recoil electron or positron (hereafter referred to as electron)
alone is detected. 
We restrict the virtuality ($Q^2$) of the untagged-side photon to be small 
by imposing a strict transverse-momentum  balance with respect to 
the beam axis for the tagged electron and the final-state 
neutral pion. In this Article, we 
refer to events tagged by an $e^+$ or an $e^-$ 
as ``p-tag'' (positron-tag) or ``e-tag'' (electron-tag), 
respectively.

\subsubsection{Belle detector}
A comprehensive description of the Belle detector is
given elsewhere~\cite{belle}. 
We mention here only the
detector components essential for this measurement.
Charged tracks are reconstructed from the drift-time information in a central
drift chamber (CDC) located in a uniform 1.5~T solenoidal magnetic field.
The $z$ axis of the detector and the solenoid are along the positron beam
direction, with the positron beam pointing in the $-z$ direction.  
The CDC measures the
longitudinal and transverse momentum components, {\it i.e.}, 
along the $z$ axis and in the $r\varphi$ plane perpendicular to
the beam, respectively.
Track trajectory coordinates near the collision point are provided by a
silicon vertex detector (SVD).  
Photon detection and energy measurements are performed with a CsI(Tl) 
electromagnetic calorimeter (ECL).
Electron identification (ID) is based on  $E/p$, the ratio of the 
calorimeter energy and the track momentum. 
We employ this rather simple variable
because the background from hadrons and muons is
small compared to the dominant
radiative Bhabha and two-photon background contributions.

\subsubsection{Data sample}
To be recorded, events must satisfy 
one of the two mutually exclusive triggers
based on ECL energy: 
the HiE (High-energy threshold) 
trigger and the CsiBB (ECL-Bhabha) trigger~\cite{ecltrig}. 
The HiE trigger requires that the sum of the energies measured 
by the ECL in an event exceed
1.15~GeV but that the event not be Bhabha-like;
the HiE trigger is vetoed by the CsiBB trigger.
The CsiBB trigger is designed to identify
back-to-back Bhabha events~\cite{ecltrig}
and is prescaled to reduce their high rate.
For the purpose of monitoring trigger performance, the events triggered 
by the CsiBB are recorded once per 50 CsiBB triggers
({\it i.e.}, are prescaled by a factor of 50).
This prescaled event sample has been extensively used in our analysis
to select a subsample of signal events, and to calibrate and test the trigger 
simulation for the CsiBB veto implemented in the HiE trigger.

We do not use information on track triggers because they
require two or more charged tracks; our signal
has only one charged track.

\subsection{Single-track skim}
\label{sub:skim}
To simplify the logistics of the data analysis, we
use a "single-track" skim whose requirements
are the following.
\newcounter{jtem}
\begin{list}%
{(S\arabic{jtem})}{\usecounter{jtem}}
\setlength{\rightmargin}{\leftmargin}
\setlength{\topsep}{0mm}
\setlength{\parskip}{0mm}
\setlength{\parsep}{0mm}
\setlength{\itemsep}{0mm}
\item 
The event is triggered by the HiE or the CsiBB trigger.

\item 
There is only one track that satisfies $p_t > 0.5$~GeV/$c$,
$-0.8660 < \cos \theta <0.9563$, $dr < 1$~cm and $|dz| < 5$~cm. 
There are no other tracks that satisfy $p_t > 0.1$~GeV/$c$ and $dr < 5$~cm,
$|dz| < 5$~cm in the entire CDC volume. Here,
$p_t$ is the transverse
momentum in the laboratory frame with respect to the positron beam axis,
$\theta$ is the polar angle of the momentum
direction with respect to the $z$ axis,
and ($dr$, $dz$) are the cylindrical coordinates of the closest approach
of the track to the beam axis. 
\item There are one or more neutral clusters in the ECL, whose
energy sum is greater than 0.5~GeV.
\end{list}

These conditions are efficient in selecting a signal process
within the kinematical regions 
of  $e^+ e^- \to (e) e \pi^0$ in which one electron escapes
detection at small forward angles.
They also select radiative Bhabha ($e^+ e^- \to (e) e \gamma$) events 
with a Virtual Compton (VC) process configuration, in which one of
the electrons escapes detection in the forward direction;
such events are used for trigger calibration.

\subsection{Selection criteria for signal candidate events}
\label{sub:selec}
A signal event consists of an energetic electron and
two photons 
as described below.
Candidate events are selected from the single-track skim
with the following selection criteria. 
The kinematical variables are calculated in the laboratory system 
unless otherwise noted; those in the $e^+e^-$ center-of-mass (c.m.) frame
are identified with an asterisk.

\setcounter{jtem}{0}
\begin{list}%
{(\arabic{jtem})}{\usecounter{jtem}}
\setlength{\rightmargin}{\leftmargin}
\setlength{\topsep}{0mm}
\setlength{\parskip}{0mm}
\setlength{\parsep}{0mm}
\setlength{\itemsep}{0mm}
\item Events are triggered by the HiE trigger. 
We also use events triggered by the CsiBB trigger, separately, for the
$Q^2 = 4 - 6$~GeV$^2$ region of  the e-tag.

\item There is exactly one track as
required in the skim condition (S2).

\item For electron ID, we require $E/p > 0.8$ for the candidate
electron track.

\item 
The absolute value of the momentum of the electron is greater than 1.0~GeV/$c$,
where the electron energy is corrected for photon radiation or 
bremsstrahlung in the following way.
In a $3^\circ$ cone around the track,
we collect all photons 
in the range 0.1~GeV$< E_\gamma < p_e/3$,
where $p_e$ is the measured absolute momentum of the electron track. 
The absolute momentum of the electron 
is replaced by $p_e+\Sigma E_\gamma$.

\item We require exactly two photons above 0.2~GeV, excluding those used 
for bremsstrahlung and final state radiation recovery in (4). 
We designate the energies of the two 
photons by $E_{\gamma1}$ for the higher and $E_{\gamma2}$ for the lower.

\item When there are one or more extra photons with energy greater 
than 0.05~GeV, we require that there be no $\pi^0$ candidate
reconstructed with a two-photon combination different from the pair
in (5) in the following way.
We search for a $\pi^0$ candidate reconstructed from two photons
that satisfies a goodness-of-fit criterion, $\chi^2<9$, in a
$\pi^0$-mass-constrained fit.
If we find such a $\pi^0$
(including the case when only one photon in (5) 
is shared in the $\pi^0$), the event is rejected.

\item 
The two-photon energy sum $E_{\gamma\gamma}$ defined as
\begin{equation}
E_{\gamma\gamma} = E_{\gamma 1}+ E_{\gamma 2} 
\end{equation}
satisfies $E_{\gamma\gamma}  > 1.0~\GeV$.

\item 
The cosines of the polar angles for the electron ($\theta_e$)
and the two photons ($\theta_{\gamma 1}$ and $\theta_{\gamma 2}$)
must be within the range $-0.6235 < \cos \theta_i < 0.9481$ ($i = $
$e$, $\gamma 1$ or $\gamma 2$).
This is the sensitive region for the HiE trigger.

\item We reject events with a polar-angle combination
($\cos \theta_e$, $\cos \theta_{\gamma\gamma})$ that falls in
the predetermined two-dimensional angular area shown in Fig.~\ref{fig:cell},
where $\theta_{\gamma\gamma}$ is
the polar angle of the momentum summed for the two photons.
This rejection is referred to as the "Bhabha-mask" rejection.
The Bhabha mask is introduced to limit the fiducial region to a range where
the efficiency of the HiE trigger is well determined.
The details of the Bhabha mask are given in Appendix~\ref{sec:A}.

\begin{figure*}
\centering
\includegraphics[width=9cm]{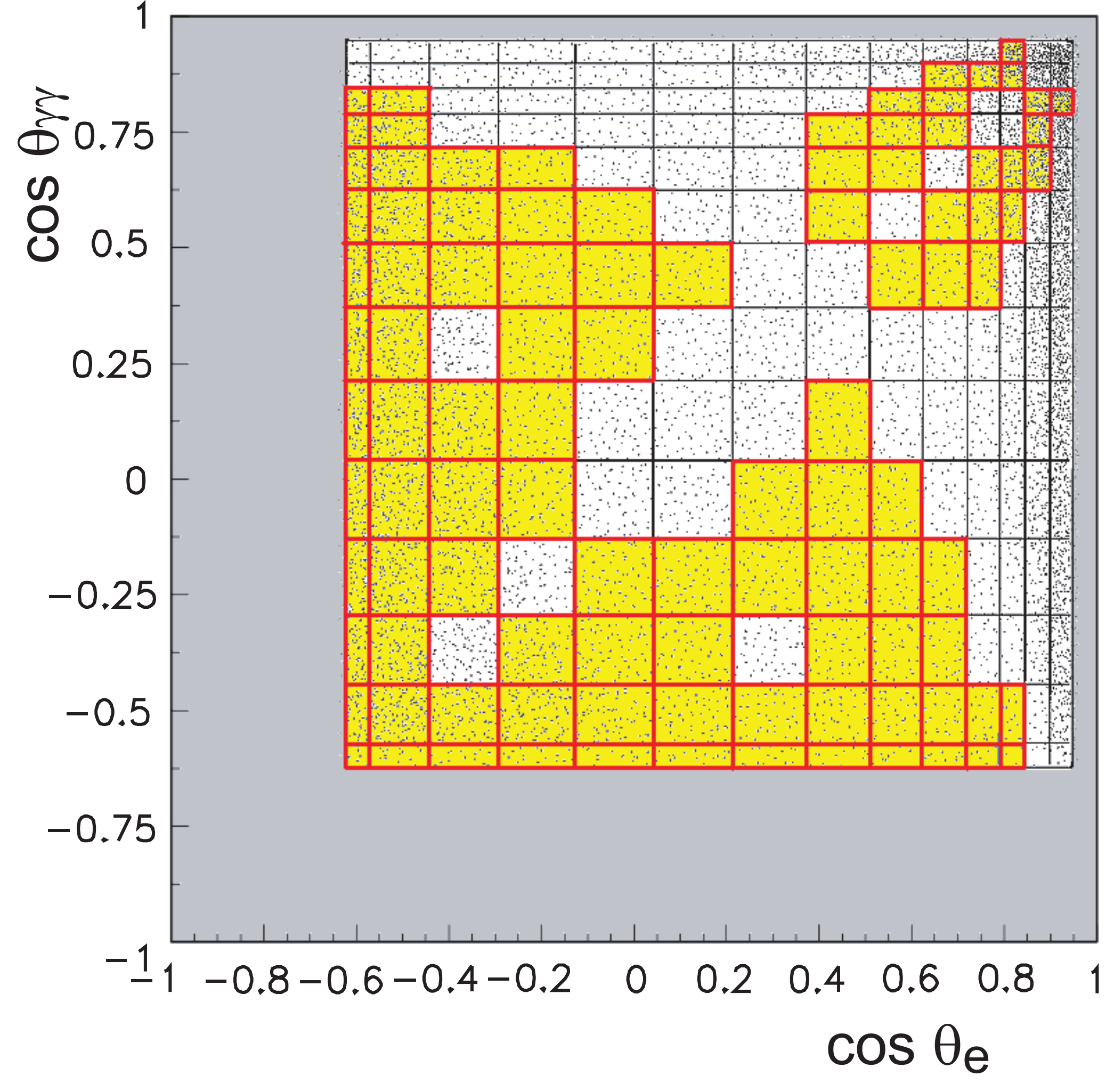}
\centering
\caption{
Two-dimensional angular regions in 
($\cos \theta_e$, $\cos \theta_{\gamma\gamma}$) 
defined as the Bhabha mask (see Appendix~\ref{sec:A} for the details). 
Events in the yellow (white) regions
are selected (rejected). 
The gray region is outside of the angular acceptance.
The mask is symmetric for $e$ and $\gamma\gamma$.
The scattered dots in the figure show signal Monte Carlo
simulation events without the Bhabha veto 
applied.
}
\label{fig:cell}
\end{figure*}

\item The energy asymmetry for the two photons, defined as
\begin{equation}
E_{\rm asym} = \frac{E_{\gamma 1}- E_{\gamma 2}}{E_{\gamma\gamma}} ,
\label{eqn:easym}
\end{equation}
satisfies $E_{\rm asym} <0.8$
to eliminate low-energy background photons.

\item 
The polar angle difference, 
$\Delta \theta = |\theta_{\gamma1} - \theta_{\gamma2}| $
for the two photons multiplied by $E_{\gamma \gamma}$ satisfies
\begin{equation}
E_{\gamma\gamma} \Delta \theta 
> 0.18~{\rm radian} \cdot \GeV .
\end{equation}
This rejects the large background 
from radiative Bhabha events with a conversion, {\it i.e.}
$e^+ e^- \to (e) e \gamma$, $\gamma \to e^+e^-$, 
where the
electron-positron pair from a photon conversion 
tends to separate in a common
plane perpendicular to the magnetic field (which is 
along the $z$ axis), and the $e^+e^-$ pair
is misreconstructed as a $\gamma \gamma$ candidate.

\item We exploit the charge-direction correlation (``right-sign'') 
for the single-tag two-photon process
between the electric charge of the 
tagged electron ($q_{\rm tag}$) and the $z$ component of the momentum
of the $e\gamma\gamma$ system in the $e^+ e^-$ c.m. frame by requiring 
\begin{equation}
q_{\rm tag} \times (p^*_{z,e} + p^*_{z,\gamma\gamma})< 0 .
\end{equation}

\item We require $0.85 < E_{\rm ratio} <1.1$, where 
\begin{equation}
E_{\rm ratio} = \frac{E^{* {\rm measured}}_{\gamma\gamma}}
{E^{* {\rm expected}}_{\gamma\gamma}} .
\end{equation}
This requirement is motivated by
a three-body kinematical calculation for  $e^+ e^- \to (e)e\pi^0$,
imposing four-momentum conservation condition
$p_{\rm initial}(e^+e^-) = p_{\rm final}((e)e \gamma\gamma)$ 
wherein the direction of the $\pi^0$ momentum is taken to be 
parallel to that of the observed $\gamma\gamma$ system in the $e^+e^-$ c.m. frame.
The expected $E^*_{\gamma\gamma}$ value is obtained
by assigning the nominal $\pi^0$ mass to the  $\gamma\gamma$ system.
This calculation is not sensitive to the input
$\gamma\gamma$ invariant mass. 

\item 
We reject events with a back-to-back configuration in the
$e^+e^-$ c.m. frame, to eliminate backgrounds from Bhabha events
in which a track is not reconstructed; we require
$\zeta^*(e,\gamma\gamma) < 177^\circ$,
where $\zeta^*(e,\gamma\gamma)$ is the opening angle between the
electron and the $\gamma\gamma$ system.

\item 
We require the electron and the $\gamma \gamma$ system to be
nearly back-to-back when their momentum
directions are projected onto the $r \varphi$ plane: 
$\alpha(e, \gamma\gamma) < 0.1$~radians, where the acoplanarity angle
$\alpha$ is defined as the difference between 
the opening angle of the two directions on the projected plane and $\pi$.

\item 
We require transverse momentum balance in the 
$e^+e^-$ c.m. frame, $|\Sigma \vec{p}_t^*| < 0.2$~GeV/$c$,
where
\begin{equation}
|\Sigma \vec{p}_t^*| =
|\vec{p}^*_{t,e} + \vec{p}^*_{t,\gamma\gamma}| .
\end{equation}
\end{list}

With the above selection criteria, we collect 
$e^+e^- \to (e)e \gamma \gamma$ candidates in the
broad $\gamma \gamma$ invariant-mass range of 
$0.07~\GeV/c^2 \simlt M_{\gamma\gamma} \simlt 0.50~\GeV/c^2$. 
The minimum $\gamma\gamma$ 
invariant mass is determined mainly by selection criteria 
(7) and (11) on 
$E_{\gamma\gamma}$ and $E_{\gamma\gamma} \Delta \theta$.
We do not apply any explicit selection on $M_{\gamma\gamma}$
for the determination of the $\pi^0$ signal yield.
Instead, we later obtain the $\pi^0$ yield by fitting the $M_{\gamma\gamma}$ 
distribution in $Q^2$ bins separately for the p-tag and e-tag
samples: the selection requirements for the two tags are very different
because the beam energies and the detector are asymmetric.

\subsection{Event distributions}
\label{sub:dist}
In this subsection, we illustrate
how the signal candidates are selected.
Figure~\ref{fig:dis2} shows the energy asymmetry ($E_{\rm asym}$,
defined in criterion (10)) of
the two photons for events in
which the electron and photons satisfy the selection
criteria for momenta, energies and angles.
Events with a large
asymmetry are suppressed by the minimum photon 
energy requirement (0.2~GeV), yet a peak at $E_{\rm asym} > 0.8$ is still visible 
in the data from low-energy photon background 
(Fig.~\ref{fig:dis2}(a)).

\begin{figure*}
\centering
\includegraphics[width=12cm]{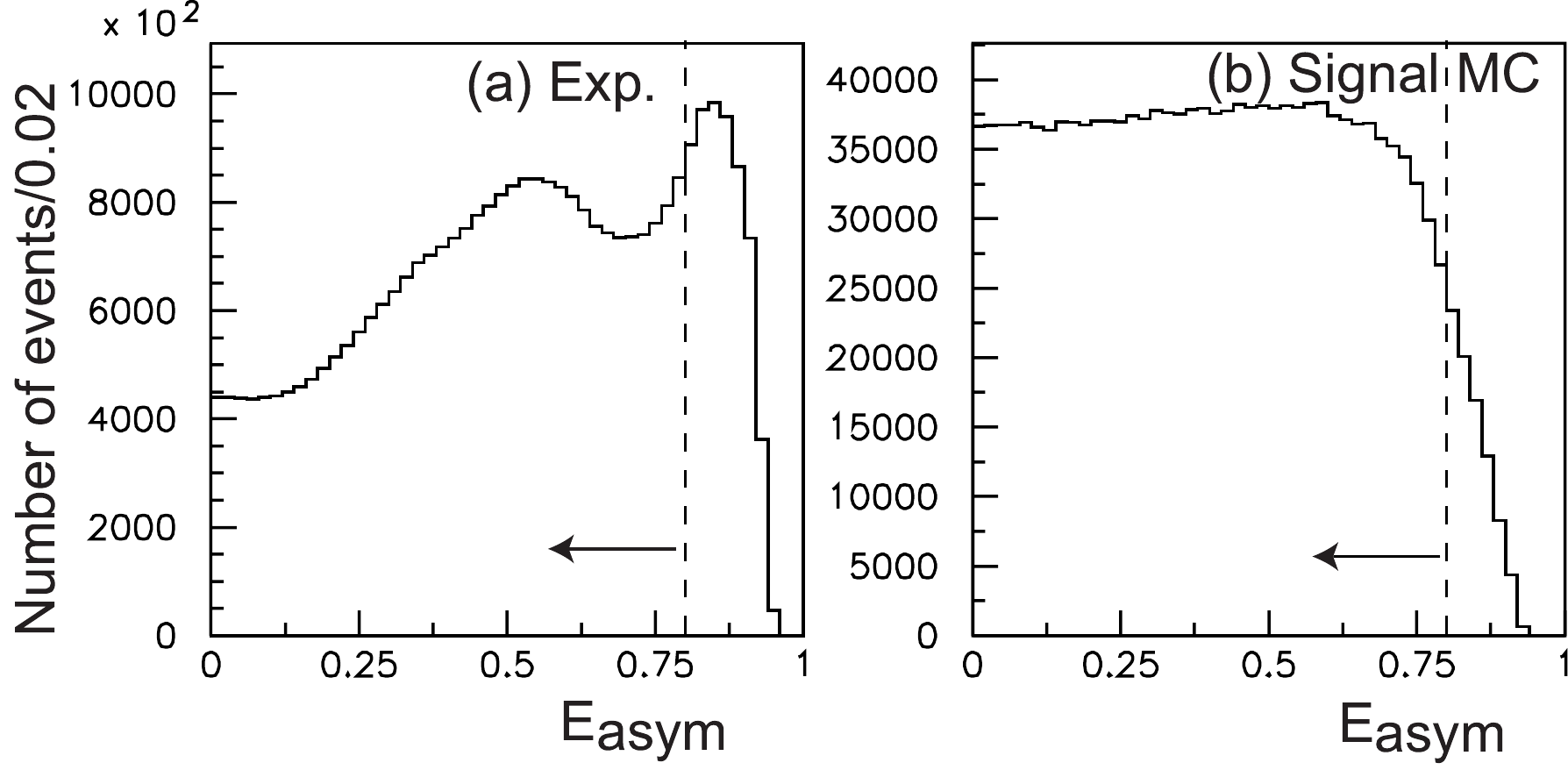}
\centering
\caption{ Energy asymmetry ($E_{\rm asym}$, defined in selection 
criterion (10)) for
the two photon candidates for events
in which $p_e$, $E_{\gamma\gamma}$ and $\theta_{\gamma\gamma}$ 
satisfy the criteria for
(a) experimental data and (b) signal Monte Carlo samples. 
The dashed lines show the selection boundaries. 
Arrows indicate the accepted regions.
}
\label{fig:dis2}
\end{figure*}

Figures~\ref{fig:dis4} and \ref{fig:dis4p} 
show $E_{\rm ratio}$ versus $M_{\gamma\gamma}$ and 
$|\Sigma \vec{p}_t^*|$, respectively,
for the samples after applying selection criteria (11) and (15) for 
$E_{\gamma \gamma} \Delta \theta$ and the 
acoplanarity angle, respectively. 
The latter is plotted only for events
in the vicinity of the pion mass, 
$0.120~\GeV/c^2 < M_{\gamma\gamma} < 0.145~\GeV/c^2$,
where the $E_{\rm ratio}$ requirement is not applied.
We find a significant enhancement in the signal region of
the experimental data near $E_{\rm ratio}$=1, $M_{\gamma\gamma} = m_\pi$
and  $|\Sigma \vec{p}_t^*|=0$, where $m_\pi$ is the nominal $\pi^0$ mass. 
A narrow enhancement in the experimental
data around $E_{\rm ratio}=1.15$ 
(Fig.~\ref{fig:dis4}(a)) 
is due to background from radiative Bhabha events and is reproduced by the
background Monte Carlo (MC). 
The signal in the experimental data is well separated
from the low-energy $\pi^0$ background 
in the region near $|\Sigma \vec{p}_t^*|=0$
(Fig.~\ref{fig:dis4p}).

\begin{figure*}
\centering
\includegraphics[width=12cm]{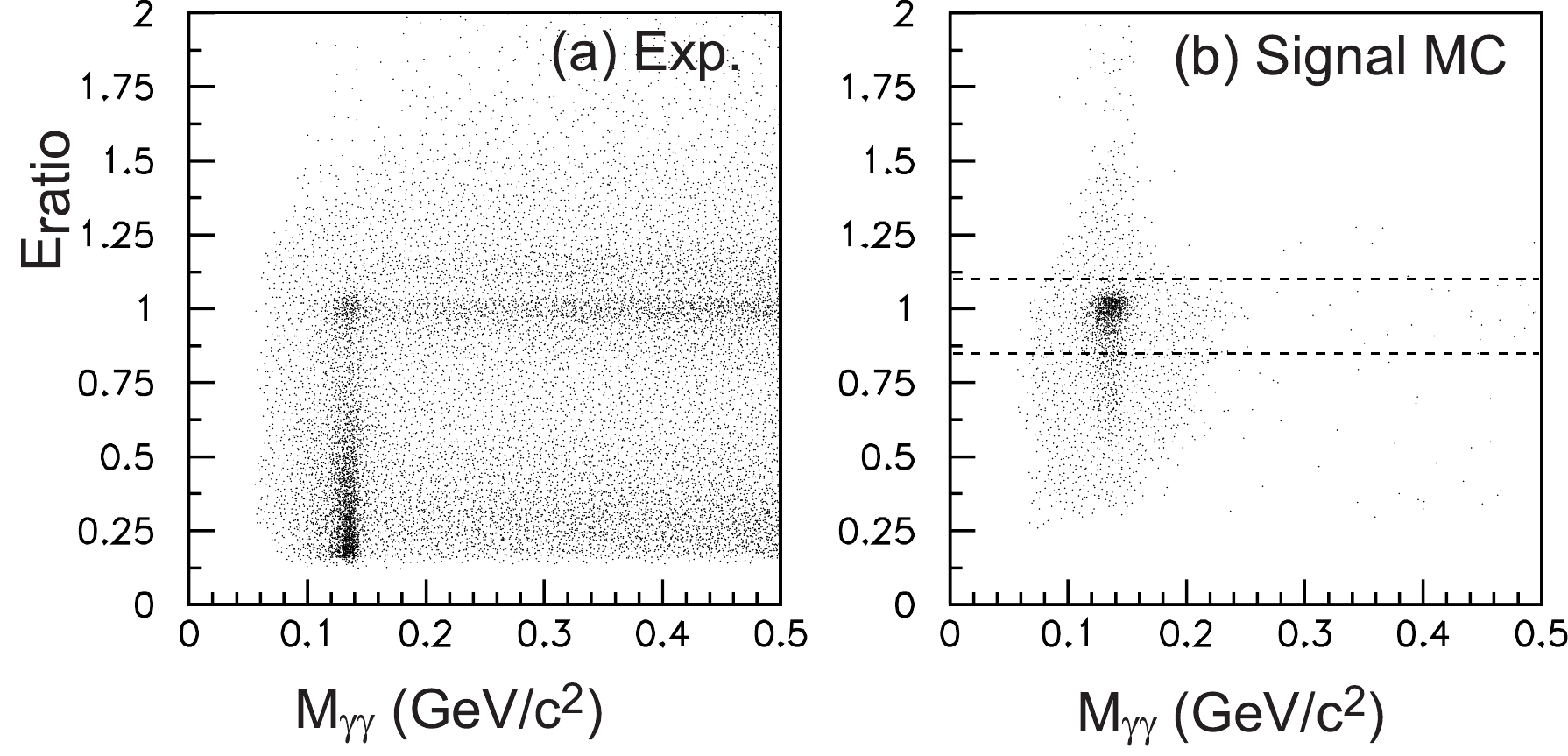}
\centering
\caption{Scatter plots 
for $E_{\rm ratio}$ vs. $M_{\gamma\gamma}$ 
for (a) experimental data and (b) signal MC samples.
See the text for the selection conditions. The horizontal
band between the dashed lines in (b) 
is the accepted region.
}
\label{fig:dis4}
\end{figure*}
\begin{figure*}
\centering
\includegraphics[width=12cm]{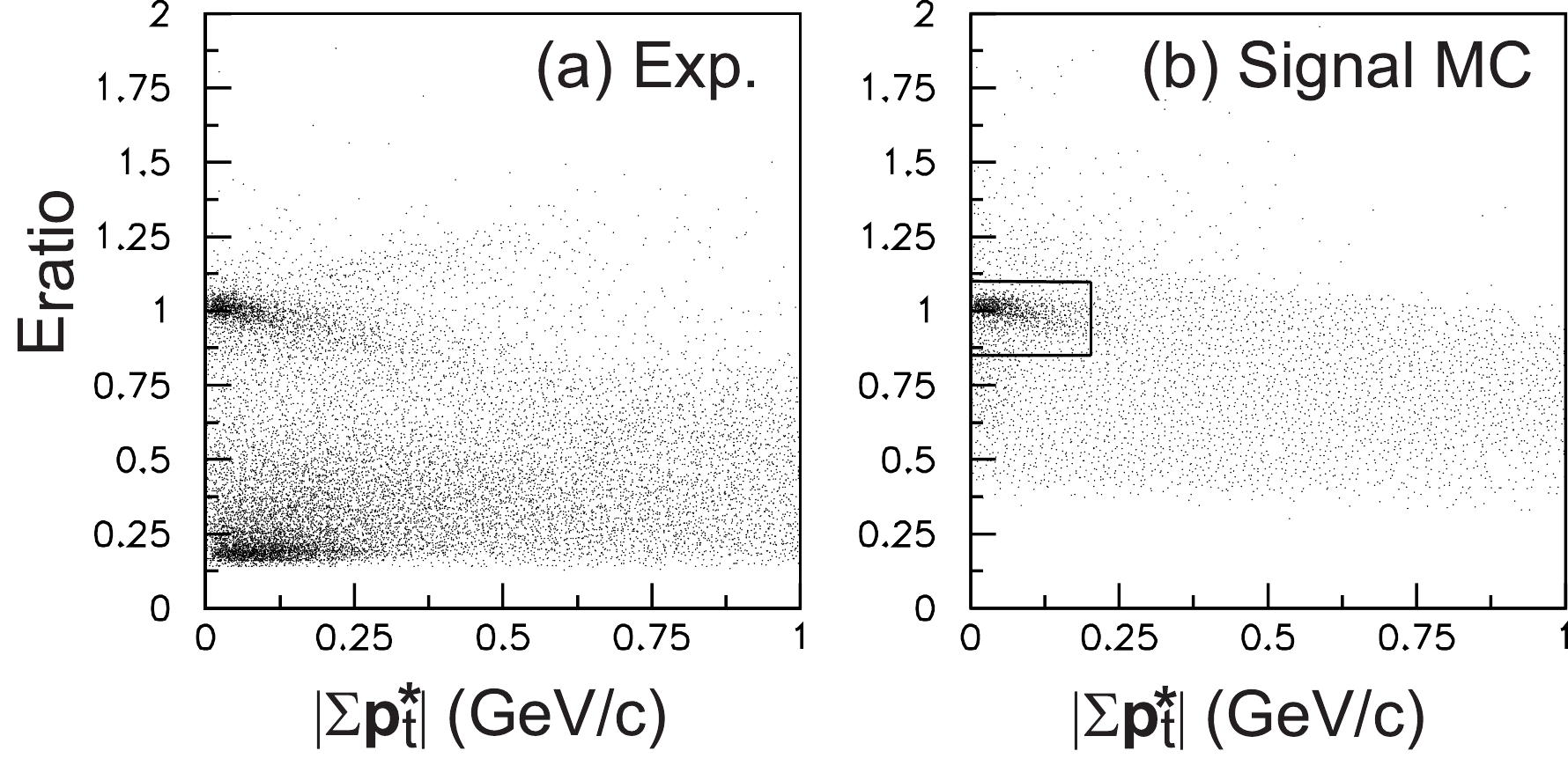}
\centering
\caption{Scatter plots 
for $E_{\rm ratio}$ vs. $|\Sigma \vec{p}_t^*|$  
for the (a) experimental data and (b) signal MC samples.
See the text for the selection requirements.
The solid box in (b) shows the selection region.
}
\label{fig:dis4p}
\end{figure*}

In Fig.~\ref{fig:dis5},
we show the scatter plot for $E_{\gamma\gamma}\Delta \theta$ 
vs. $M_{\gamma\gamma}$
for the final samples in the range 15~GeV$^2<Q^2<40$~GeV$^2$, 
where only the corresponding selection criterion
is not applied.
The large background
from radiative Bhabha events is present primarily
in the range $0.05~\GeV/c^2 \simlt M_{\gamma \gamma} \simlt 0.20~\GeV/c^2$
below the selected boundary in $E_{\gamma\gamma}\Delta \theta$, 
with no contamination seen above the boundary. 
This background is discussed 
in more detail in Sec.~\ref{sub:bkg}.
\begin{figure*}
\centering
\includegraphics[width=12cm]{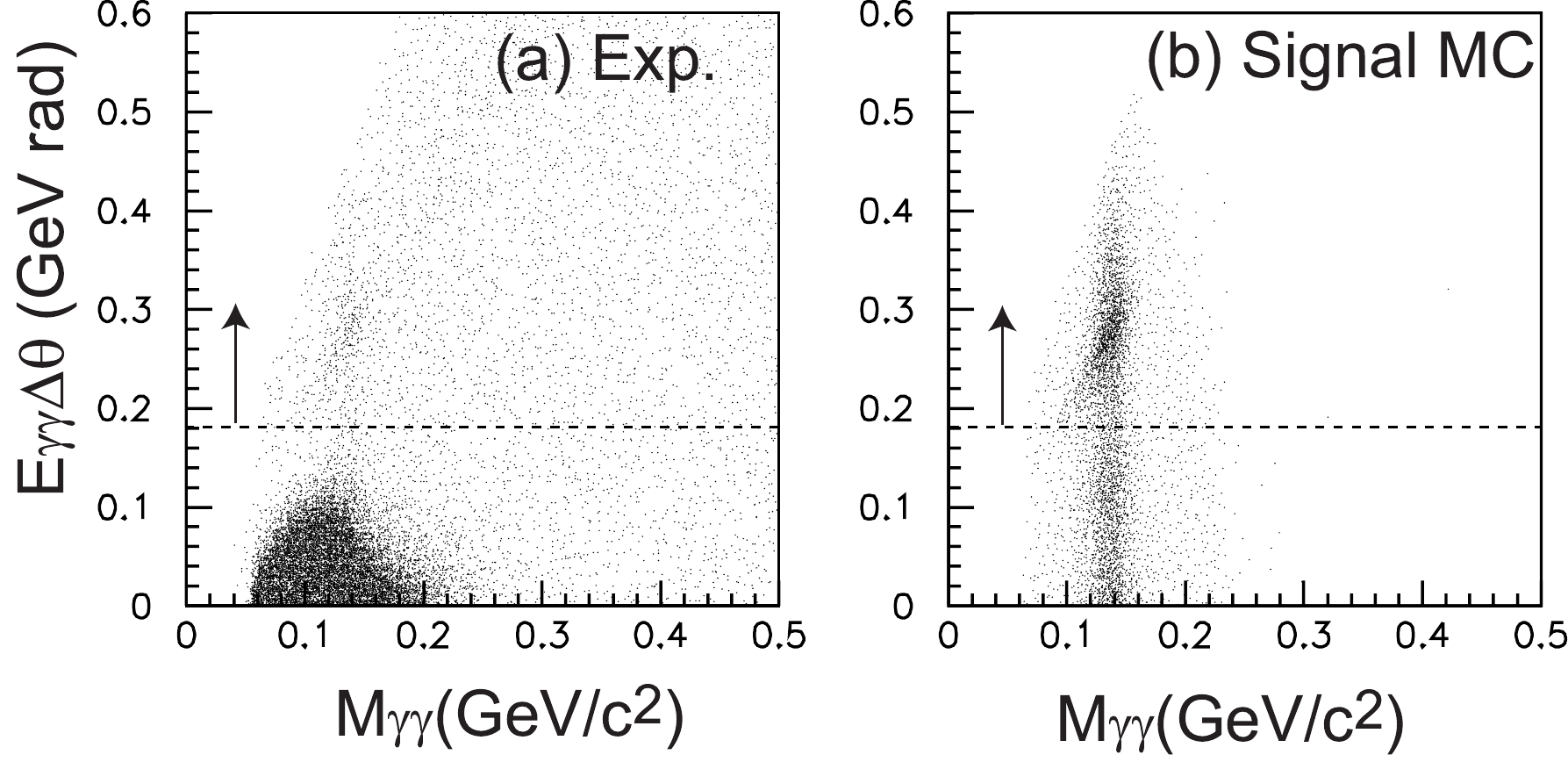}
\centering
\caption{Scatter plots 
for $E_{\gamma\gamma} \Delta \theta$ vs. $M_{\gamma\gamma}$ 
for (a) the experimental data and (b) signal MC samples.
This distribution is for the final samples in the range
 15~GeV$^2<Q^2<40$~GeV$^2$ where only 
the $E_{\gamma\gamma} \Delta \theta$ criterion is not applied.
The dashed lines show the selection boundaries. 
Arrows show the accepted regions.  
}
\label{fig:dis5}
\end{figure*}

\section{Signal Monte Carlo and Estimation of the Detection Efficiencies}
\label{sec:mc}
In this analysis, MC programs play an essential role in defining 
the selection 
criteria and determining the detection efficiencies.
In this section, a specially developed signal MC program 
is described. 
The efficiency determination is then described.
A detailed discussion of the trigger simulator tuning
is given in Appendix~\ref{sec:B}, where 
radiative Bhabha events are compared with
MC events generated by Rabhat~\cite{rabhat}.
This MC is a specially developed program to deal with the VC process
in which an electron is scattered forward and remains undetected.

\subsection{Signal Monte Carlo, TREPSBST}
\label{sub:calmet}
To calculate the conversion factor between the transition form factor 
and the differential cross section, to estimate the overall efficiency 
using signal MC events and to perform other studies,
an MC program, TREPSBST, was developed for the 
signal process $e^+ e^- \to (e) e \pi^0$.

TREPSBST implements the formulae of Eqs.~(2.1) and (4.5) 
in Ref.~\cite{bkt} and is based on MC program TREPS~\cite{treps},
which was modified 
to match the condition that the equivalent 
photon approximation (EPA) is not applied.
In TREPSBST, we always retain the
condition $Q^2_1 > Q^2_2$ for virtuality of the two
colliding photons 
by requiring $Q^2_1 > 3.0$~GeV$^2$
and $Q^2_2 < 1.0$~GeV$^2$ for the ranges of
integration and event generation. 

This $Q^2_2$ range is sufficient to generate signal events over
the kinematical region for the $|\Sigma \vec{p}_t^*|$
selection criterion, and is also used to define the differential cross
section $d\sigma/dQ^2$ for the signal process in this analysis
(see Sec.~\ref{sub:crosc}).

The program calculates the differential cross 
section $d\sigma/dQ^2_1$ for the given $Q^2_1$ points
and, separately, generates MC events with a specified 
$Q^2_1$ distribution. 
The shape of the distribution,
in principle,
does not affect the efficiency determination, 
which is evaluated for each $Q^2_1$ bin with a bin width of 1~GeV$^2$.
However, we assign a systematic error from
the uncertainty in the $Q^2$-dependence effect for some bins in the
lowest $Q^2$ region (see Sec.~\ref{sub:syserr}).

The differential cross section for the signal process
is calculated at tree level assuming a $g^2$ coupling
parameter derived from $\Gamma_{\gamma\gamma}=8$~eV (defined 
for real photons as in Eq.~(\ref{eqn:fq0}) below).
Both the transition form factor squared $|F(Q^2)|^2$ and  
differential cross section $d\sigma/dQ^2$
are proportional to the parameter $g^2$, and thus 
their ratio 
\begin{equation}
2A(Q^2) = \frac{d\sigma/dQ^2}{|F(Q^2)|^2}, 
\label{eqn:aq2}
\end{equation}
which has to be used to convert
the measured differential cross section to the form factor,
is not affected by the assumed $\Gamma_{\gamma\gamma}$ 
value~\cite{bkt}.  
In Eq.(\ref{eqn:aq2}), we explicitly  include a factor of two to
show the contributions from both p- and e-tags
to the coefficient. 
The transition form factor $F(Q^2)$ is defined as 
\begin{equation}
|F(Q^2)|^2 = \lim_{Q_2^2 \to 0} |F(Q^2,Q_2^2)|^2. 
\label{eqn:fq2}
\end{equation}
The normalization of the form factor $F(Q^2,Q_2^2)$
is defined based on the relation 
in Ref.~\cite{cleo}, where it is extrapolated to the real
two-photon coupling ($\gamma\gamma \to \pi^0$), 
\begin{equation}
|F(0,0)|^2 = \frac{1}{(4\pi\alpha)^2}
\frac{64\pi\Gamma_{\gamma\gamma}}{m_\pi^3};
\label{eqn:fq0}
\end{equation}
here $\alpha$ is the fine structure constant and
$\Gamma_{\gamma\gamma}$ 
is the two-photon decay width for the neutral pion.

A calculation of the conversion factor 
$2A(Q^2)$ requires a 3-fold numerical integration,
which is performed over the following variables: 
$Q^2_2$, the photon energy of the small-$Q^2$ side ($\omega_2$),
and the acoplanarity for the two colliding virtual photons ($\Delta \varphi$).
The choice of kinematical variables is discussed in Ref.~\cite{shuler}.

We use the following assumption to extend the form factor 
to non-zero $Q^2_2$:
\begin{equation}
|F(Q^2_1,Q^2_2)| = \frac{|F(Q^2_1,0)|}
{1 + \frac{Q^2_2}{m^2_\rho}},
\end {equation}
where $m_\rho = 0.77$~GeV/$c^2$. The ratio $2A(Q^2)$ is calculated
for the $Q^2$ range of the present measurement 
with a step size of 1~GeV$^2$.
Convergence and stability of the numerical integration 
have been carefully verified to an accuracy of $\sim 10^{-3}$.
We have applied spline interpolation for 
$2A(Q^2)$ when necessary.
The results obtained for $2A(Q^2)$ are plotted in Fig.~\ref{fig:par2a}
for $\sqrt{s} = 10.58$~GeV,
$Q^{2 {\rm max}}_2 = 1.0$~GeV$^2$ and to leading order in QED. 

The event generation is performed using the same
integrand but including initial-state radiation (ISR)
effects from the tag-side electron. 
Inclusion of ISR changes the kinematics and $Q^2_1$ significantly. 
Meanwhile, ISR from the untagged side has little effect
because ISR is nearly parallel not only
to the initial-state electron but also to the final-state
untagged electron. 
We use an exponentiation technique~\cite{expo}
for the photon emission 
based on the parameter 
$\eta = (2\alpha/\pi)(\log(Q^2_1/m_e^2)-1)$ and
the probability density for the photon energy distribution,
$dP(r_k) \propto r_k^{\eta-1} dr_k$, where 
\begin{equation}
r_k \equiv \frac{E^*_{\rm ISR}}{E_{\rm beam}^*}.
\label{eqn:rk1}
\end{equation}
As an approximation, the photon is always 
emitted along the electron direction.
We set the maximum fractional energy of radiation $r_k^{\rm max}=0.25$.
Because events with a large $r_k$, typically above 0.1, 
are rejected by selection criterion (13) using $E_{\rm ratio}$ 
(see Sec.~\ref{sub:check} and Fig.~\ref{fig:rkdisnew}),
this value of $r_k^{\rm max}$ in MC generation is large enough
to generate MC events over the selected range.
In this configuration, 
the correction factor $1+\delta$ to the tree-level 
cross section from the untagged side
is close to unity~\cite{radcor1}.
Radiation of low-$Q^2$ virtual photons is strongly suppressed
by the emission of ISR in the untagged side.

We generate events with a virtuality of 
the tagged-side photon $Q^2_1$ distributed with a constant form factor.
The $Q^2_1$ of each event is modified by ISR. 
We use the momentum of the ISR photon to estimate
the true $Q^2_1$ value in signal MC events
and to study the $Q^2$ dependence of the detection efficiency.
\begin{figure}
\centering
\includegraphics[width=7cm]{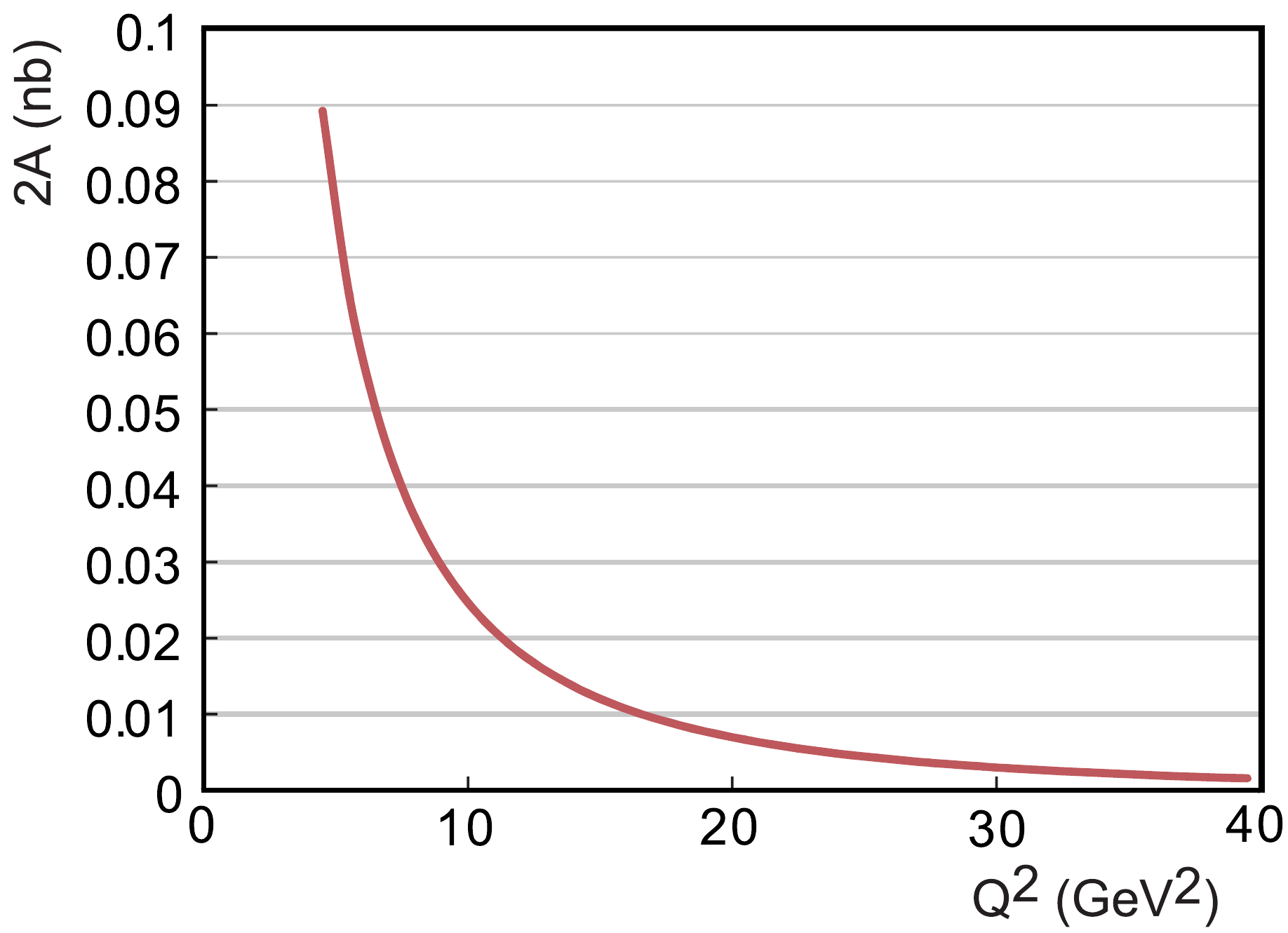}
\centering
\caption{Calculated $2A(Q^2)$ function
for $\sqrt{s}=10.58$~GeV.}
\label{fig:par2a}
\end{figure}

\subsection{Efficiency determination}
\label{sub:effdet}
The selection efficiency is calculated using the signal MC,
the detector simulation (GSIM) and the trigger simulation (TSIM).
The efficiency is estimated for each $Q^2$ bin for p- and e-tags separately. 
We show the $Q^2$ dependence of the efficiency for 
a typical run at the $\Upsilon(4S)$ energy and 
another at the $\Upsilon(5S)$ energy in Figs.~\ref{fig:toteff} and
\ref{fig:trigeff}. 
The overall detection efficiency shown in Fig.~\ref{fig:toteff} 
includes the trigger efficiency and the Bhabha-mask selection.
The trigger efficiency (Fig.~\ref{fig:trigeff})
is shown within the acceptance region
after the Bhabha-mask selection is applied.  

A measurement corresponding to these efficiencies
is reported only in the region 6~GeV$^2 < Q^2 < 40$~GeV$^2$,
where the HiE trigger with the Bhabha mask is used.
The complicated shapes are due to the structure of the Bhabha mask.
The efficiencies are obtained for each $Q^2$ bin,
where the value of $Q^2$ is evaluated using generated
four-momentum vectors of signal MC events with ISR
taken into account.
We generated MC samples of $8 \times 10^6$ events
at $\Upsilon(4S)$ and   $2 \times 10^6$ events
at $\Upsilon(5S)$; these samples correspond, approximately,
to an integrated luminosity
a few dozen times larger than our data.
The efficiency obtained in bins of $Q^2$ is summarized in
Table~\ref{tab:res1}.

To validate the efficiency calculation,
we compare the distributions for the $\pi^0$ laboratory 
angle in signal candidate events with those from the signal
MC samples in different $Q^2$ regions in Fig.~\ref{fig:pi0angcomp}.
Data events are shown in the range of two-photon invariant mass
$0.120~\GeV/c^2 < M_{\gamma\gamma} < 0.145~\GeV/c^2$
after subtracting the background that was estimated  using the $\pi^0$-mass
sideband region $0.170~\GeV/c^2 < M_{\gamma\gamma} < 0.195~\GeV/c^2$
using the fitting procedure described in Sec.~\ref{sub:yield}.
Qualitative agreement between the data and MC in Fig.~\ref{fig:pi0angcomp} 
confirms the validity of the efficiency calculation based on the signal MC.
The validity of this calculation has also been
checked in data using a sample of radiative-Bhabha events,
which have no trigger bias.
The details are described in Appendix~\ref{sec:B}.

\begin{figure*}
\centering
\includegraphics[width=12cm]{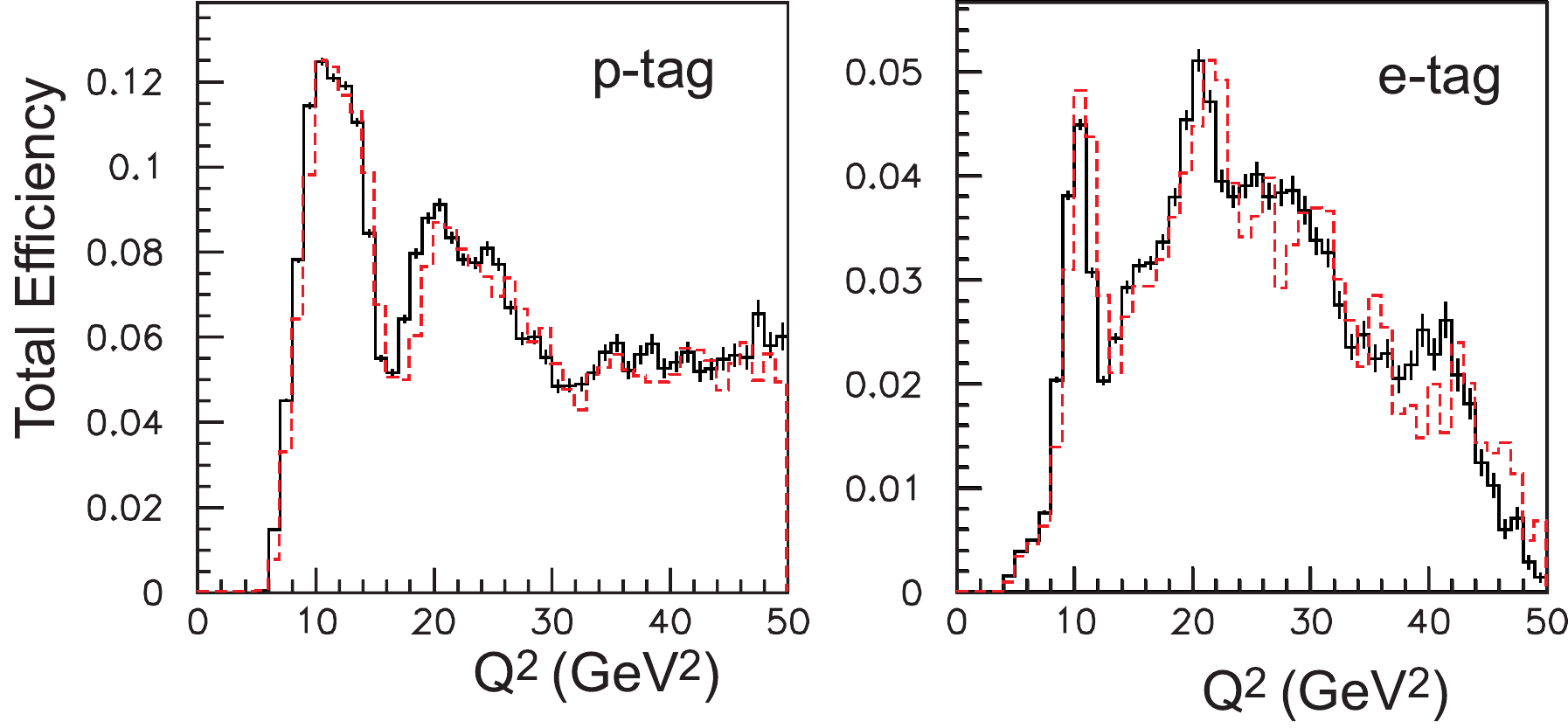}
\centering
\caption{The total selection efficiencies as a function of $Q^2$
determined by MC for runs at the $\Upsilon(4S)$  
(solid histogram, where the error bars show statistical errors)
and $\Upsilon(5S)$ energies (dashed histogram). }
\label{fig:toteff}
\end{figure*}

\begin{figure*}
\centering
\includegraphics[width=12cm]{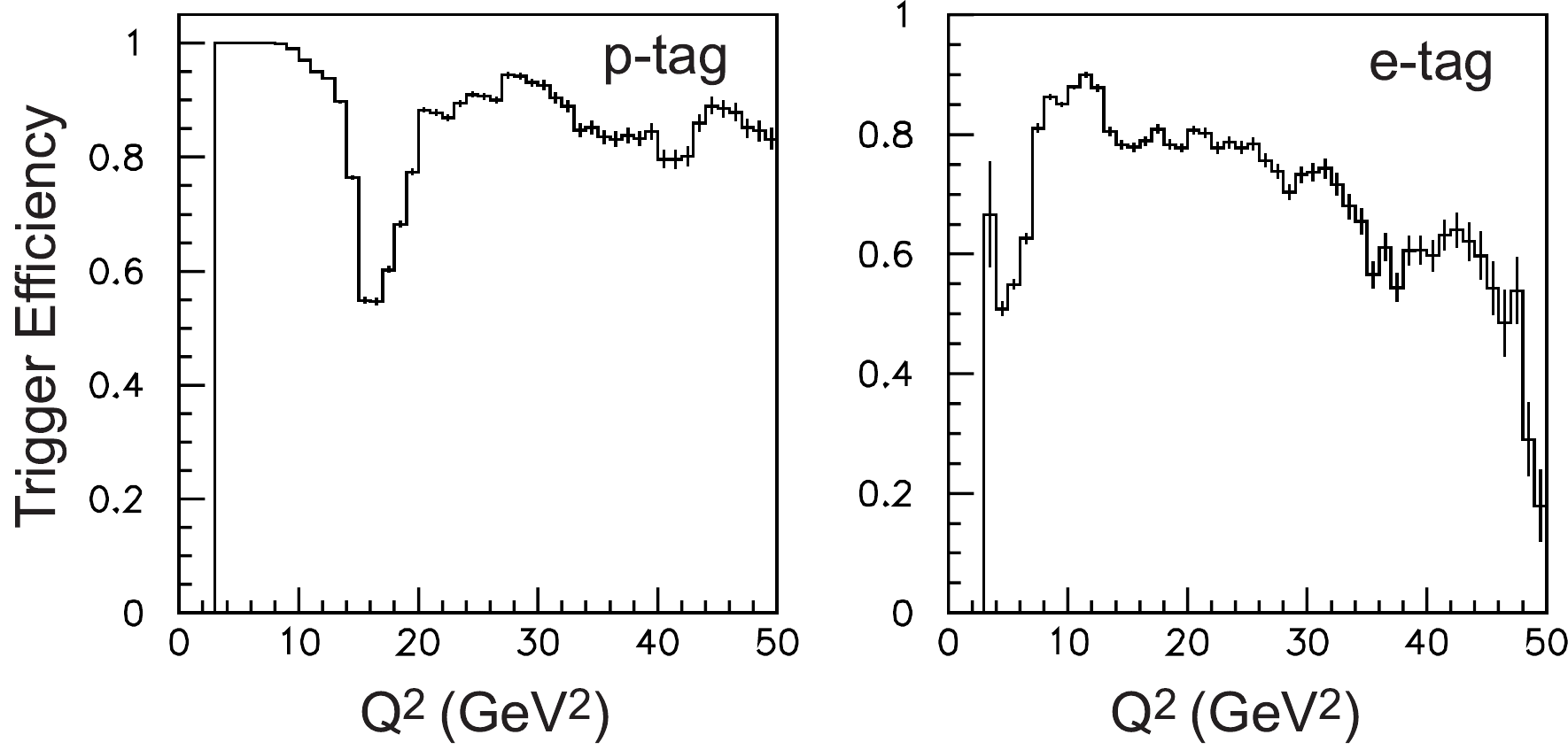}
\centering
\caption{The trigger efficiencies determined
from MC for
a typical run at the $\Upsilon(4S)$ energy. 
The error bars show statistical errors.
}
\label{fig:trigeff}
\end{figure*}

\begin{figure}
\centering
\includegraphics[width=6cm]{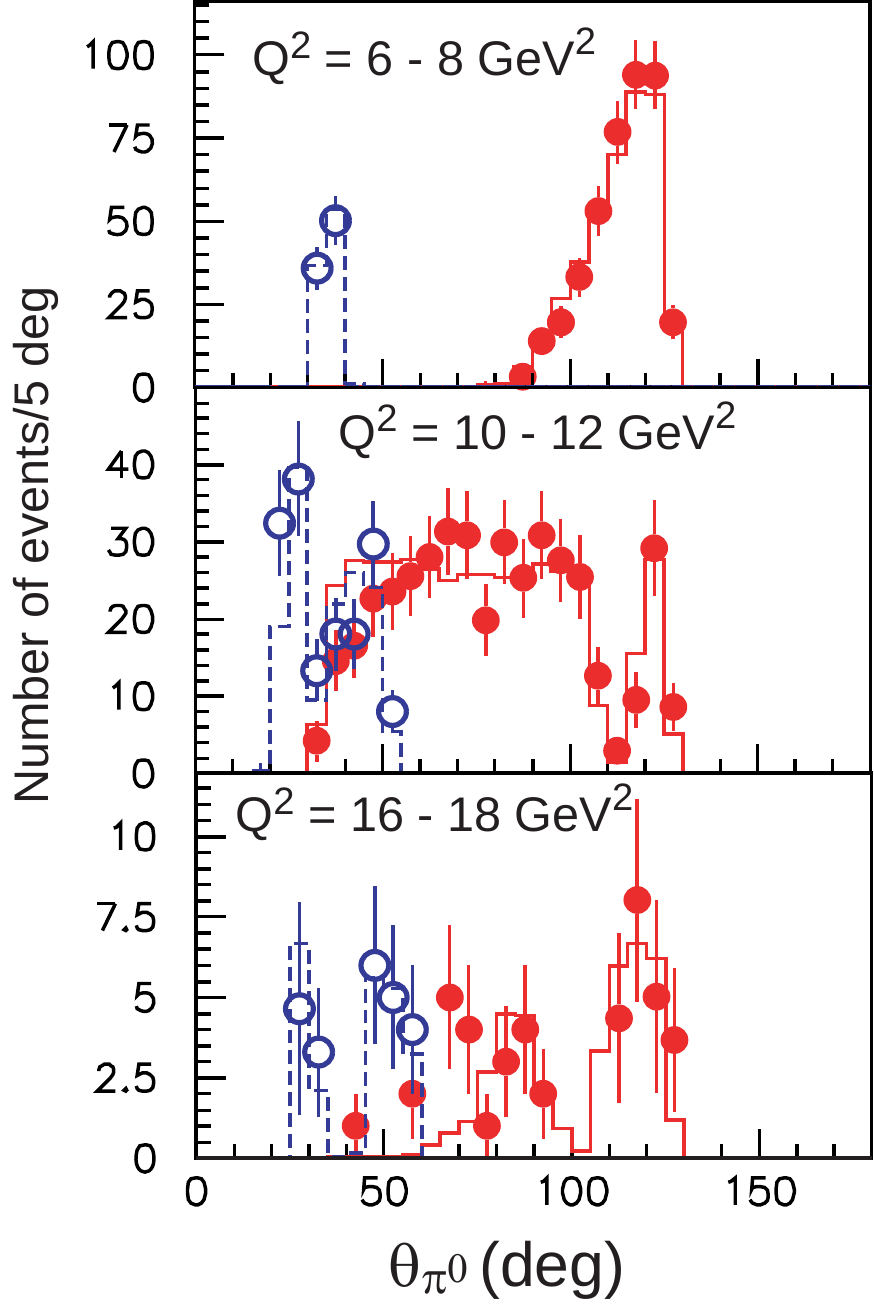}
\centering
\caption{
Comparison between the $\pi^0$ angular distributions in signal
candidate events in data (circles) and signal MC
(histograms) for different $Q^2$ regions. 
Closed (open) circles and solid (dashed) histograms are 
from events of p-tag (e-tag).
In the data, the background contribution is subtracted.
The MC events are normalized to the experimental 
yield after the background subtraction 
in each $Q^2$ region. 
} 
\label{fig:pi0angcomp}
\end{figure}

\section{Extraction of the $\pi^0$ yield and background estimation}
\label{sec:yield}
In deriving the differential cross section, $d \sigma/d Q^2$,
we first extract the $\pi^0$ yield in each $Q^2$ bin by fitting 
the $M_{\gamma \gamma}$ distribution.
Possible backgrounds are then identified, evaluated and subtracted.
These analysis steps are described below.

\subsection{Extraction of the $\pi^0$ yield}
\label{sub:yield}
The $\pi^0$ yield is extracted 
in each $Q^2$ bin for the p-tag and e-tag samples.
For each signal candidate event, $Q^2$
is calculated using the measured four-momentum of the detected
electron ($p_e$) by 
\begin{eqnarray}
Q^2_{\rm rec} &=& - (p_{\rm beam} - p_e)^2 \nonumber \\
&=& 2 E_{\rm beam}^* E_e^*(1 + q_{\rm tag} \cos \theta_e^*),
\nonumber \\ 
\label{eqn:q2meas}
\end{eqnarray}
where $p_{\rm beam}$ is the
nominal four-momentum of the beam particle with the same charge
as the detected electron; the right-hand side is given by
the beam energy $E_{\rm beam}^*$ and the observables of the tagged
electron in the $e^+e^-$ c.m. frame.
We do not apply a correction for 
ISR on an event-by-event basis; instead, this 
effect is taken into account in the $Q^2$ unfolding
described in Sec.~\ref{sub:unfold}.

Figure~\ref{fig:q2dis} shows scatter plots of $Q^2$ vs.
$M_{\gamma\gamma}$ for the final experimental samples. 
After dividing them into $Q^2$ bins,
we fit the $M_{\gamma\gamma}$ distribution in the region
0.08~GeV/$c^2 < M_{\gamma\gamma} < 0.30$~GeV/$c^2$ to determine
the $\pi^0$ yield. 

To extract this yield,
we fit the two-photon invariant mass distribution
with a double Gaussian describing the signal
and a quadratic polynomial approximating the background, namely
\begin{eqnarray}
f(x) &=& a + bx + cx^2 + \frac{A}{\sqrt{2\pi}\sigma}\{r 
e^{-\frac{(x-m)^2}{2\sigma^2}} \nonumber \\
&& {}+ \frac{1-r}{k} e^{-\frac{\{x-(m+\Delta m)\}^2}{2(k\sigma)^2}} \},
\label{eqn:fitfn}
\end{eqnarray}
where $x \equiv M_{\gamma\gamma}$, $a$, $b$ and $c$ are the background parameters, 
$A$ is signal yield, and 
$m$ and $\sigma$ are the mass and width ({\it i.e.}, mass resolution),
respectively, for the signal. For $Q^2$ bins with poor statistics, the
last two fit parameters are fixed to values estimated from signal MC, 
as described below. 
The parameters $r, k$ and $\Delta m$ approximate the
relative size, width and position of the two Gaussian
components, respectively, and are fixed to
values from MC simulation in a $Q^2$-dependent manner. 

In $Q^2$ bins with low signal statistics, the determination of the 
width is poor and is significantly correlated with the yield.
For such bins, we fix the mass and width parameters to values
that are expected from the signal MC, taking into account a
small deviation measured using the
data: the difference between data and MC is
$\delta m \simeq -1$~MeV/$c^2$, whereas
the mass resolutions show good agreement.
These parameters are fixed in the fit for $Q^2 > 20~\GeV^2$
for the p-tag, and $Q^2 = 7-8~\GeV^2$ or $Q^2 > 12~\GeV^2$ for the HiE
and $Q^2 < 6~\GeV^2$ for the CsiBB for the e-tag.
The systematic uncertainty induced by fixing these parameters is 
dominated by the large statistical error on
the yield difference between the constrained and non-constrained fits. 
We estimate this uncertainty by comparing the results obtained
using two different sets of fit
functions, as described in Sec.~\ref{sub:syserr}. 

The following values for the fixed parameters were obtained from the MC: 
$m = 134 - 137$~MeV/$c^2$, $\sigma = 6 - 9$~MeV/$c^2$,
$\Delta m = -5.5 - 0.0$~MeV/$c^2$, $1-r = 0.11 - 0.42$
depending on $Q^2$ and the p- or e-tag; $k$ is fixed at 2.4.
The absolute values for these parameters tend to be larger 
when $Q^2$ is large.
There are no significant anomalous tails 
in the signal $M_{\gamma\gamma}$ distribution down to 
the ${\cal O}(10^{-2})$ level of the peak, according to the signal MC,
as shown in Fig.~\ref{fig:pi0mc}.
Figure~\ref{fig:fit1}
shows the results of the fit to data in five representative $Q^2$ bins.
The yields of the fits are summarized in Table~\ref{tab:res1}
and are labeled $N_{\rm rec}$.

\begin{center}
\begin{table*}
\caption{Numbers of reconstructed events ($N_{\rm rec}$)
obtained from the fit to the experimental $M_{\gamma\gamma}$
distribution (Sec.~\ref{sub:yield}),
the number of events after the unfolding ($N_{\rm cor}$),
background fractions and efficiencies for 
each $Q^2$ bin, listed separately for p- and e-tags.
The first column is the representative $Q^2$ for that bin 
(see Eq.~(\ref{eqn:q2form})).
$N_{\rm rec}$ and $N_{\rm cor}$
include both signal and background yields. 
Note that the ``${\rm HiE}+50*{\rm CsiBB}$'' samples are used for
the $Q^2 = 4-6~\GeV^2$ bins in the e-tag case.
}
\label{tab:res1}
\begin{tabular}{ccc|cccc} \hline \hline
$Q^2$ (GeV$^2$) & $Q^2$ bin range (GeV$^2$) & tag & $N_{\rm rec}$ &  
$N_{\rm cor}$& background (\%) & Efficiency \\
\hline
6.47	 & 	6.0 - 7.0	 & 	p	 & $	140 	 \pm	14 	$ & $	167 	 \pm	17 	$ & 	2.8 	 & 	0.0134 	 \\
7.47	 & 	7.0 - 8.0	 & 	p	 & $	320 	 \pm	21 	$ & $	323 	 \pm	25 	$ & 	2.8 	 & 	0.0434 	 \\
8.48	 & 	8.0 - 9.0	 & 	p	 & $	418 	 \pm	23 	$ & $	426 	 \pm	27 	$ & 	2.8 	 & 	0.0763 	 \\
9.48	 & 	9.0 - 10.0	 & 	p	 & $	379 	 \pm	21 	$ & $	364 	 \pm	25 	$ & 	2.8 	 & 	0.1120 	 \\
10.48	 & 	10.0 -11.0	 & 	p	 & $	286 	 \pm	19 	$ & $	279 	 \pm	22 	$ & 	2.8 	 & 	0.1260 	 \\
11.48	 & 	11.0 - 12.0	 & 	p	 & $	222 	 \pm	17 	$ & $	222 	 \pm	21 	$ & 	2.8 	 & 	0.1220 	 \\
12.94	 & 	12.0 - 14.0	 & 	p	 & $	259 	 \pm	19 	$ & $	249 	 \pm	22 	$ & 	3.0 	 & 	0.1150 	 \\
14.95	 & 	14.0 - 16.0	 & 	p	 & $	105 	 \pm	13 	$ & $	98 	 \pm	14 	$ & 	3.0 	 & 	0.0724 	 \\
16.96	 & 	16.0 - 18.0	 & 	p	 & $	49 	 \pm	9 	$ & $	49 	 \pm	11 	$ & 	3.0 	 & 	0.0571 	 \\
18.96	 & 	18.0 - 20.0	 & 	p	 & $	52 	 \pm	9 	$ & $	51 	 \pm	12 	$ & 	3.0 	 & 	0.0819 	 \\
22.29	 & 	20.0 - 25.0	 & 	p	 & $	82 	 \pm	13 	$ & $	78 	 \pm	15 	$ & 	3.5 	 & 	0.0826 	 \\
27.33	 & 	25.0 - 30.0	 & 	p	 & $	44 	 \pm	9 	$ & $	44 	 \pm	10 	$ & 	3.5 	 & 	0.0648 	 \\
34.46	 & 	30.0 - 40.0	 & 	p	 & $	8.4	 \pm	6.8	$ & $	8.4 	 \pm	6.8 	$ & 	5.0 	 & 	0.0550 	 \\
\hline
4.46	 & 	4.0 - 5.0	 & 	e	 & $	656 	 \pm	190 	$ & $	680 	 \pm	197 	$ & 	2.8 	 & 	0.0231 	 \\
5.47	 & 	5.0 - 6.0	 & 	e	 & $	1066 	 \pm	230 	$ & $	1072 	 \pm	231 	$ & 	2.8 	 & 	0.0630 	 \\
6.47	 & 	6.0 - 7.0	 & 	e	 & $	60 	 \pm	9 	$ & $	58 	 \pm	8 	$ & 	2.8 	 & 	0.0049 	 \\
7.47	 & 	7.0 - 8.0	 & 	e	 & $	38.0 	 \pm	7.0 	$ & $	36.9 	 \pm	9.1 	$ & 	2.8 	 & 	0.0073 	 \\
8.48	 & 	8.0 - 9.0	 & 	e	 & $	84 	 \pm	11 	$ & $	93 	 \pm	15 	$ & 	2.8 	 & 	0.0191 	 \\
9.48	 & 	9.0 - 10.0	 & 	e	 & $	149 	 \pm	16 	$ & $	159 	 \pm	21 	$ & 	2.8 	 & 	0.0365 	 \\
10.48	 & 	10.0 -11.0	 & 	e	 & $	115 	 \pm	13 	$ & $	104 	 \pm	16 	$ & 	2.8 	 & 	0.0447 	 \\
11.48	 & 	11.0 - 12.0	 & 	e	 & $	53 	 \pm	9 	$ & $	45 	 \pm	12 	$ & 	2.8 	 & 	0.0322 	 \\
12.94	 & 	12.0 - 14.0	 & 	e	 & $	39 	 \pm	8 	$ & $	37 	 \pm	10 	$ & 	3.0 	 & 	0.0224 	 \\
14.95	 & 	14.0 - 16.0	 & 	e	 & $	41 	 \pm	8 	$ & $	41 	 \pm	10 	$ & 	3.0 	 & 	0.0296 	 \\
16.96	 & 	16.0 - 18.0	 & 	e	 & $	30.6 	 \pm	7.2 	$ & $	30.5 	 \pm	8.9 	$ & 	3.0 	 & 	0.0319 	 \\
18.96	 & 	18.0 - 20.0	 & 	e	 & $	31.0 	 \pm	7.8 	$ & $	31.7 	 \pm	9.8 	$ & 	3.0 	 & 	0.0406 	 \\
22.29	 & 	20.0 - 25.0	 & 	e	 & $	32.4 	 \pm	9.0 	$ & $	29.2 	 \pm	9.9 	$ & 	3.5 	 & 	0.0426 	 \\
27.33	 & 	25.0 - 30.0	 & 	e	 & $	13.6 	 \pm	6.4 	$ & $	13.1 	 \pm	7.2 	$ & 	3.5 	 & 	0.0375 	 \\
34.46	 & 	30.0 - 40.0	 & 	e	 & $	14.0 	 \pm	5.0 	$ & $	13.4 	 \pm	4.8 	$ & 	5.0 	 & 	0.0262 	 \\
\hline
\hline
\end{tabular}
\end{table*}
\end{center}

\begin{figure*}
\centering
\includegraphics[width=13cm]{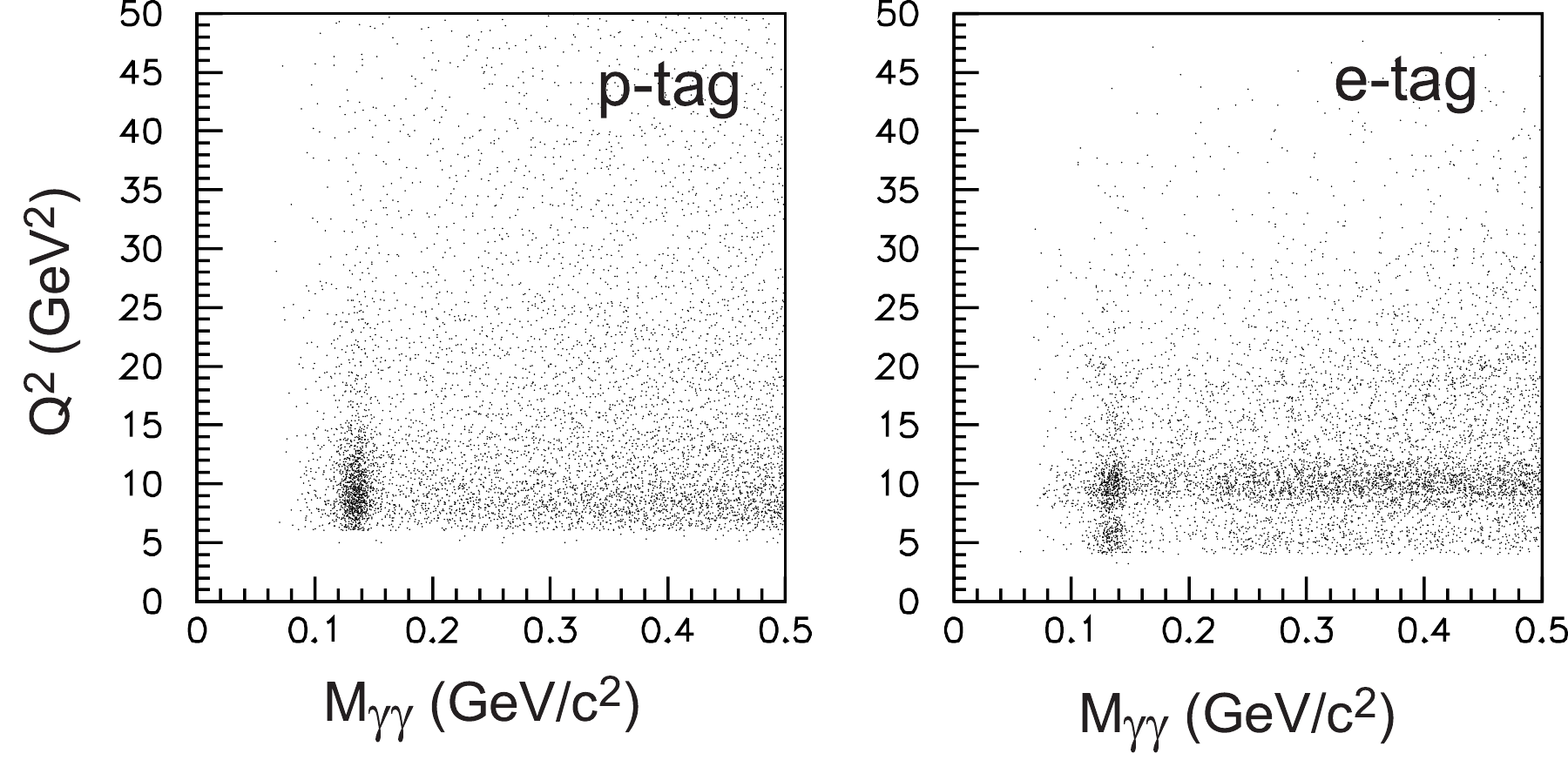}
\centering
\caption{Scatter plots of $Q^2$ vs. $M_{\gamma \gamma}$ 
for the final experimental samples, shown separately
for the p-tag (left) and e-tag (right) events.}
\label{fig:q2dis}
\end{figure*}

\begin{figure*}
\centering
\includegraphics[width=13cm]{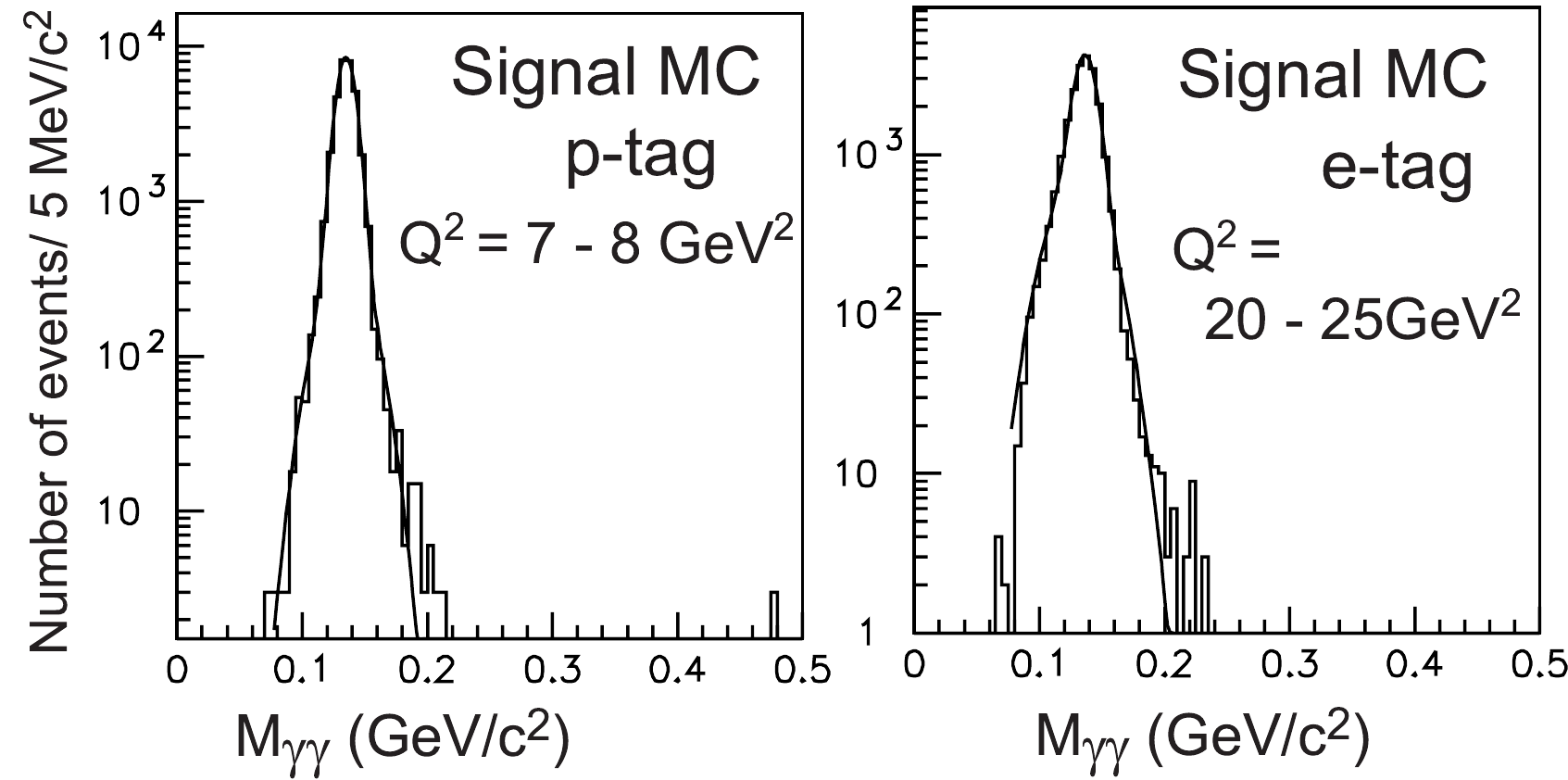}
\centering
\caption{$M_{\gamma \gamma}$ distributions for the signal
MC events for the selected $Q^2$ ranges for 
the p-tag (left, 7~GeV$^2 < Q^2 < 8$~GeV$^2$) 
and e-tag (right, 20~GeV$^2 < Q^2 < 25$~GeV$^2$) samples.
The curves are fits with a double-Gaussian function
without background terms.
}
\label{fig:pi0mc}
\end{figure*}

\begin{figure*}
\centering
\includegraphics[width=14cm]{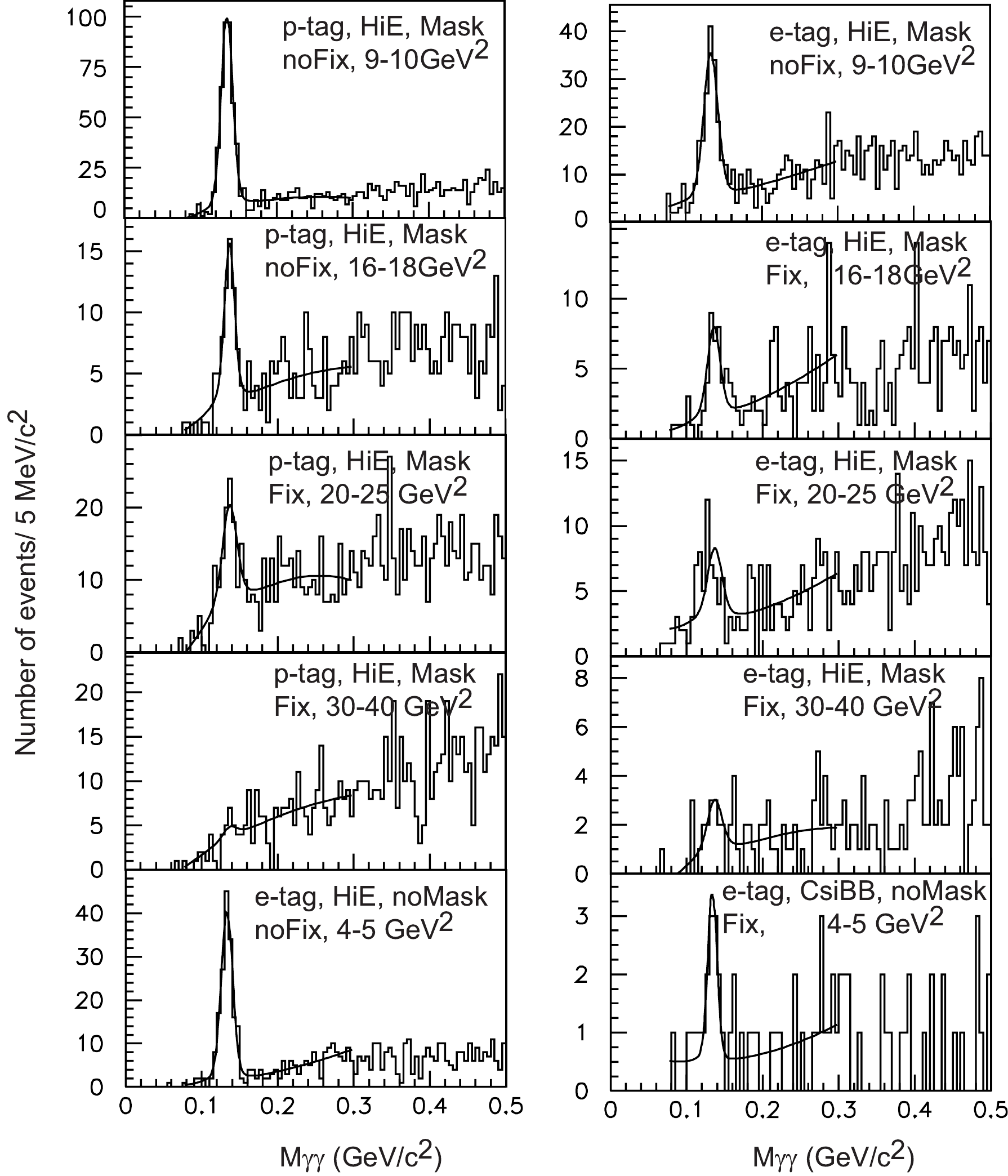}
\centering
\caption{Results of the fit to the experimental $M_{\gamma\gamma}$
distributions for representative $Q^2$ bins. 
The caption in each subfigure shows the type of tag (p- or e-tag), 
the required trigger,
whether or not the Bhabha mask was applied (upper line), whether or not
the mass and resolution parameters ($m$ and $\sigma$) are fixed in the fit
and the $Q^2$ range (lower line).
}
\label{fig:fit1}
\end{figure*}

\subsection{Estimation of background contamination}
\label{sub:bkg}
The measured $\pi^0$ yield could contain some contribution
from background processes in which real $\pi^0$'s are present
in the final state.
In this subsection, we explain how these resonant backgrounds
are estimated and subtracted.
First, possible backgrounds from $e^+ e^-$ annihilation processes 
are examined.
A potentially large background from radiative Bhabha events 
is also checked.
Finally, we study the background contribution from the processes
$\gamma \gamma^* \to \pi^0 \pi^0$ and $e^+ e^- \to (e)e \rho^0/\omega ,
\; \rho^0/\omega \to \pi^0 \gamma$.

\subsubsection{Backgrounds from $e^+ e^-$ annihilation processes}
 Backgrounds from $e^+e^-$ annihilation are 
estimated by examining the yield of wrong-sign events,
in which the tagged electron has the opposite sign to
that expected in 
selection criterion (12). 
The electron charge in the final state of the annihilation processes 
should have no significant forward-backward asymmetry. 
The contamination of this kind of 
background in the signal sample is expected to be
the same as the number of events found in the wrong-sign samples.

We have studied the wrong-sign events by examining the parameter
$E_{\rm ratio}$, whose distribution is shown in Fig.~\ref{fig:eratiobkg}(a).
We find that a small number of events peak near $E_{\rm ratio}=1$
for the wrong-sign sample.
However, there is no visible $\pi^0$ peak in the $M_{\gamma\gamma}$ 
distribution in any $Q^2$ regions as shown in Fig.~\ref{fig:eratiobkg}(d).  
The yield in the $\pi^0$ peak is $1.2 \pm 0.9$~events in the sum of 
p- and e-tags with $Q^2<40$~GeV$^2$ for the total sample.
We conclude that
the neutral pions from $e^+e^-$ annihilation processes
do not contaminate the signal region $E_{\rm ratio} \simeq 1$, 
and the main background near $E_{\rm ratio} = 1$
is from the QED process $e^+e^- \to e^+ e^- \gamma \gamma$,
which is of order $\alpha^4$. 
Although this process has a charge asymmetry, 
this is not a problem for background estimation
since it does not peak at the pion mass.

We also conclude that any hadron or muon misidentified as a tagged electron 
is not accompanied by neutral pions and does not contribute to the
peaking backgrounds.
Hadrons or muons would appear with the same
probability in the wrong-sign sample 
if they were significant in the right-sign signal sample.
As seen in  Fig.~\ref{fig:eratiobkg}(b,c), the tail of
the $E_{\rm ratio}$ distribution near the lower selection
boundary (0.85) is dominated by the ISR tail of the signal process, which
is accounted for in the signal MC down to $r_k=0.25$. 
Some contributions from the peaking backgrounds ($(e)e\pi^0 X$) 
discussed in Sec.~\ref{sub:bkg}.3 and 4 will also be present.

\begin{figure*}
\centering
\includegraphics[width=9cm]{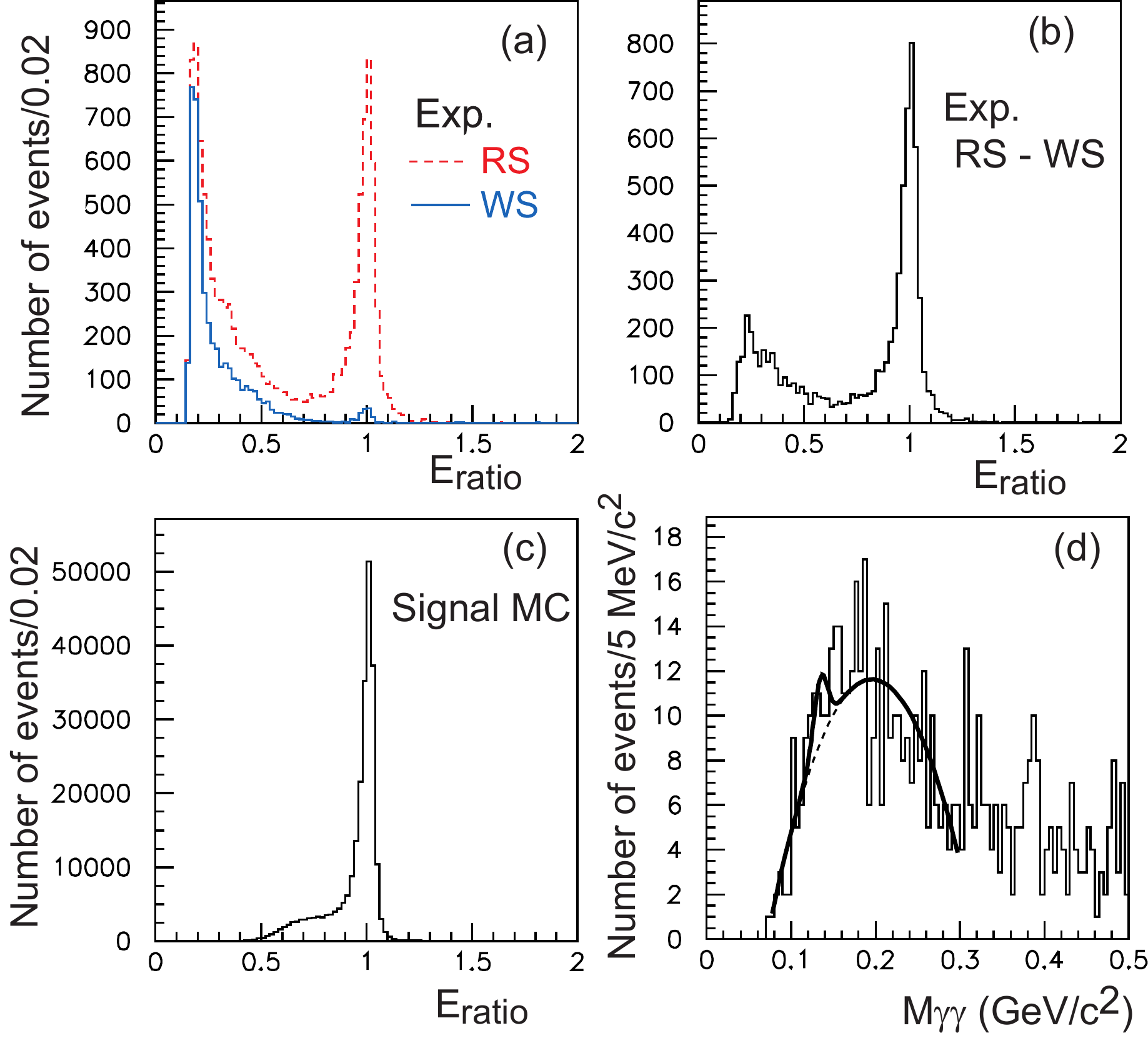}
\centering
\caption{
(a) The $E_{\rm ratio}$ distribution for signal candidates 
for right-sign (RS, dashed line) and
wrong-sign (WS, solid line) events. 
(b) The difference between
RS events and WS events. 
(c) The $E_{\rm ratio}$ distribution for the signal MC sample.
(d) The $M_{\gamma\gamma}$ distribution of wrong-sign events with 
the best fit superimposed (solid line) where the $\pi^0$ peak  
and the background contribution (dashed line) are taken into account
in the fit.
}
\label{fig:eratiobkg}
\end{figure*}

\subsubsection{Radiative Bhabha backgrounds}
Estimation of the radiative Bhabha backgrounds where the photon
is misreconstructed as a neutral pion is challenging.
While the probability of such misreconstruction is small, the cross
section of this background process
is much greater than that of the signal process. 
A misidentification rate as small as ${\cal O}(10^{-6})$ can
lead to background contamination.
However, as described below, even in a conservative estimate, the background
is negligibly small, safely less than one event
in each high $Q^2$ bin. 

The photon conversion process, $\gamma \to e^+ e^-$, is the main source
of radiative Bhabha background, but
almost all such events are rejected by 
selection criterion (11). 
Moreover, conversions can only give a broad distribution in 
$M_{\gamma \gamma}$. 
We estimate the background
level to be negligibly small for $Q^2<20$~GeV$^2$ by extrapolating
the $E_{\gamma\gamma}\Delta \theta$ distribution.
Figure~\ref{fig:radbhatail} shows this distribution for the 
experimental data and for signal and background MC, where
the background contribution falls exponentially
while the signal has a nearly flat distribution. 

To check for additional background arising from radiative Bhabha events,
we study the possibility that some pions from a secondary interaction
or beam background overlap with a radiative Bhabha event
where a photon escapes detection
($e^+ e^- \to (e)e (\gamma) + \pi^0({\rm overlap})$). 
Note that there can be a solution satisfying the required 
three-body signal kinematics even if the photon
is emitted at a forward angle. 
However, in this case, 
the background would have a broad $E_{\rm ratio}$ distribution
concentrated at zero because the secondary $\pi^0$'s are
dominantly of low energy.
There is a depletion of 
events around 0.8 in the $E_{\rm ratio}$ distribution as shown, {\it e.g.}, in
Fig.~\ref{fig:eratiobkg}(a), even for the high $Q^2$ regions;
we conclude that the contribution from this type
of background is negligible.

There is a small probability that a photon from a radiative Bhabha
converts to a neutral pion at the beam pipe or the central 
part of the detector, where the material thickness is kept to a minimum,
via a soft nuclear interaction $\gamma A \to \pi^0 A'$.
The pion is
produced at forward angles with a small momentum transfer;
the nucleus $A$ can break up with an energy
slightly larger than the binding energy.  
This cross section is poorly known, but we estimate that it is 
smaller than that for the elementary process 
$\gamma p \to \pi^0 p$ per nucleon. 
According to the previous measurements~\cite{lbbook}, the
cross section for scattering
angles below 10$^\circ$ is about 0.04~$\mu$b 
for incident photon energies between 2 and 9~GeV.
This cross section corresponds to a probability of $1.0 \times 10^{-7}$
 for a photo-nuclear conversion to a neutral pion with a similar energy.
From the observed number of radiative Bhabha events,
we estimate that the contamination is at most 0.5~event
in the highest $Q^2$ bin and negligible for all other bins.
We thus conclude that the background from this source in all $Q^2$ bins
is small, and assign  3\% and 1\% uncertainties only to  the two 
highest $Q^2$ bins.

\begin{figure*}
\centering
\includegraphics[width=13cm]{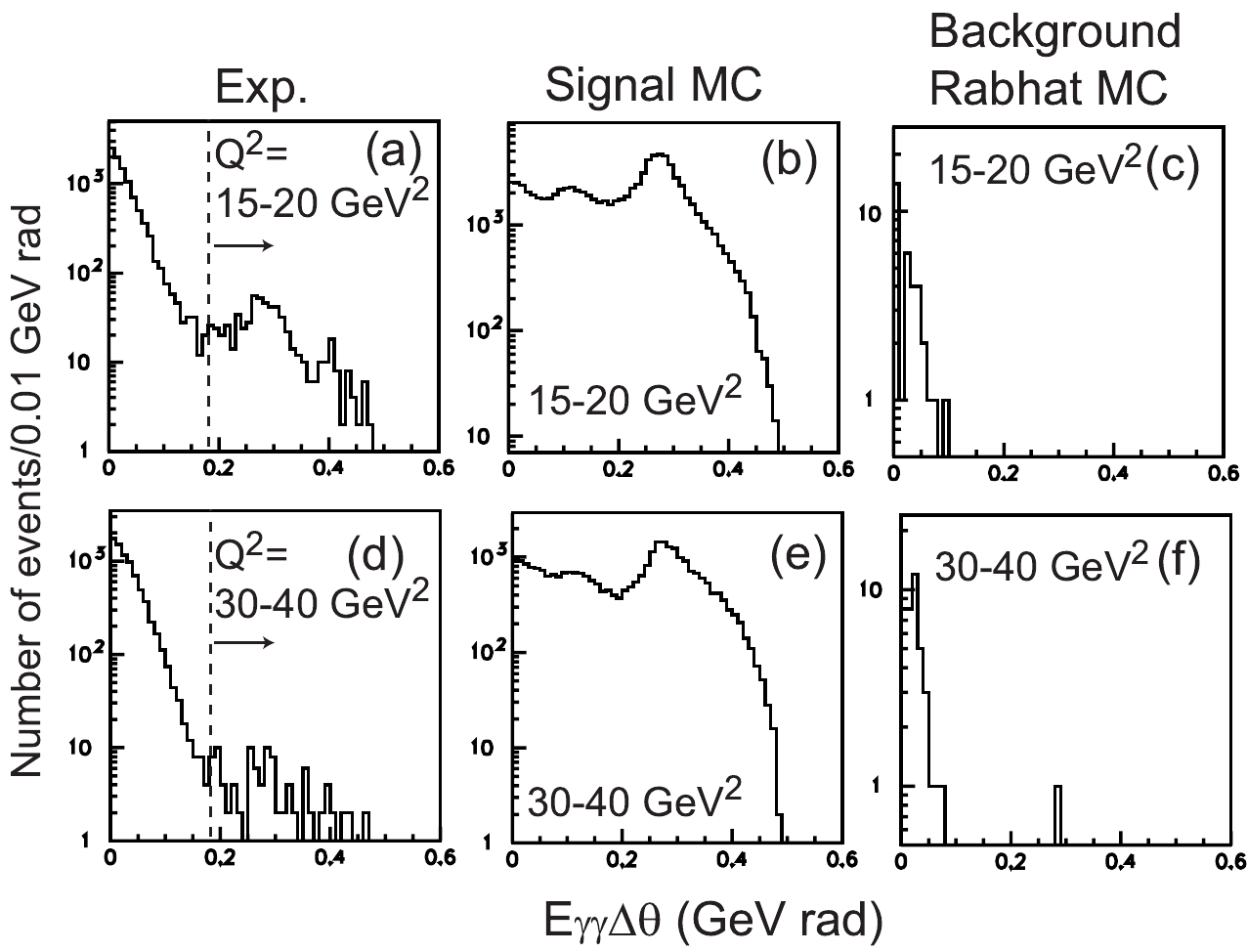}
\centering
\caption{
The $E_{\gamma\gamma}\Delta \theta$ distributions
for the (a,d) experimental, (b,e) signal MC, and (c,f)
radiative Bhabha background MC samples, for two different 
$Q^2$ regions indicated in each subfigure. 
The dashed line shows the selection boundary.
Arrows show the accepted regions.
The background MC sample corresponds to an integrated 
luminosity 140 times smaller than the experimental data.
}
\label{fig:radbhatail}
\end{figure*}

\subsubsection{Background from $\gamma \gamma^* \to \pi^0 \pi^0$}
Processes with one extra particle, {\it i.e.},
$e^+e^- \to (e) e \pi^0 X$,
where $X$ is another $\pi^0$ or photon, could also be
a source of resonant background that would have to be
subtracted from the measured $\pi^0$ yield in each bin of $Q^2$.

This type of background may have both a charge asymmetry that 
favors the right-sign and a peak at the $\pi^0$ mass. 
For large $X$ invariant masses, this background is suppressed 
by the $E_{\rm ratio}$ selection criterion.
However, if the invariant
mass and the momentum of $X$ are small, $E_{\rm ratio}$ 
approaches unity, 
and we cannot use it to distinguish between signal and 
background processes.

To study such background, we use a rather loose set of selection 
criteria to positively identify $e^+e^- \to (e) e \pi^0 \pi^0$ candidates
that contribute in a wide kinematical range.
There should be four or more photons in the final state; 
we accept the extra pion with a loose energy constraint.
We apply a selection requirement on the energy of 
the $\pi^0 \pi^0$ system using an energy
ratio similar to $E_{\rm ratio}$ to enhance three-body kinematical
configurations.  
We take the ratio of the observed energy to the expected energy calculated
using the observed invariant mass for the $\pi^0 \pi^0$ system.

An MC generator was developed for this background process
with EPA to reproduce the $W (\equiv M_{\pi\pi})$ and
c.m.-scattering-angle distributions. 
We assume that the angular distribution is the same as that in the no-tag
process~\cite{pi0pi0}, inspired by the observed distribution. 
The $W$ distribution is tuned to the experimental data. 
Figure~\ref{fig:pi0pi0bkga} shows
the experimental data and the background MC
in a two-dimensional plot of
the $\pi^0 \pi^0$ invariant mass and cosine of the scattering
angle ($\theta^*$) of the $\pi^0$ in the $\pi^0 \pi^0$ rest frame
with respect to the  $\gamma\gamma^*$ collision axis.
The $Q^2$ dependence of the cross section is 
tuned to reproduce the data before determining
the background contribution for each $Q^2$ bin. 
We normalize the background size
to the observed background yield.
We find that events from the signal process,
$(e)e \pi^0$ with misidentified photons, produce a background
in the sample of $(e)e \pi^0 \pi^0$ at extremely
forward or backward scattering angles 
near an invariant mass of 1~GeV/$c^2$.
We subtract this component in the evaluation of
the background. In total, 1305 events are found for the process
$e^+ e^- \to (e)e \pi^0 \pi^0$ with
$|\cos \theta^*| < 0.9$.

Using $e^+ e^- \to (e)e \pi^0 \pi^0$ MC, the background
contribution from this process
is estimated to be about 2\% of the signal yield, with 
a negligible $Q^2$ dependence. 
This is accounted for as a background
factor in the calculation of the cross section for the signal process.

\begin{figure}
\centering
\includegraphics[width=7.9cm]{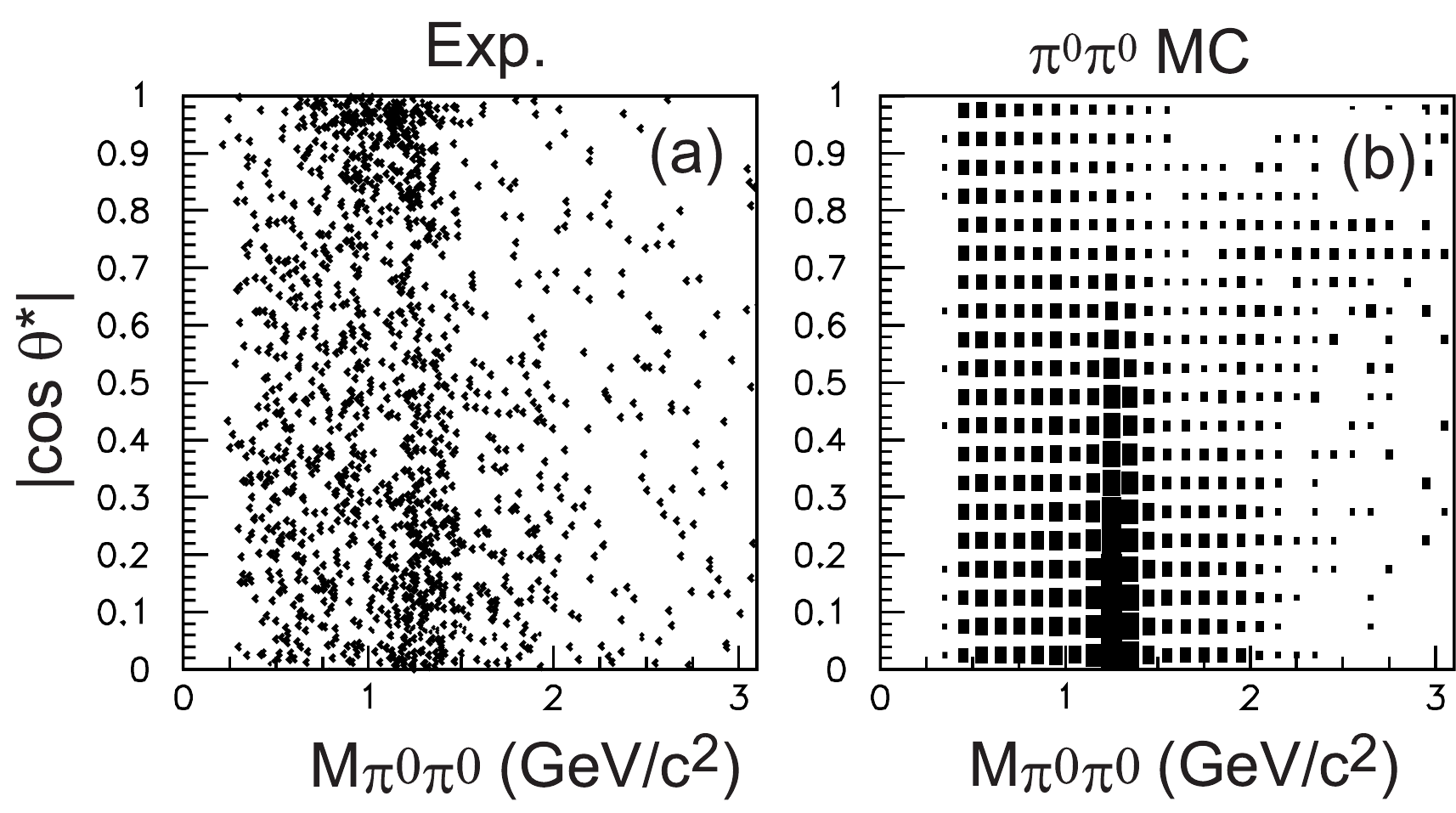}
\centering
\caption{
Two dimensional plots of 
$|\cos \theta^*|$ versus invariant mass of the $\pi^0\pi^0$ system 
for the selected $(e)e \pi^0 \pi^0$ events
(a) for the experimental data and (b) for the $(e)e \pi^0 \pi^0$ MC samples.
}
\label{fig:pi0pi0bkga}
\end{figure}

\subsubsection{Background from 
$e^+e^- \to (e)e\rho^0/\omega, \;  \rho^0/\omega \to \pi^0\gamma$}
A similar background from
$e^+e^- \to (e)e\rho^0/\omega$, $\rho^0/\omega \to \pi^0 \gamma$
is also studied.
This process can contribute to background through a virtual 
pseudo-Compton process, where the final-state photon is replaced 
by a neutral vector meson.
We have searched for events that could be described by
this process by requiring an extra detected photon
with a loose energy constraint.
We find about 180 events with the
$\pi^0 \gamma$ invariant mass peaking at the nominal $\omega$
mass.  Figure~\ref{fig:gpi0bkga} shows
the experimental data in a two-dimensional plot of
the $\pi^0 \gamma$ invariant mass versus the cosine of the scattering
angle ($\theta^*$) of the $\pi^0$ in the $\pi^0 \gamma$ rest frame
with respect to the  $\pi^0 \gamma$ momentum direction.
Events with an  $\omega$ dominate,
although it is difficult to separate them from a possible
$\rho^0$ contribution.
In Fig.~\ref{fig:gpi0bkga}, the horizontal band seen at 
$|\cos \theta^*| >0.9$ is from
the signal process ($e^+ e^- \to (e) e \pi^0$) with a misidentified photon.
This contribution is removed when we estimate the 
$\rho^0 / \omega \to \pi^0 \gamma$ background.

We simulate this background in
MC by replacing the $e\gamma$ system in Rabhat by an
$e\omega$ system with the same invariant mass and let them
scatter at the same angle in the $e\gamma$ (thus, $e\omega$) 
rest frame. 
We normalize the MC sample to the observed
number of background events and estimate the background
to the signal process. 
The result is $Q^2$ dependent
because the background process is not a two-photon process; it
varies from 0.8\% for the low $Q^2$ bin to 3\% for the
highest $Q^2$ bin.
 This is accounted for as a background
factor in the derivation of the cross section for
the signal process.

The background yield can be 
estimated by a GVDM (generalized vector-meson dominance model) factor
in which a photon is replaced by a neutral vector meson, with 
branching fractions (BF's) for 
$\rho^0/\omega \to \pi^0\gamma$ decays. 
The GVDM factors for the $\rho^0$ and $\omega$ are assumed to be 1/300 
and 1/3000, respectively~\cite{sakurai}.
Multiplying them by the BF's to $\gamma\pi^0$,  
we obtain a  factor of 1/32000 for the probability of the background
process compared to the VC process.
With the estimated efficiency of $1.5 \times 10^{-4}$ 
from the background MC, the expected background yield from the GVDM
is 120 events.
This estimate is within a factor of two of the observed background.

\begin{figure}
\centering
\includegraphics[width=6.5cm]{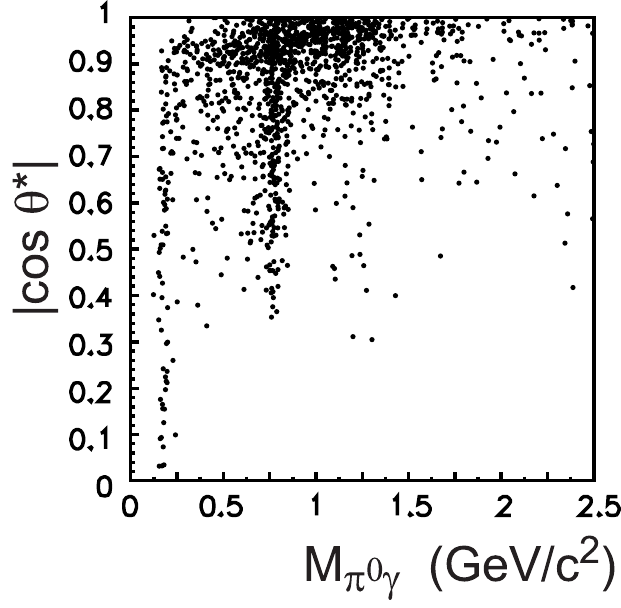}
\centering
\caption{
 Scatter plot for the selected $(e)e\pi^0\gamma$
events: $|\cos \theta^*|$ vs. invariant mass of the $\pi^0\gamma$ 
(see the text for the definition of the $\theta^*$).
}
\label{fig:gpi0bkga}
\end{figure}

We conclude that
the backgrounds discussed in subsections B.1 and B.2 are negligible.
The total contribution from the backgrounds described
in B.3 and B.4 is shown in column 6 of Table~\ref{tab:res1}.

\section{Measurement of differential cross section and 
transition form factor}
\label{sec:dsdq}
In this section,
we first describe the method of $Q^2$ unfolding and determination
of the representative $Q^2$ value for each $Q^2$ bin.
We then outline how the cross section is measured.
This is followed by the evaluation of the systematic uncertainties 
and some cross-checks.
The transition form factor for $\pi^0$ is then determined.
Finally, the observed TFF is compared with previous experiments and 
empirical parameterizations.

\subsection{Procedure for $Q^2$ unfolding} 
\label{sub:unfold}
The  $\pi^0$ yield is measured 
in bins of reconstructed momentum transfer $Q^2_{\rm rec}$.
We approximate the true value of $Q^2$ with the corrected value 
$Q^2_{\rm cor}$.
We use signal MC events to calculate the folding matrix $A_{ij}$ 
for a total of 15 bins in $Q^2$, of 1~GeV$^2$ width for $5 - 12\GeV^2$, 
2~GeV$^2$ width for $12 - 20~\GeV^2$,  5~GeV$^2$ width for $20-30~\GeV^2$,
and 10~GeV$^2$ width for $30-50~\GeV^2$.
This is done separately  for the p- and e-tags.
In our algorithm, we include 
bins that are not used to calculate the TFF.
The generated events are weighted to have a realistic $Q^2$ dependence 
$d\sigma/dQ^2 \propto Q^{-7}$, which is consistent with the data.

The number of events in the same bin before and after the unfolding
is expressed by
\begin{equation}
n_{i,{\rm rec}} = \sum_j A_{ij} n_{j,{\rm cor}},
\label{eqn:eqn:unf}
\end{equation}
where $n_{i,{\rm rec}}$ is the number of events in the $i$-th bin before 
the unfolding, which is based on the measured $Q^2$,
and $n_{j,{\rm cor}}$ is the number of events in the $j$-th bin after the 
unfolding,
which is based on the true $Q^2$.
By definition, $\sum_i A_{ij}$ is normalized to unity.
$A_{ij}$ is close to a unit matrix, and the majority of the
off-diagonal elements are close to zero except for the components
adjacent to the diagonal.  
We calculate the inverse matrix,
which is also close to diagonal. 
The edge bins in the inverse matrix
are irregular because no experimental measurements are
available.
To avoid bias, we do not derive $n_{j,{\rm cor}}$ near the edge bins,
and use the unfolding method described above
to obtain results only for the range $7-30$~GeV$^2$.

For the other bins near the edge, we 
correct the measurement by the factor 
$f_i = \sum_j N_{ji}/\sum_j N_{ij}$, 
where $N_{ij}$ is the unnormalized
original matrix obtained from the $Q^{-7}$-dependent
MC simulation;
this method yields correct results
when MC events are generated
with a realistic $Q^2$ dependence.
Figure~\ref{fig:yield} shows the number of events
in each bin before and after the unfolding.
The effect of the unfolding is small, but the
statistical uncertainty increases slightly.
The resulted numbers after the unfolding are tabulated as $N_{\rm cor}$
in Table~\ref{tab:res1}.

\begin{figure}
\centering
\includegraphics[width=7.5cm]{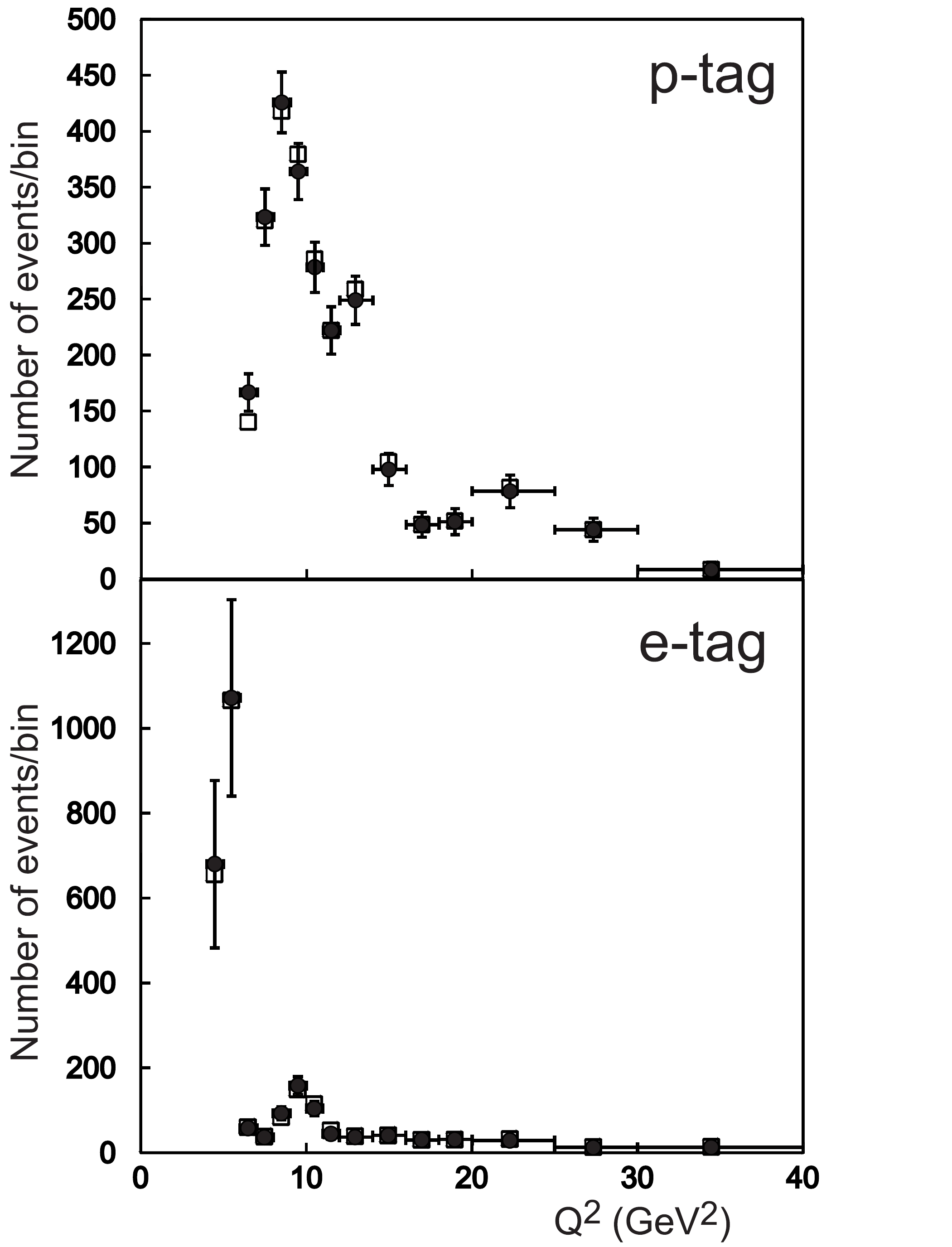}
\centering
\caption{The number of events for the signal candidates
in $Q^2$ bins before and after the unfolding. Note that
the yield is not normalized to the bin width, and 
the background yield is not subtracted. 
The open squares (dots with error bars) are the
yields before (after) the $Q^2$ unfolding.
The vertical error bars are statistical 
and are only shown for the plots after the unfolding.
The data points at $Q^2 < 6$~GeV$^2$ for the e-tag are the results from
the ``HiE+50*CsiBB'' sample.
}
\label{fig:yield}
\end{figure}

\subsection{Representative $\bar{Q^2}$ values for each bin}
\label{sub:q2val}
We define the representative $Q^2$ for each bin 
with a finite bin width,  $\bar{Q^2}$, using the formula
\begin{equation}
\frac{d\sigma}{dQ^2}(\bar{Q^2}) = \frac{1}{\Delta Q^2} 
\int_{\rm bin} \frac{d\sigma}{dQ^2}(Q^2) dQ^2 ,
\label{eqn:q2form}
\end{equation}
where $\Delta Q^2$ is the bin width. 
We assume an approximate dependence 
of  $d\sigma/dQ^2 \propto Q^{-7}$ for the calculation.
The representative values are shown in the leftmost column of
Table~\ref{tab:res1}.

\subsection{Extraction of differential cross section}
\label{sub:crosc}
The number of signal events is obtained by the
fit procedure shown in Sec.~\ref{sub:yield} in each $Q^2$
bin, separately for the e-tag and p-tag samples.
To take into account event migration among bins due to finite resolution,
the number of events observed in each $Q^2$ bin is corrected using the 
unfolding method described in Sec.~\ref{sub:unfold}.
The number of events is then converted to
the cross section for  $e^+ e^- \to (e) e \pi^0$
using the following formula:
\begin{equation}
\frac{d\sigma}{dQ^2} = \frac{N(1-b)}{\eta {\cal B} \int{\cal L}dt (1+\delta)
\Delta Q^2},
\label{eqn:cross}
\end{equation}
where $N$ is the number of signal events in the bin, $b$ is the fraction
of the estimated $\pi^0$ peaking background in the yield, 
and $\eta$ is the efficiency
corrected for the beam energy difference between the $\Upsilon(4S)$
and  $\Upsilon(5S)$, which is the mean of the
efficiencies weighted by the ratio of 
the integrated luminosities and the $2A(Q^2)$ parameter
for different beam energies. 
Using this procedure,
the differential cross section is obtained 
at the $e^+e^-$ c.m. energy of the $\Upsilon(4S)$. 
Here, ${\cal B}$ is the branching fraction for $\pi^0 \to \gamma\gamma$,
$\int {\cal L}dt$ is the total integrated luminosity, and
$1+\delta$ is the radiative correction for $r_k^{\rm max}=0.25$
used to determine the cross section for the tree-level diagrams.
The value of $\delta$ is $-1\%$ or +2\% without or with a correction
for hadronic vacuum polarization~\cite{radcor2}, respectively;
we adopt the latter value. 
We include its $Q^2$ dependence, which is around
1\% in the kinematical region of the present measurement,
into the systematics.
The quantity $\Delta Q^2$ is the bin width in $Q^2$.

We extract  the cross section of $e^+e^- \to (e)e \pi^0$
defined in the range $Q_2^2< 1~\GeV^2$,
where $Q_2^2$ is the $Q^2$ of the photon with smaller virtuality.
Our measurement is performed in a 
small $Q^2_2$ region, mainly less than 0.01~GeV$^2$, 
and then the upper bound is extrapolated to 1.0~GeV$^2$.
The requirement for the incident photon with the smaller virtuality 
(which is the $Q^2$ requirement in this paper) 
depends on the experiment and hence 
the definition of the cross section is somewhat 
dependent on the experiment.
The result of this procedure is shown in
Fig.~\ref{fig:dcsa}.
The differential cross sections are shown separately
for the p- and e-tags in Table~\ref{tab:res2}.

\begin{figure}
\centering
\includegraphics[width=7.5cm]{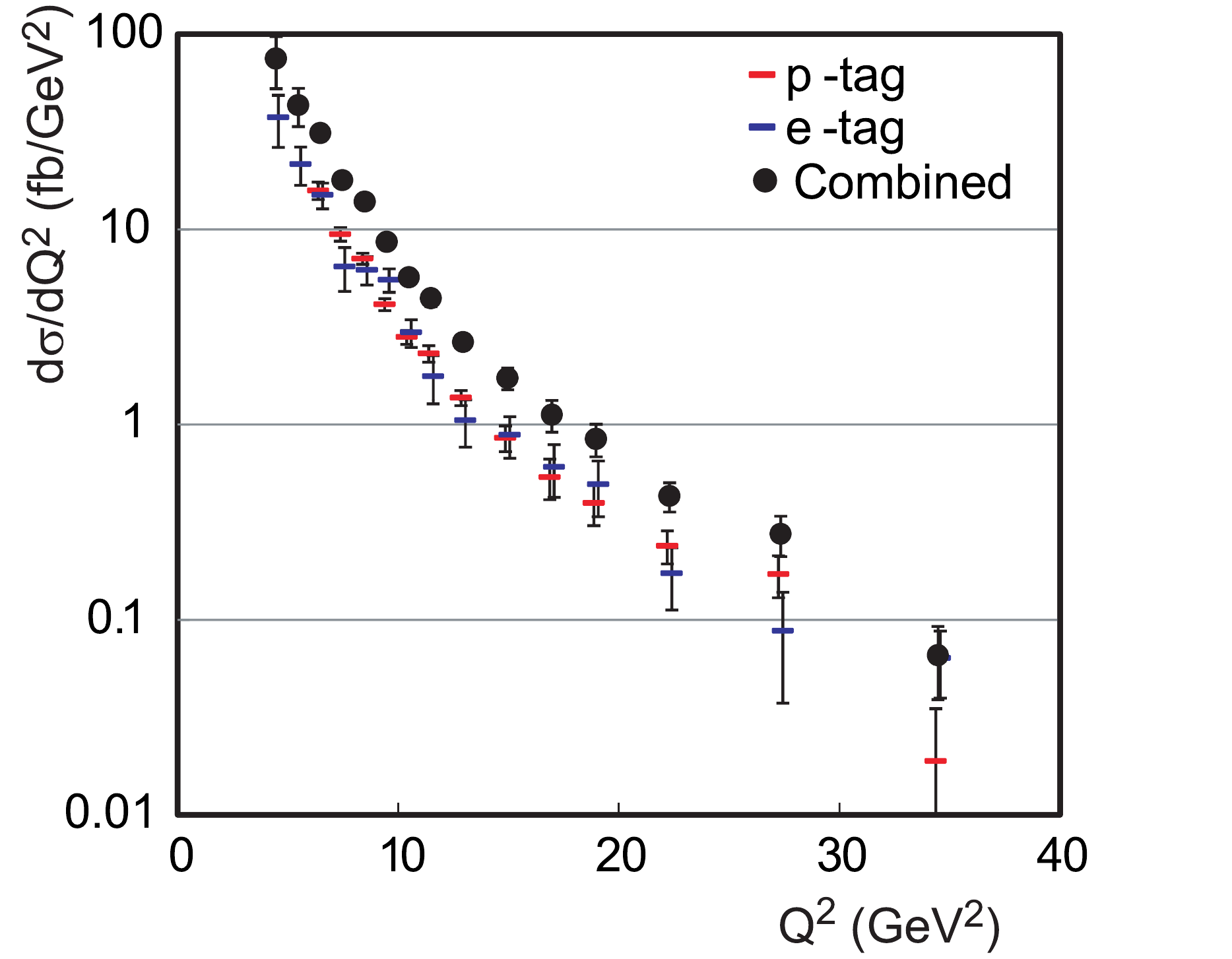}
\centering
\caption{
Results for the $e^+ e^- \to (e) e \pi^0$ differential cross section.
``Combined'' corresponds to twice the weighted average of
the p-tag and e-tag results.
Error bars show statistical uncertainties only.
The horizontal
positions in the p-tag and e-tag plots are shifted slightly
relative to each other to make it easier to see the sizes
of the statistical uncertainties.
}
\label{fig:dcsa}
\end{figure}

\begin{center}
\begin{table*}
\caption{
$e^+ e^- \to (e) e \pi^0$ differential cross section
and systematic uncertainties in bins of
$Q^2$ summarized for the p- and e-tags separately. 
The ``$Q^2$-indep.~syserr'' 
and  ``$Q^2$-dep.~eff.~syserr'' uncertainties
are described in Sec.\ref{sub:syserr}.}
\label{tab:res2}
\begin{tabular}{cc|c|cccc|c} \hline \hline
 & $Q^2$ & $d\sigma/dQ^2$  & $Q^2$-indep. &  $Q^2$-dep.~eff.
& Fit & Peaking-bkg  & Total\\
Tag & (GeV$^2$) & (fb/GeV$^2$) & syserr(\%) & syserr(\%) & syserr(\%) & 
syserr(\%) & syserr(\%) \\
\hline
&6.47	 & $	15.85 	 \pm	1.63 	$ & 	6 	 & 	4 	 & 	 4	 & 	2	 & 	8 	 \\
&7.47	 & $	9.47 	 \pm	0.76 	$ & 	6 	 & 	2	 & 	4	 & 	2	 & 	8 	 \\
&8.48	 & $	7.09 	 \pm	0.47 	$ & 	6 	 & 	2 	 & 	4	 & 	2	 & 	8 	 \\
&9.48	 & $	4.13 	 \pm	0.29 	$ & 	6 	 & 	2 	 & 	4	 & 	2	 & 	8 	 \\
&10.48	 & $	2.81 	 \pm	0.23 	$ & 	6 	 & 	2 	 & 	4	 & 	2	 & 	8 	 \\
&11.48	 & $	2.31 	 \pm	0.23 	$ & 	6 	 & 	3 	 & 	4	 & 	2	 & 	8 	 \\
p &12.94	 & $	1.37 	 \pm	0.12 	$ & 	6 	 & 	10 	 & 	4	 & 	2	 & 	12 	 \\
&14.95	 & $	0.857 	 \pm	0.129 	$ & 	6 	 & 	12 	 & 	4	 & 	2	 & 	14 	 \\
&16.96	 & $	0.539 	 \pm	0.127 	$ & 	6 	 & 	10	 & 	4	 & 	2	 & 	12 	 \\
&18.96	 & $	0.397 	 \pm	0.093 	$ & 	6 	 & 	10 	 & 	4	 & 	2	 & 	12 	 \\
&22.29	 & $	0.240 	 \pm	0.046 	$ & 	6 	 & 	5 	 & 	10	 & 	2	 & 	13 	 \\
&27.33	 & $	0.172 	 \pm	0.042 	$ & 	6 	 & 	5 	 & 	10	 & 	2	 & 	13 	 \\
&34.46	 & $	0.019 	 \pm	0.016 	$ & 	6 	 & 	5 	 & 	10	 & 	4	 & 	13 	 \\
\hline
&4.46	 & $	37.5 	 \pm	11.2 	$ & 	6 	 & 	2 	 & 	7	 & 	2	 &      10 	 \\
&5.47	 & $	21.6 	 \pm	4.8 	$ & 	6 	 & 	1 	 & 	7	 & 	2	 & 	9 	 \\
&6.47	 & $	15.03 	 \pm	2.26 	$ & 	6 	 & 	8 	 & 	7	 & 	2	 & 	12 	 \\
&7.47	 & $	6.45 	 \pm	1.63 	$ & 	6 	 & 	5 	 & 	7	 & 	2	 & 	11 	 \\
&8.48	 & $	6.21 	 \pm	1.02 	$ & 	6 	 & 	5 	 & 	7	 & 	2	 & 	11 	 \\
&9.48	 & $	5.53 	 \pm	0.76 	$ & 	6 	 & 	5 	 & 	7	 & 	2	 & 	11 	 \\
&10.48	 & $	2.97 	 \pm	0.48 	$ & 	6 	 & 	5 	 & 	7	 & 	2	 & 	11 	 \\
e&11.48	 & $	1.77 	 \pm	0.49 	$ & 	6 	 & 	5 	 & 	7	 & 	2	 & 	11 	 \\
&12.94	 & $	1.05 	 \pm	0.29 	$ & 	6 	 & 	8 	 & 	7	 & 	2	 & 	12 	 \\
&14.95	 & $	0.886 	 \pm	0.213 	$ & 	6 	 & 	10 	 & 	7	 & 	2	 & 	14 	 \\
&16.96	 & $	0.607 	 \pm	0.183 	$ & 	6 	 & 	10 	 & 	7	 & 	2	 & 	14 	 \\
&18.96	 & $	0.495 	 \pm	0.157 	$ & 	6 	 & 	10 	 & 	7	 & 	2	 & 	14 	 \\
&22.29	 & $	0.173 	 \pm	0.061 	$ & 	6 	 & 	10 	 & 	10	 & 	2	 & 	15 	 \\
&27.33	 & $	0.088 	 \pm	0.051 	$ & 	6 	 & 	10 	 & 	10	 & 	2	 & 	15 	 \\
&34.46	 & $	0.064 	 \pm	0.024 	$ & 	6 	 & 	10 	 & 	10	 & 	4	 & 	16 	 \\
\hline
\hline
\end{tabular}
\end{table*}
\end{center}
}

\begin{figure}
\centering
\includegraphics[width=7.5cm]{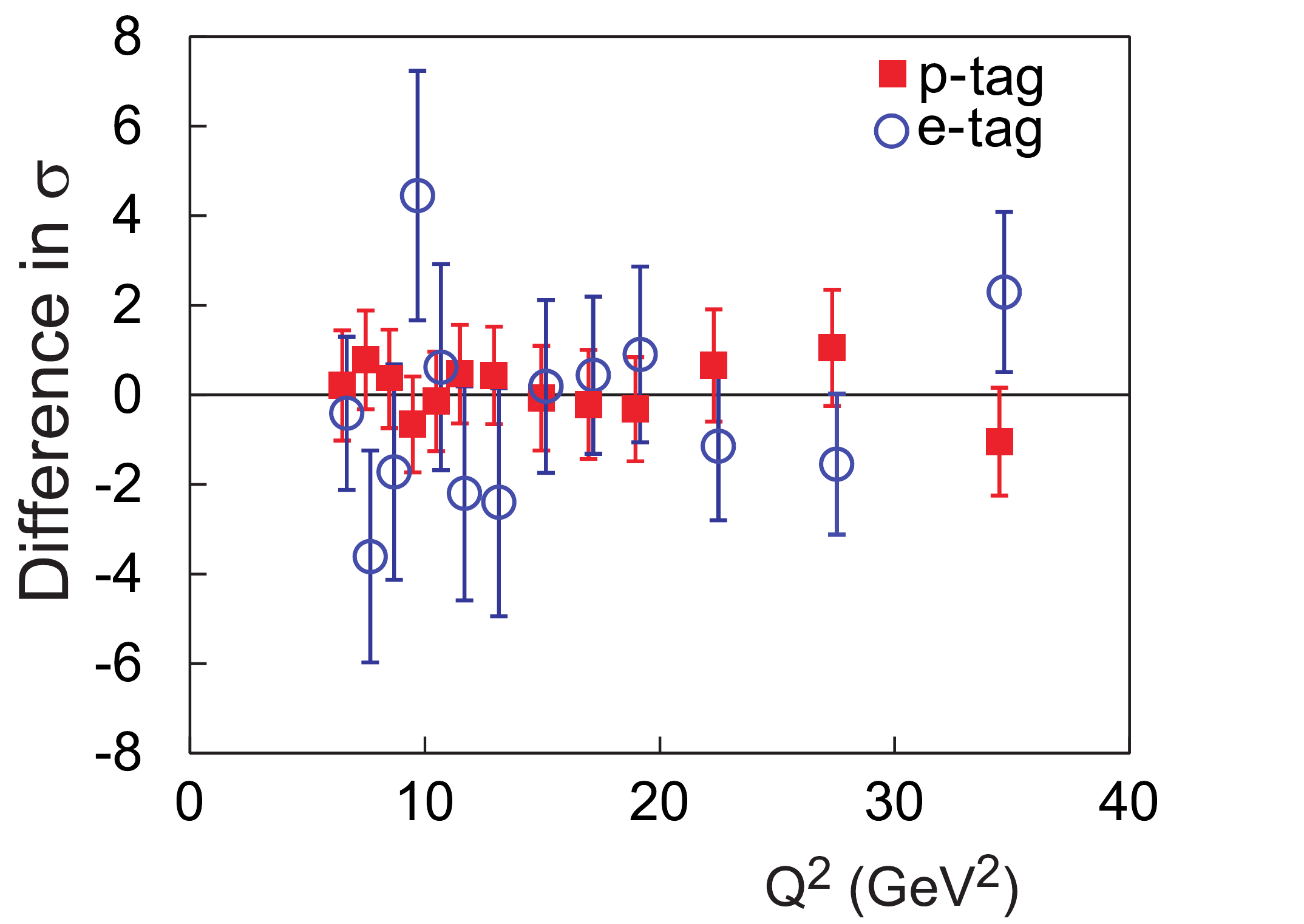}
\centering
\caption{ 
Differences between the $e^+ e^- \to (e) e \pi^0$ differential cross section  
measured by the p-tag (squares) and
e-tag (circles) from their weighted mean 
(equal to a half of the combined cross section). 
The vertical scale is normalized by the error of the weighted mean 
(which is the same as 
half of the error for the combined cross section). The horizontal
positions of the p-tag and e-tag plots are displaced intentionally so that
the error bars are visible.
}
\label{fig:dcsb}
\end{figure}

We find that the differential cross sections are consistent
between the p- and e-tag measurements. 
The difference between the two measurements in the same $Q^2$
region is smaller than $1.8\sigma$, where $\sigma$ is the combined
statistical error, for each $Q^2$ bin as shown in Fig.~\ref{fig:dcsb}. 
We combine the p- and e-tag results for the same $Q^2$
bin, if available, by taking the weighted average
of the results with a weight that is the inverse square 
of the statistical error. 
We then double the average. 
Where only the result from the e-tag is available 
(for the $Q^2$ range of $4-6$~GeV$^2$), 
we simply double the value.
The differential cross sections combined 
for the p- and e-tags are shown in Table~\ref{tab:res3}.

\begin{figure}
\centering
\includegraphics[width=7.5cm]{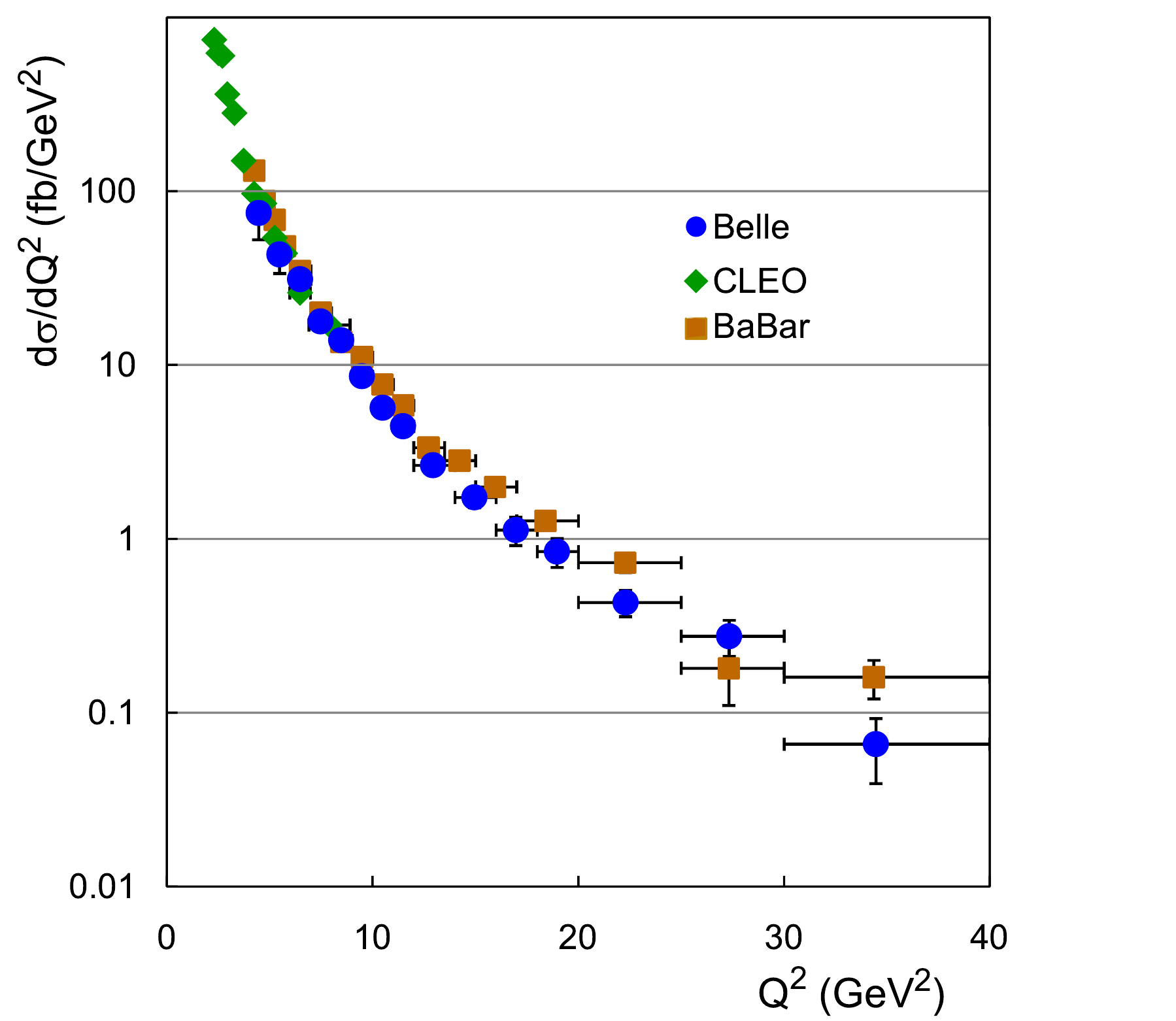}
\centering
\caption{ 
Comparison of the results for the $e^+ e^- \to (e) e \pi^0$ 
differential cross sections among different experiments. 
The vertical error bars are statistical only. 
The algorithms used to measure
the cross section differ between the experiments.
}

\label{fig:dcscomp}
\end{figure}

The obtained differential cross sections are compared
with those from the previous experiments~\cite{babar1,cleo}
in Fig.~\ref{fig:dcscomp}.
Although the definitions of the differential
cross sections differ slightly between the experiments, 
the difference should be compensated 
in the algorithms used to extract the transition form factor.
Later in this section, we compare in more detail the
$\pi^0$-TFF results obtained by different experiments.

\subsection{Determination of the transition form factor}
\label{sub:tff}
\begin{figure}
\centering
\includegraphics[width=7.5cm]{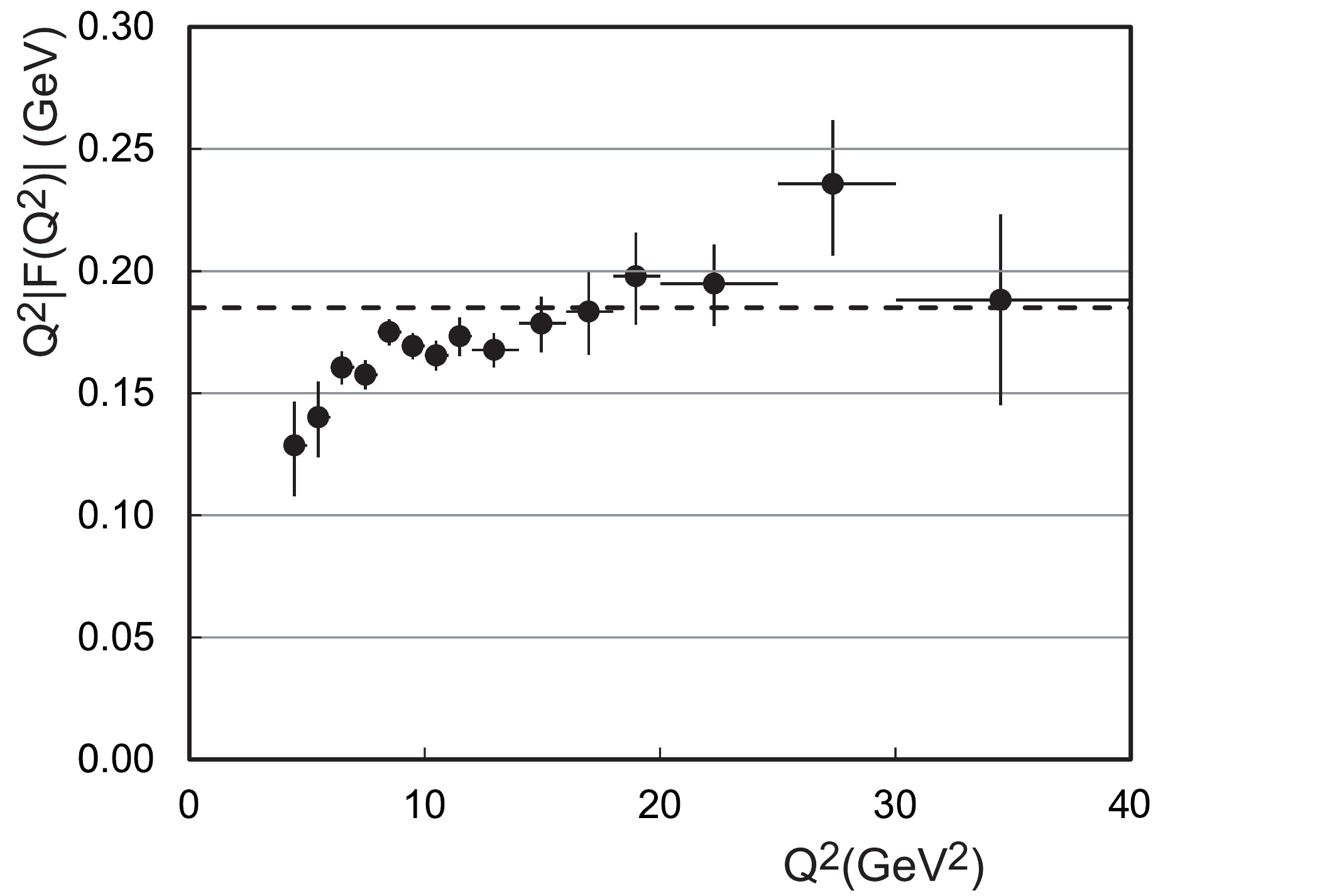}
\centering
\caption{Results for the $\pi^0$ transition form factor
from the present measurement multiplied by $Q^2$. 
The error bars are statistical only. 
The dashed line shows the asymptotic limit from pQCD ($\sim 0.185$~GeV).
}
\label{fig:pi0ff}
\end{figure}

The $e^+ e^- \to (e) e \pi^0$ cross section is converted to the 
transition form factor
squared by dividing it by $2A(\bar{Q^2})$.
Finally, we obtain
\begin{equation}
Q^2|F(Q^2)| = Q^2 \sqrt{\frac{d\sigma/d Q^2 (Q^2)}{2A(Q^2)}},
\label{eqn:tff}
\end{equation}
where $Q^2$ replaces the symbol $\bar{Q^2}$. 
The results for this product are shown in
Fig.~\ref{fig:pi0ff} and  Table~\ref{tab:res3}.

\begin{center}
\begin{table}
\caption{
$e^+ e^- \to (e) e \pi^0$ differential cross section 
combined for the p- and e-tags with
systematic uncertainties ($\epsilon_{\rm sys}$) and the
transition form factor $Q^2|F(Q^2)|$.
The first and second uncertainties for
$Q^2|F(Q^2)|$ are statistical and systematic, respectively.
}
\label{tab:res3}
\begin{tabular}{c|cc|c} \hline \hline
$Q^2$  & $d\sigma/dQ^2$ & ${\epsilon}_{\rm sys}$ & $Q^2|F(Q^2)|$  \\
(GeV$^2$) & (fb/GeV$^2$) & (\%) & (GeV) \\
\hline
4.46	 & $	75.0 	 \pm	22.3 	$ & 	10 	 & $	0.129 	 \pm	0.020 		 	 \pm	0.006 	 $ \\
5.47	 & $	43.3 	 \pm	9.6 	$ & 	9 	 & $	0.140 	 \pm	0.016 		 	 \pm	0.007 	 $ \\
6.47	 & $	31.15 	 \pm	2.64 	$ & 	10  	 & $	0.161 	 \pm	0.007 		 	 \pm	0.008 	 $ \\
7.47	 & $	17.86 	 \pm	1.38 	$ & 	8 	 & $	0.158 	 \pm	0.006 		 	 \pm	0.007 	 $ \\
8.48	 & $	13.88 	 \pm	0.85 	$ & 	8 	 & $	0.175 	 \pm	0.005 		 	 \pm	0.007 	 $ \\
9.48	 & $	8.62 	 \pm	0.55 	$ & 	8 	 & $	0.169 	 \pm	0.005 		 	 \pm	0.007 	 $ \\
10.48	 & $	5.68 	 \pm	0.42 	$ & 	8 	 & $	0.165 	 \pm	0.006 		 	 \pm	0.007 	 $ \\
11.48	 & $	4.44 	 \pm	0.41 	$ & 	9 	 & $	0.173 	 \pm	0.008 		 	 \pm	0.007 	 $ \\
12.94	 & $	2.65 	 \pm	0.23 	$ & 	12 	 & $	0.168 	 \pm	0.007 		 	 \pm	0.010 	 $ \\
14.95	 & $	1.73 	 \pm	0.22 	$ & 	14 	 & $	0.179 	 \pm	0.012 		 	 \pm	0.013 	 $ \\
16.96	 & $	1.123 	 \pm	0.208 	$ & 	13 	 & $	0.183 	 \pm	0.017 		 	 \pm	0.012 	 $ \\
18.96	 & $	0.845 	 \pm	0.160 	$ & 	13 	 & $	0.198 	 \pm	0.019 		 	 \pm	0.013 	 $ \\
22.29	 & $	0.431 	 \pm	0.074 	$ & 	14 	 & $	0.195 	 \pm	0.017 		 	 \pm	0.013 	 $ \\
27.33	 & $	0.275 	 \pm	0.064 	$ & 	14 	 & $	0.236 	^{ +	0.026 	}_{ -	0.029 	 } \pm 	0.016 	 $ \\
34.46	 & $	0.066 	 \pm	0.027 	$ & 	14 	 & $	0.188 	^{ +	0.035 	}_{ -	0.043 	 } \pm 	0.013 	 $ \\

\hline
\hline
\end{tabular}
\end{table}
\end{center}

\subsection{Systematic uncertainties and some checks}
\label{sub:syserr}
The estimates of systematic uncertainties for the $Q^2$ differential 
cross sections are described in this subsection.
They can be divided into $Q^2$-dependent and -independent contributions. 

\subsubsection{Estimates of systematic uncertainties}
The $Q^2$-independent contributions in the
systematic uncertainty include
tracking uncertainty (1\%), 
electron-ID (1\%), $\gamma\gamma$-pair 
reconstruction efficiency (3\%), kinematical selection (2\%), 
geometrical selection including
the trigger (2\%), beam-background effects
that affect the event reconstruction (2\%), 
and integrated luminosity (1.4\%). 
These are correlated $Q^2$-independent uncertainties, resulting in a 
total of 5\% in the quadratic sum 
for the systematic uncertainty arising from the efficiency.
Note that we do not use a conventional $\pi^0$ reconstruction
from a mass-constrained fit, as we fit
experimental $\gamma\gamma$ mass spectra.
The peaking-background contamination (1\%-4\%), fit error in
the $\pi^0$-yield extraction (5\%-10\%), and the trigger efficiency
(2\%-12\%) have significant $Q^2$ dependence. 
We estimate the fit uncertainty by taking the difference between 
the central values and a simpler fit with a single Gaussian
and a linear background applied for the narrower $M_{\gamma\gamma}$
fit range, $0.08-0.20$~GeV/$c^2$. 
We evaluate the systematic uncertainty 
due to this $M_{\gamma \gamma}$ fitting by averaging the 
deviations in the fit results in each of two $Q^2$
regions (below and above $Q^2 = 20~\GeV^2$)
for each of the p- and e-tags to compensate for the statistical fluctuations, 
as listed in column 6 of Table~\ref{tab:res2}.

To estimate the systematic uncertainty in the trigger efficiency,
we study the efficiency using a radiative Bhabha process; 
we find agreement between the data and the Rabhat
MC at the $5-10$\% level, as shown in Fig.~\ref{fig:compeff_rabhati}(c,d) 
(Appendix~\ref{sub:compar}).
In addition, the comparison of the absolute yields of the
radiative Bhabha events shows consistency between the data
and MC also within 10\% level, which is comparable to the
estimated size of systematic uncertainty
without including the uncertainty from the trigger efficiency of 
$\sim 8\%$ (see Fig.~\ref{fig:rbmc_q}(c,d)
and the text in Appendix~\ref{sub:compar}). 
Therefore, we conclude that
the systematic uncertainty in the trigger efficiency does not exceed
the deviations seen in studies using radiative Bhabha
events: 12\% for the p-tag and 10\% for the e-tag,
where the values are taken from the overall tendency of the
deviations in the $Q^2$ dependence shown in Figs.~\ref{fig:rbmc_q}(c,d)
and \ref{fig:compeff_rabhati}(c,d).
However, we do expect the systematic uncertainty to be smaller 
for many individual $Q^2$ bins, 
{\it e.g.}, in bins where the effect of the Bhabha-veto threshold is relatively
small and the trigger efficiency is not very sensitive to it.
To obtain the $Q^2$-dependent systematic error,
we have varied the Bhabha-veto trigger threshold by $\pm 0.2$~GeV
in the trigger simulation applied to the signal MC events and
taken the change of the efficiency for each bin if it is smaller
than the maximum values mentioned above.  
In practice, we have found that the variation has a large $Q^2$ dependence, 
ranging from 2\% to 25\%, in each of the p- and e-tag samples.
 We replace the uncertainty determined by the variation of
the trigger threshold in the signal MC
by 12\% (10\%) for the p-tag (e-tag) case
if the former is greater than 12\% (10\%) in an individual $Q^2$ bin.
We replace the systematic uncertainty in the
bin $Q^2=14-16~\GeV^2$ for the p-tag sample and in the six bins
in the $Q^2$ range $14-40~\GeV^2$ for the e-tag sample.

We assign a $\pm (1\%-3\%)$ systematic uncertainty 
from a possible deviation of the $Q^2$ dependence in the signal MC
from the data distribution for the lowest two bin results in
each of the p- and e-tags. 
In other $Q^2$ regions, this uncertainty is
negligibly small. We take the difference between the calculated 
efficiencies with two assumed $Q^2$ dependences:
$|F(Q^2)|^2 \sim 1$ and
$|F(Q^2)|^2 \sim 1/Q^2$. 
The combined uncertainties
from this source and from the trigger efficiency are shown 
in Table~\ref{tab:res2} as a $Q^2$-dependent systematic 
uncertainty in the efficiency.
The evaluation of uncertainties is performed separately 
for the p- and e-tags.

In addition, the uncertainties from the calculation of the radiative 
corrections (3\%) and assumed form factor effects for the photon 
with smaller virtuality (1.0\%) enter
when determining the differential cross section.
The correlated $Q^2$-independent uncertainty is in total
6\%, of which 5\% is from the efficiency determination as described above, 
in a quadrature sum for the uncertainty in the differential cross section.

The total systematic uncertainties for the combined differential
cross section and for the transition form factor 
are shown in Tables~\ref{tab:res2} and ~\ref{tab:res3}.
We assume that the systematic uncertainties in the p- and e-tags
correlate maximally.
Because the transition form factor is derived from
the square root of the cross section, 
the relative systematic uncertainty is half of the above. 
The calculation uncertainty for the conversion factor $2A(Q^2)$, 2\%, is 
small compared to other sources of  uncertainties.
The uncertainty in the value of $\bar{Q}^2$ in each bin is also negligible
($\Delta \bar{Q}^2/\bar{Q}^2 \sim 0.3\%$) 
because $Q^2 |F(Q^2)|$ varies slowly with $Q^2$.

\subsubsection{Check for ISR in the signal process}
\label{sub:check}
\begin{figure}
\centering
\includegraphics[width=6.5cm]{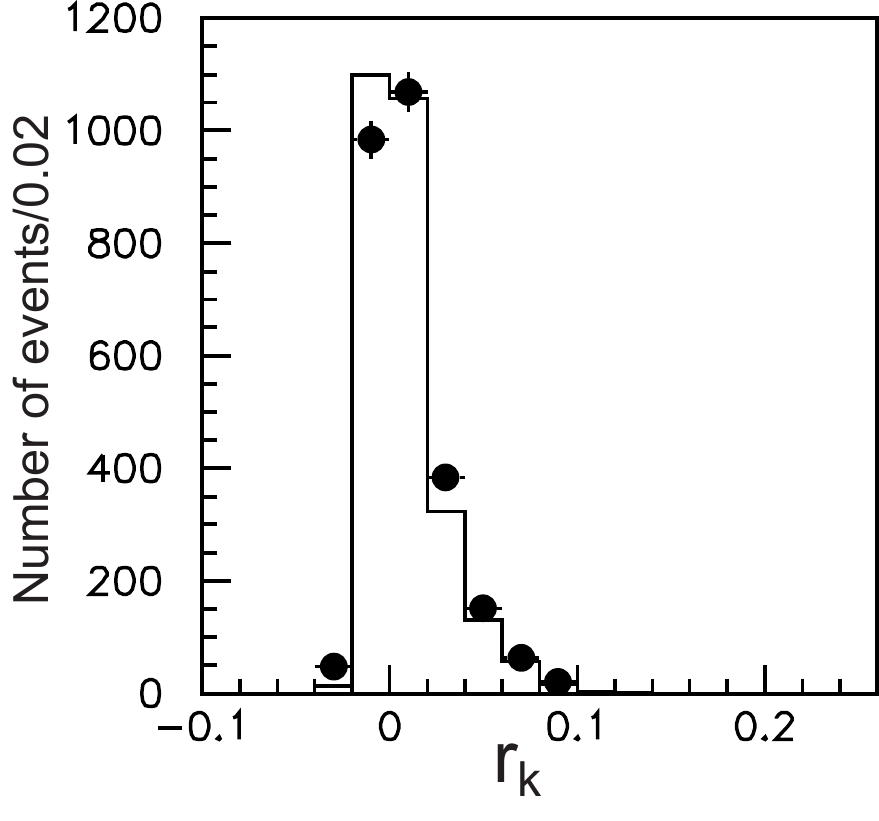}
\centering
\caption{
Distribution of the ISR energy fraction ($r_k$) calculated from
observables in data (solid dots) and in the signal MC samples (histogram). 
The data distribution is shown after background subtraction.  
The MC events are normalized to the experimental yield after
background subtraction. 
Note that there are uncertainties in the value of
$r_k$ originating from the measurement and kinematical approximation.
}
\label{fig:rkdisnew}
\end{figure}

We have included the effect of ISR in the signal MC simulation but
only for the tagged electron side with exponentiation.
This is a reasonable approximation because only the tag side shifts
$Q^2$ and kinematics significantly, while
low-$Q^2$ radiation from the untagged side does not change any observable. 
The corrections in $\delta$ are small for both cases (at the 2\% level).
We compare the $r_k$ distributions in data
after background subtraction using the $\pi^0$-mass 
sideband events and the signal MC samples 
after the selection as shown in Fig.~\ref{fig:rkdisnew}. 
The fraction of energy from radiation relative to
the beam energy, $r_k$ 
(defined in Eq.(\ref{eqn:rk1})), is calculated by an approximate kinematical relation
from the observed variables,
\begin{equation}
r_k = \frac{\sqrt{s} - (E^*_\pi+E^*_e) - | \vec{p}^*_\pi 
+  \vec{p}^*_e|}{\sqrt{s}}. 
\label{eqn:rk}
\end{equation} 
In the BaBar analysis, this parameter is directly used as a 
selection~\cite{babar1}.
In our analysis, 
instead of $r_k$ we use the variable $E_{\rm ratio}$.  
These two variables
are strongly correlated, as, 
kinematically, $E_{\rm ratio}=1$ when $r_k=0$.
The peak and the core part 
of the 
$r_k$ distribution ($-0.02 < r_k < 0.06$) are safely contained by our 
selection criteria for $E_{\rm ratio}$, so the consistency between the data 
and MC after this selection permits a reliable determination of 
the signal efficiency.
As the signal MC events are generated with $r_k^{\rm max} = 0.25$, 
no bias is introduced in the signal region
by our selection criteria.

\subsection{Comparison with previous measurements and 
empirical fits}
\label{sub:fit}
\begin{figure}
\centering
\includegraphics[width=7cm]{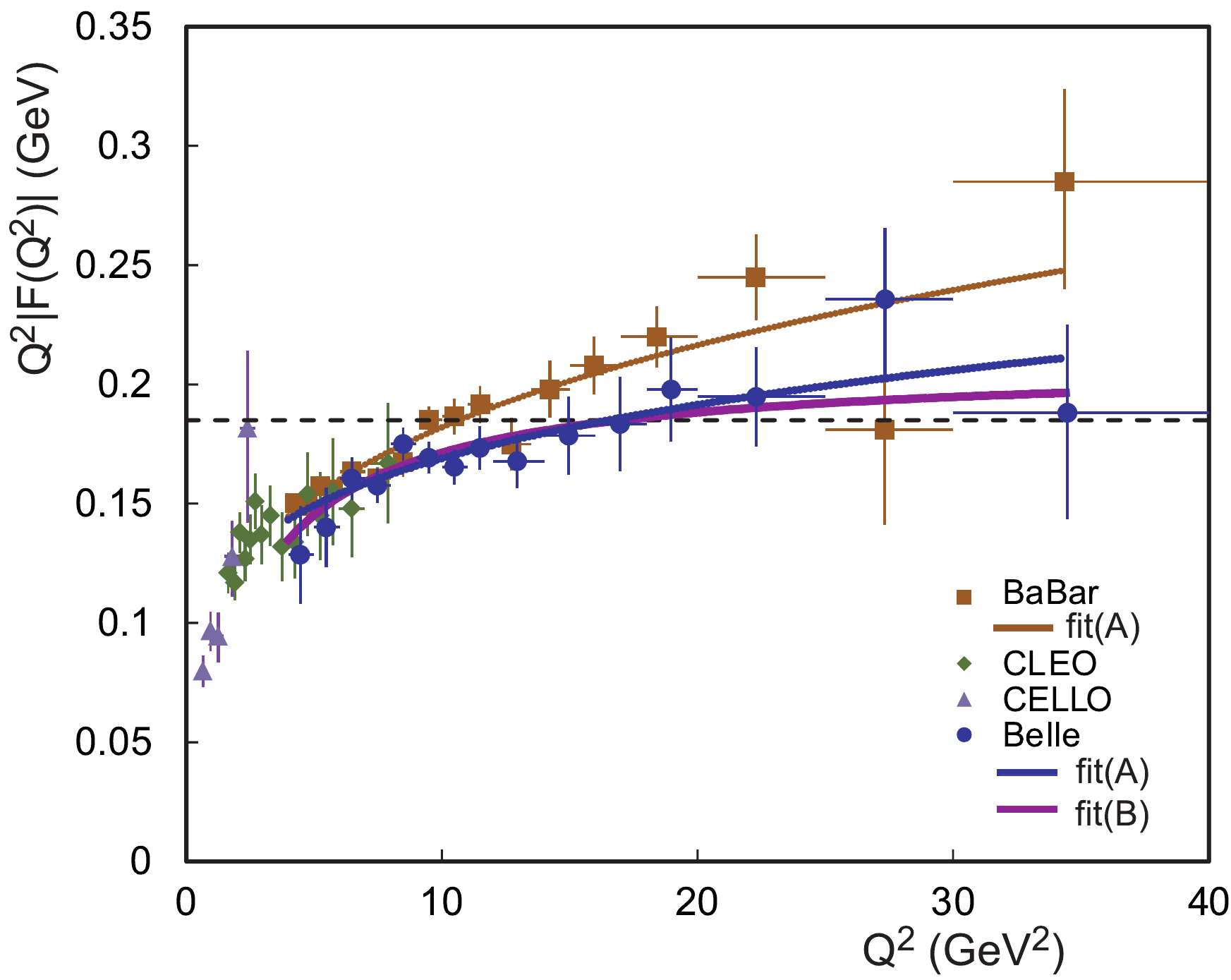}
\centering
\caption{Comparison of the results for the product
$Q^2|F(Q^2)|$ for the $\pi^0$ from different
experiments. 
The error bars are a quadratic sum of
statistical and systematic uncertainties. For the
Belle and BaBar results, only a $Q^2$-dependent
systematic-error component is included. 
The two curves denoted fit(A) 
use the BaBar parameterization while the curve denoted fit(B) 
uses Eq.(\ref{eqn:aspfit}) (see the text).
The dashed line shows
the asymptotic prediction from pQCD ($\sim 0.185$~GeV).}
\label{fig:pi0ff_aspfit}
\end{figure}

The measured product $Q^2 |F(Q^2)|$ is compared with the
previous measurements~\cite{cello, cleo, babar1} in 
Fig.~\ref{fig:pi0ff_aspfit}.
Our results agree with previous measurements within uncertainties 
for $Q^2$ below $9~\GeV^2$; in the higher $Q^2$ region,
they do not show the rapid growth observed by BaBar~\cite{babar1}.

BaBar has suggested a parameterization fit
with the functional form
\begin{equation}
Q^2|F(Q^2)| = A\left( \frac{Q^2}{10~{\rm GeV}^2} \right)^\beta,
\label{eqn:tfffit}
\end{equation}
where $A$ and $\beta$ are fit parameters. 
BaBar reported $A=0.182 \pm 0.002$~GeV and $\beta =
0.25 \pm 0.02$;
however, the uncertainty for $A$ 
does not include the $Q^2$-independent systematic
uncertainty that amounts to 2.3\% (namely $\pm 0.0041$~GeV
as this error component for $Q^2|F(Q^2)|$). 
Taking this into account, the uncertainty on $A$ reported by BaBar could 
be replaced by $A =0.182 \pm 0.005$~GeV.
To compare our results with BaBar's, we use the same parameterization
in our fit procedure and assume a $Q^2$-independent systematic 
uncertainty (thus, the total normalization error) of
3.2\% for $Q^2|F(Q^2)|$; we remove this component 
from the combined statistical and systematic errors
and instead add it in quadrature 
to the uncertainty in $A$.
The fit results from Belle are $A=0.169 \pm 0.006$~GeV 
and $\beta = 0.18 \pm 0.05$. 
The goodness of the fit is 
$\chi^2/ndf = 6.90/13$, where
$ndf$ is the number of degrees of freedom. 
The fit results are also shown in Fig.~\ref{fig:pi0ff_aspfit}.
The fit of the Belle data to the function is good, 
and we find a difference of $\sim 1.5\sigma$ 
between the Belle and BaBar results
in both $A$ and $\beta$. 

We then try another parameterization in which $Q^2 |F(Q^2)|$ approaches
an asymptotic value, namely
\begin{equation}
Q^2 |F(Q^2)| = \frac{B Q^2}{Q^2 + C} .
\label{eqn:aspfit}
\end{equation}
The fit gives $B=0.209 \pm 0.016$~GeV
and $C=2.2 \pm 0.8~\GeV^2$ 
with $\chi^2/ndf = 7.07/13$,
and is also shown in Fig.~\ref{fig:pi0ff_aspfit}.
The fitted asymptotic value, $B$, is
slightly larger than 
the pQCD 
value of $\sim 0.185~\GeV$ but is consistent.

For a simple 
estimate of the consistency between the
Belle and BaBar results, we compare the data 
from individual experiments with a reference curve
obtained by fitting the data from both experiments together
using parameterization of Eq.~(23).
The seven data points from Belle (from BaBar) for the range
9~GeV$^2 < Q^2 <20$~GeV$^2$, where the two measurements seem 
to be systematically shifted, deviate from the reference 
curve by $(-6.1 \pm 3.8)\%$ (by $(+4.8 \pm 3.0)\%$) in average
relative to it. We incorporate the $Q^2$-independent uncertainty 
in each measurement in the above error. The difference 
between the Belle and BaBar deviations, $(10.9 \pm 4.8)\%$, 
corresponds to a $2.3\sigma$ significance. 
This result does not depend on choice of 
the reference curve.

\section{Summary and Conclusion}
\label{sec:summ}
We have presented a measurement of the neutral pion
transition form factor for the process $\gamma \gamma^* \to \pi^0$
in the region $4~\GeV^2 \simlt Q^2 \simlt 40~\GeV^2$
with a 759~fb$^{-1}$ data sample collected with the Belle detector
at the KEKB asymmetric-energy $e^+e^-$ collider.
The measured values of $Q^2 |F(Q^2)|$ agree with the previous 
measurements~\cite{cello, cleo, babar1} for $Q^2 \simlt 9~\GeV^2$.
In the higher $Q^2$ region, in contrast to BaBar, our results 
do not show a 
rapid growth with $Q^2$ and are closer 
to theoretical expectations~\cite{LB}.

\acknowledgments
We are indebted to Drs. Y.~Kurihara and K.~Tobimatsu for their 
helpful suggestions in the tuning of the Rabhat and Bases/Spring programs 
to match our configuration.
Valuable discussions with Dr. V.~Chernyak are acknowledged.
We are also grateful to the BabaYaga team for their efforts in developing 
and providing a modified version of BabaYaga suitable for 
radiative Bhabha event generation for our kinematical configuration,
although we could not make use of it properly.
We thank the KEKB group for the excellent operation of the
accelerator, the KEK cryogenics group for the efficient
operation of the solenoid, and the KEK computer group and
the National Institute of Informatics for valuable computing
and SINET4 network support.  We acknowledge support from
the Ministry of Education, Culture, Sports, Science, and
Technology (MEXT) of Japan, the Japan Society for the 
Promotion of Science (JSPS), and the Tau-Lepton Physics 
Research Center of Nagoya University; 
the Australian Research Council and the Australian 
Department of Industry, Innovation, Science and Research;
the National Natural Science Foundation of China under
contract No.~10575109, 10775142, 10875115 and 10825524; 
the Ministry of Education, Youth and Sports of the Czech 
Republic under contract No.~LA10033 and MSM0021620859;
the Department of Science and Technology of India; 
the BK21 and WCU program of the Ministry Education Science and
Technology, National Research Foundation of Korea,
and NSDC of the Korea Institute of Science and Technology Information;
the Polish Ministry of Science and Higher Education;
the Ministry of Education and Science of the Russian
Federation and the Russian Federal Agency for Atomic Energy;
the Slovenian Research Agency;  the Swiss
National Science Foundation; the National Science Council
and the Ministry of Education of Taiwan; and the U.S.\
Department of Energy and the National Science Foundation.
This work is supported by a Grant-in-Aid from MEXT for 
Science Research in a Priority Area (``New Development of 
Flavor Physics''), and from JSPS for Creative Scientific 
Research (``Evolution of Tau-lepton Physics'').

\appendix
\section{Bhabha-veto and Bhabha-mask selection}
\label{sec:A}
The Bhabha veto in the HiE trigger of the Belle trigger system rejects
a significant portion of the signal~\cite{footn1}.
An accurate determination of the efficiency from this veto is essential
for a reliable measurement of the signal cross section.
In this section we provide a detailed description of 
the Bhabha-mask rejection that
selection criterion (9) is based on.

\subsection{Bhabha-veto trigger}
\label{sub:bhtrg}
A  detailed description of the Bhabha trigger (CsiBB) at Belle is 
given elsewhere~\cite{ecltrig}.
It fires 
when the energy sum over 
one- or two-``$\theta$-ring'' groups among eleven possible combinations 
(referred to hereafter as
"Logic") is higher than a given threshold.
A $\theta$-ring is an array of CsI(Tl) crystals in a certain polar angular
range covering $2 \pi$ in azimuth.
A $\theta$-ring combination corresponds to back-to-back 
Bhabha electrons in that angular range. 

The sum of energies over two $\theta$-rings 
is required but the energy can concentrate in either ring. 
Thus, an event with a single cluster can satisfy the trigger logic
even if it has a
non-back-to-back topology in the $e^+e^-$ c.m. frame. 
This vetoes some fraction of the signal in the
two-dimensional angular acceptance plane 
for ($\cos \theta_e$, $\cos \theta_{\gamma\gamma})$,
especially in the forward region where the energies of
the electron and photons are relatively high.

\subsection{Bhabha-mask rejection}
\label{sub:bhrej}
As mentioned in selection criterion (9), we reject events for
which the angular combination ($\cos \theta_e$, $\cos \theta_{\gamma\gamma}$) 
falls in the predetermined two-dimensional angular area.
We divide the polar angle region for the selection of
an electron and photon system,  $-0.6235 < \cos \theta <0.9481$,
that is, $18.53^\circ< \theta < 128.57^\circ$, 
into 14 subregions with 13 boundaries~\cite{footn2}
according to the boundaries of the CsI crystals that belong to different
trigger segments. 
The acceptance plane 
($\cos \theta_e$, $\cos \theta_{\gamma\gamma}$) is then subdivided
into $14 \times 14$  rectangular cells
(see Fig.~\ref{fig:cell}).

We estimate the efficiency of the Bhabha veto 
in each cell 
using the trigger simulator, TSIM. 
The cells that are sensitive to the veto are removed from the selection.
The accepted cells are shown in Fig.~\ref{fig:cell}
in yellow.
In this figure, we overlap the distribution from the
signal MC simulation (shown with the scattered dots), where 
the Bhabha veto is not applied,
but the high-$Q^2$ contribution is enhanced as compared to the
actual distribution.
The two-dimensional angular regions for the p-tag and e-tag
are separated into the left-lower region and the right-upper region, 
respectively. 
 
\subsection{Unbiased sample}
\label{sub:unbias}
The HiE trigger is vetoed by the CsiBB trigger.
Hence, the  ``HiE OR CsiBB'' data sample, which is triggered by 
either of them, is completely free from the Bhabha-veto logic.  
The events triggered by the Bhabha (CsiBB) trigger are recorded
after prescaling by a factor of 1/50.
These prescaled events are used in the analysis of the $\pi^0$
TFF in the region of $Q^2$ between 4 and 6 GeV$^2$.

In addition, this sample is very useful to calibrate the energy threshold
of the CsiBB trigger as well as  to test
the trigger simulation of the HiE trigger in which the 
 Bhabha veto is included.
These calibrations and the test of the trigger simulation have been
carried out by using radiative Bhabha events ($e^+ e^- \to   (e) e \gamma$)
in the special configuration where  one of
the final state electrons is scattered into the beam pipe.
This configuration is usually referred to as Virtual Compton scattering (VC).
The kinematics of VC are very similar to the signal process
($e^+ e^- \to (e)  e  \pi^0$) so
we can use this sample as a calibration signal to validate the
trigger simulation using data, which is one of the most important aspects of
this analysis.

To compensate for prescaling, we multiply by 50 the number of
 events recorded by the CsiBB trigger.
Thus, to emphasize that the effect of the Bhabha veto is statistically
compensated in the complete dataset, it is referred to as
``HiE+50*CsiBB''.

\begin{figure*}
\centering
\includegraphics[width=9cm]{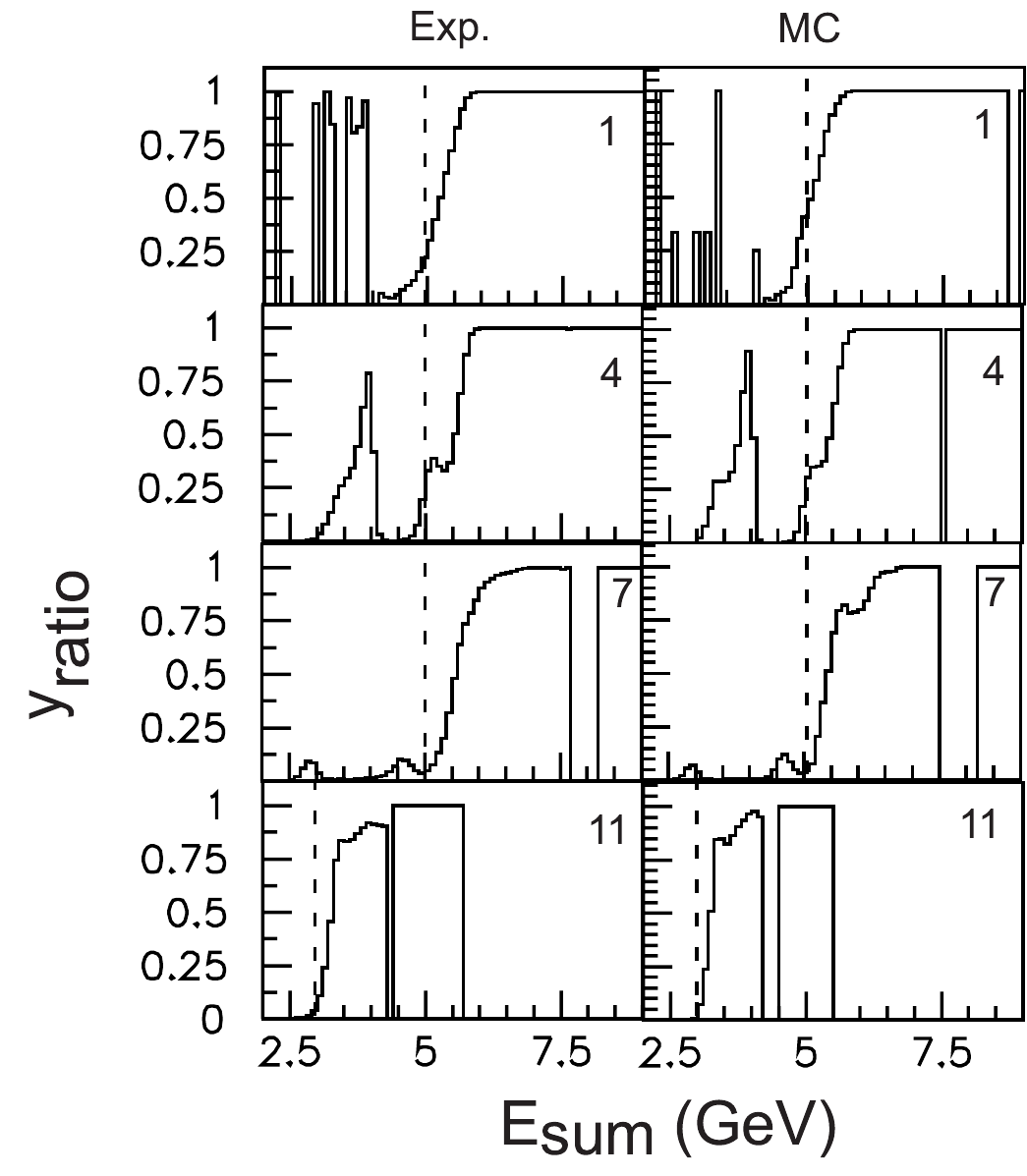}
\centering
\caption{The threshold behavior for the four selected Logics
of the CsiBB trigger (identified in each subfigure) 
obtained from the experimental samples (left column)
of radiative Bhabha events and from the MC events,
where the trigger simulator is tuned (right column). 
The dashed lines show the position of the nominal 
trigger thresholds. }
\label{fig:trigexpthreb}
\end{figure*}

\section{Calibration of the trigger threshold}
\label{sec:B}
It is necessary to calibrate the trigger thresholds
for the Bhabha veto because the inefficiency due to
the trigger for the signal is significant. 
In contrast, the threshold for the HiE trigger is efficient
because it is low (around 1.15~GeV) compared to our analysis
selection requirement.

There are eleven trigger thresholds for the Bhabha veto
(the same as the CsiBB trigger) and their nominal
value is 5.0~GeV, except for Logic\#2 (5.5~GeV) and
Logic\#11 (3.0~GeV)~\cite{ecltrig}. 
However, the thresholds determined from Bhabha events in data indicate 
that the actual values of the thresholds are 
larger, typically by 0.2~GeV.
In addition, the threshold implemented in the TSIM trigger simulator 
is a step function while the data exhibit the turn-on curves typical
for such triggers.
We correct for these differences between data and MC.

The trigger efficiency is studied and tuned using
radiative Bhabha ($e^+e^- \to (e) e \gamma$) events
with the VC configuration as described in the next subsection.
The radiative Bhabha cross section 
in the VC configuration is very large;
it is ${\cal O}(10^4 - 10^5)$ times larger than the
signal process. 
Significant and sufficient statistics for radiative Bhabha events
are available even in the prescaled CsiBB data samples.

In Sec.~\ref{sub:radbha}, we describe tuning of the Bhabha-veto threshold
using an unbiased sample of radiative Bhabha events, which are 
compared to
events generated by Rabhat MC together with TSIM and GSIM
(detector simulator).
The Rabhat MC is described in detail in Sec.~\ref{sub:rabhat}.
Finally, in Sec.~\ref{sub:compar}, various distributions including 
$Q^2$ in data and MC are compared to check the validity of the tuned MC,
first for unbiased data and then for HiE triggered data.

\subsection{Tuning of CsiBB thresholds with radiative Bhabha events}
\label{sub:radbha}
We compare the radiative Bhabha events in data with the MC events 
from Rabhat, collected with  
selection criteria similar to those for the signal 
process. 
To select the radiative Bhabha events, we relax the selection requirements
by allowing events with a single photon.
The energy requirement for both the electron and photon is slightly 
more restrictive, greater than 1.5~GeV, in order to collect 
$(e) e \gamma$ samples with negligible background contamination.
The conditions for the angles and the other selection requirements are
the same  as those in 
the analysis of the signal process.
The three-body kinematical selection is applied assuming
a single real photon, $0.85< E_{\rm ratio} <1.1$. 
We define the p- and e-tags
and $Q^2$ in the same manner as in the measurement of
the single-tag two-photon process from the detected electron
for each event, although the 
physics mechanism is different. 

We measure the threshold curve for each CsiBB Logic
by counting the yields from the two triggers
and by forming the ratio
\begin{equation}
y_{\rm ratio} = \frac{50*N_{\rm CsiBB}}{N_{\rm HiE} + 50*N_{\rm CsiBB}}
\end{equation}
as a function of the energy sum ($E_{\rm sum}$) over the 
electron and photon deposited in a pattern of the Logic
for both experimental data and MC, 
where $N_{\rm CsiBB}$
($N_{\rm HiE} + 50*N_{\rm CsiBB}$) is the number of events triggered
by the CsiBB (HiE + CsiBB multiplied by 50) trigger(s) and
the factor of 50 corrects for the prescaling in the CsiBB trigger.
Note that, in matching $y_{\rm ratio}$ between data and MC,
tuning of the TSIM threshold is essential
while the details of the physics simulated by Rabhat are not.
We obtain the energy dependence of $y_{\rm ratio}$
for each Logic. 
We use the energy sum
($E_{\rm sum}$) over the electron or the photon that
enters the relevant ECL $\theta$-ring(s) to the Logic.
In each Logic, only the events in which the sum of the 
deposited energy to the corresponding Logic exceeds
$E_{\rm max} - 1~\GeV$ are used, where $E_{\rm max}$ is
the greatest energy deposit in a given Logic (among
the eleven) in each event.

We obtain the threshold curves for each 
Logic for data and MC simulation and tune
the latter to match the former. 
We tune the threshold behavior of the MC so that it agrees
well with that in data as shown
in Fig.~\ref{fig:trigexpthreb}.
The thresholds are thus adjusted with an
accuracy better than $\pm 0.2$~GeV.
The non-zero efficiency below the nominal threshold 
is due to an artifact from events 
that are triggered by a different Logic.
The real thresholds are near 5--6~GeV for Logics\#$1-10$ and near
3~GeV for Logic\#11. 
The gaps near 7.5~GeV, when present, correspond to a region
almost prohibited by the kinematics of $(e)e\gamma$.

 We confirm the result of the
tuning and its validity by comparing
distributions of the radiative Bhabha
events between the experimental data and MC
in Sec.~\ref{sub:compar}.
We use the event yields for this purpose because the radiative Bhabha 
cross section is well known.

\subsection{Rabhat, MC code for radiative Bhabha events}
\label{sub:rabhat}
We use the MC simulation program Rabhat~\cite{rabhat} to generate 
events with a virtual Compton configuration.
Rabhat simulates the radiative
Bhabha process with a $t$-channel mass singularity.
This singularity occurs in a topology
in which the virtual photon that is emitted at an extremely forward
angle from one of the incident electrons is 
hard scattered off the other electron. 
Although the cross section
is finite and is not very large, our requirement of
an energetic final-state photon at a finite angle
results in a relatively sharp 
mass singularity for an acoplanarity 
angle of order $E_\gamma/m_e$. 
Rabhat adopts a careful choice of integration
variables and a proper technique for numerical integration.
We used Rabhat to simulate approximately $2 \times 10^8$
($3.6 \times 10^7$) MC events for $\sqrt{s}=10.58~\GeV$ (10.88~GeV).
A limitation of Rabhat is that it calculates the lowest order
$e^+ e^- \to e^+ e^- \gamma$ process (${\cal O}(\alpha^3$)). 
There could be an uncertainty from radiative corrections,
typically of order 5\%, which
is estimated from the size of the radiative correction
$1+\delta$ for the Compton process, $e \gamma \to e \gamma$,
where $\delta \sim -6\%$ in the similar
$e \gamma$ c.m. energy region (around 3--10~GeV)~\cite{radcor3}.

\begin{figure}
\centering
\includegraphics[width=6.5cm]{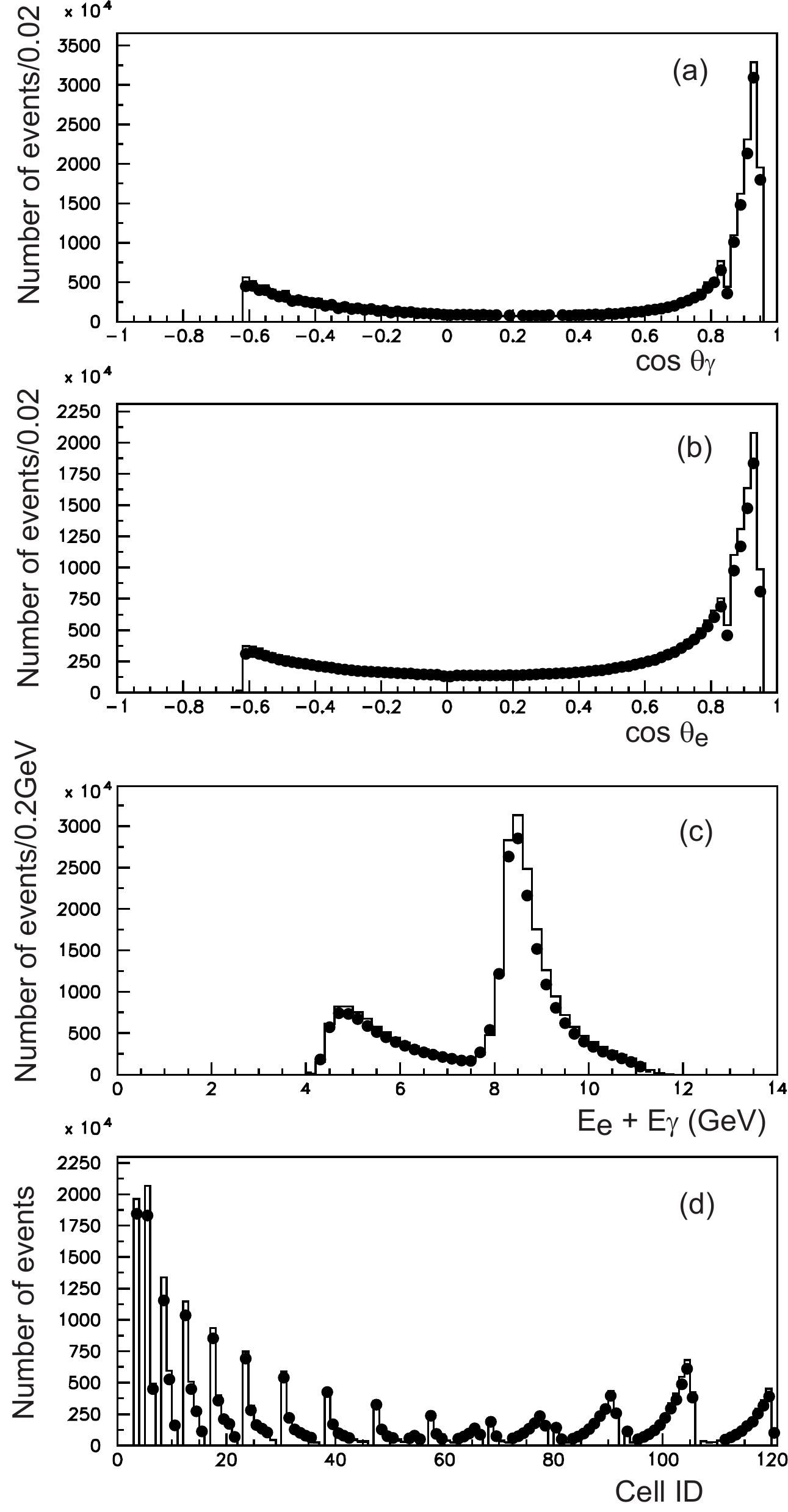}
\centering
\caption{Comparison of the experimental
radiative-Bhabha samples (dots) and the corresponding
MC (Rabhat, histogram) normalized to the integrated luminosity, triggered
by HiE or CsiBB, where the Bhabha mask is not applied.
The distributions are (a) polar angle of the photon,
(b) polar angle of the electron, (c) sum of the photon
and electron energies and (d) identification number of 
the cell in the ($\cos \theta_e$, $\cos \theta_\gamma$) plane. 
}
\label{fig:rbmcsum_a}
\end{figure}

\begin{figure}
\centering
\includegraphics[width=6.5cm]{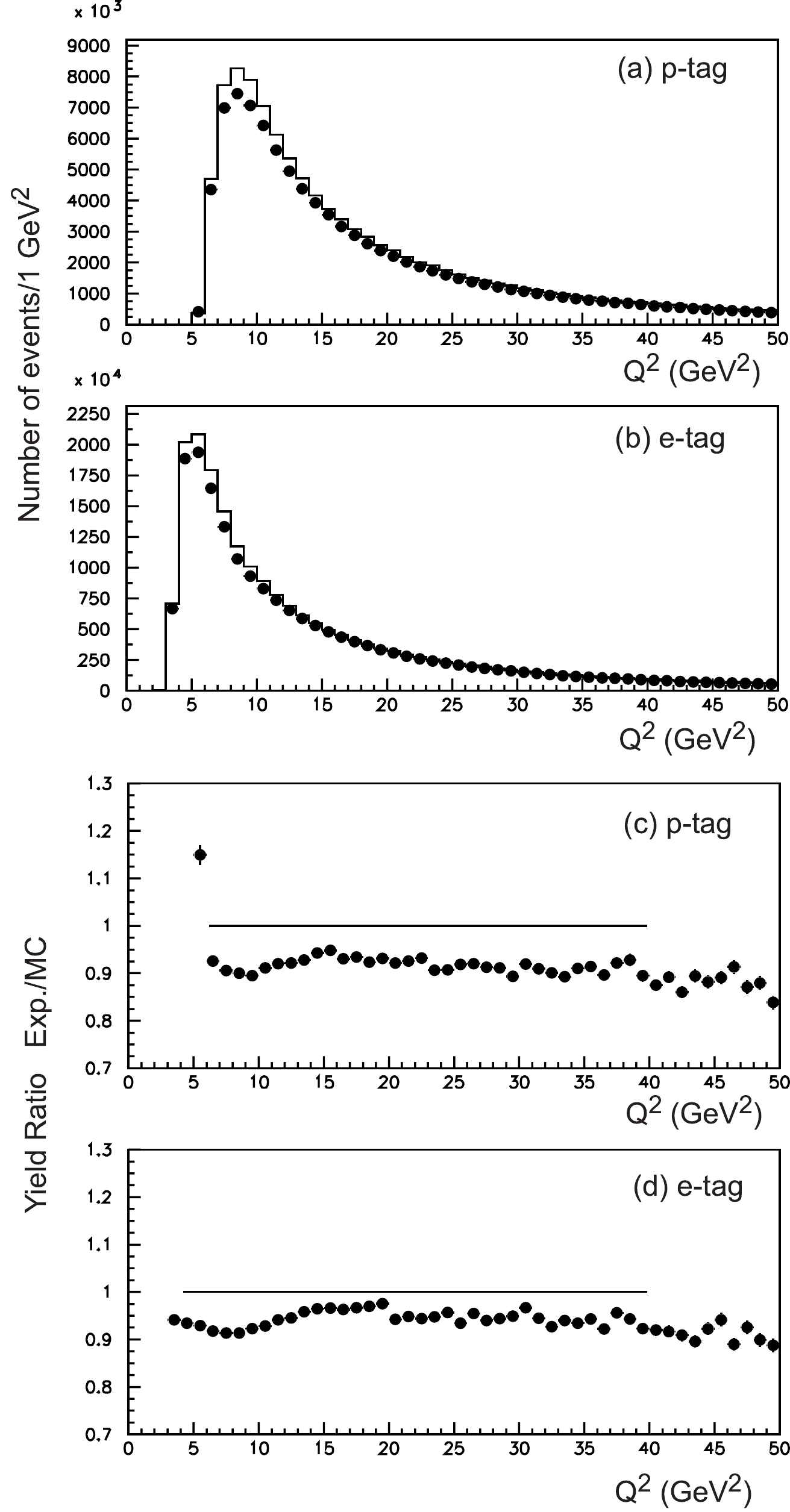}
\centering
\caption{(a,b) Comparison of the experimental
radiative-Bhabha samples (dots) and the corresponding
MC (Rabhat, histogram) normalized to the integrated luminosity, triggered
by HiE or CsiBB, where the Bhabha mask is not applied,
for the $Q^2$ distributions of the detected electron,
where the two selection criteria related to the $p_t$ balance
are loosened (see the text).
(c,d) The ratio of the experimental yield to the normalized
MC yield. 
The sizes of the vertical error bars correspond
to the statistical errors
in both experimental and MC samples.
The horizontal lines at 1.0 show the $Q^2$ region where the
measurement of the $\pi^0$ TFF is performed.
}
\label{fig:rbmcsum_q}
\end{figure}

\begin{figure}[t]
\centering
\includegraphics[width=6.5cm]{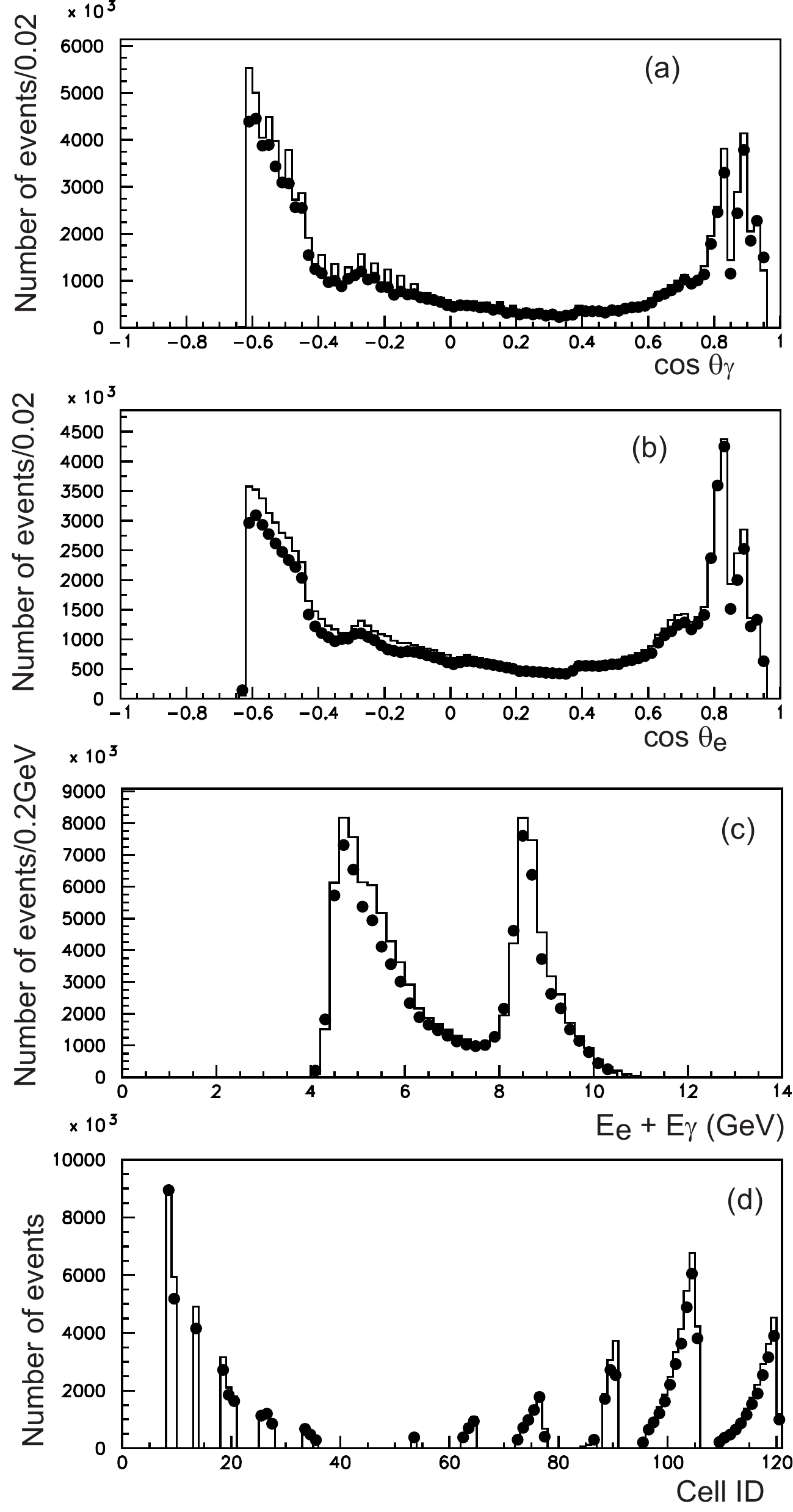}
\centering
\caption{Comparison of the experimental
radiative-Bhabha samples (dots) and the corresponding
MC (Rabhat, histogram) normalized to the integrated luminosity, triggered
by  HiE, where the Bhabha-mask and -veto are applied.
See the caption of Fig.~\ref{fig:rbmcsum_a} for more information.
}
\label{fig:rbmc_a}
\end{figure}

\begin{figure}[t]
\centering
\includegraphics[width=6.5cm]{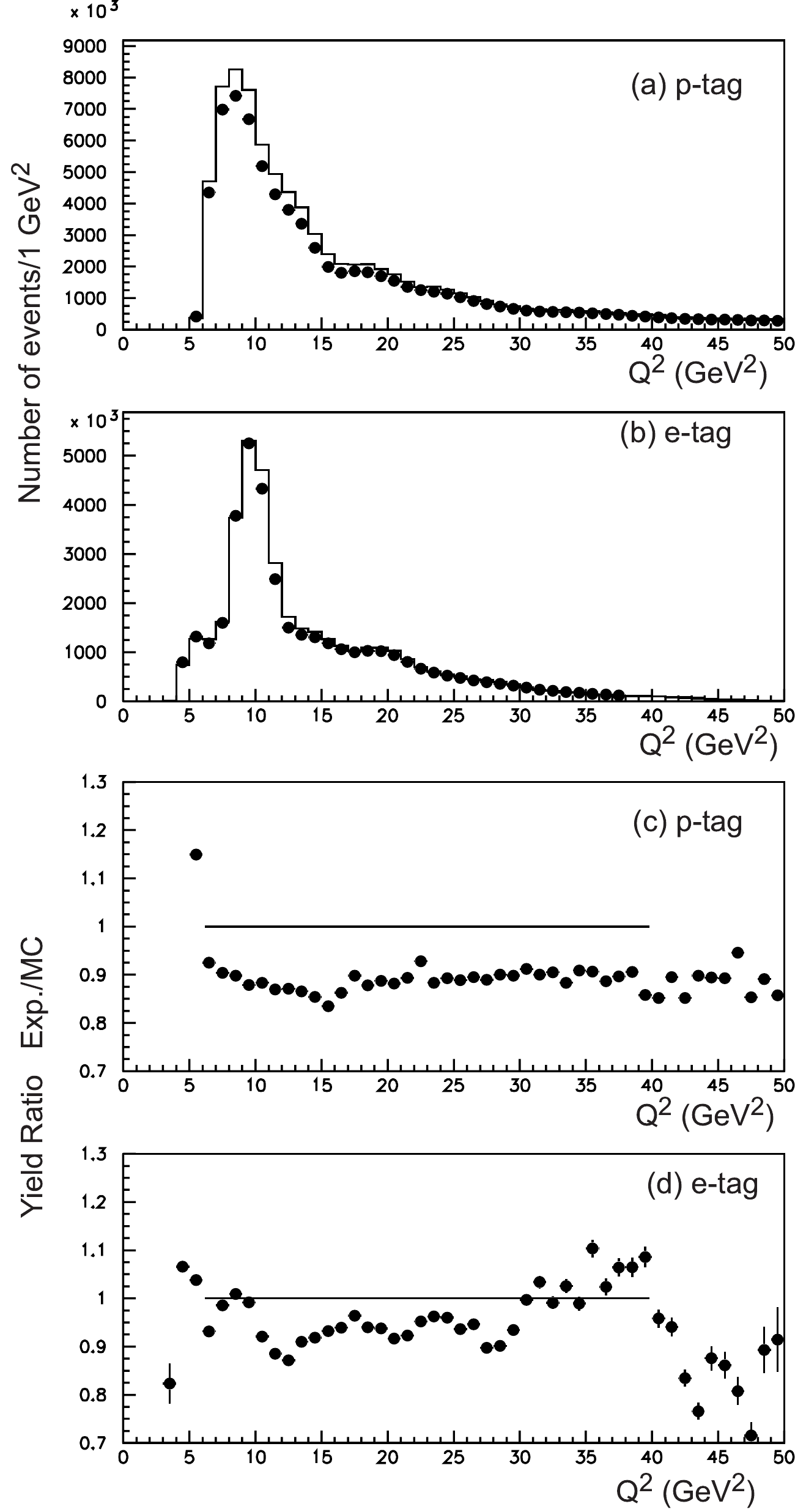}
\centering
\caption{Comparison of the experimental
radiative-Bhabha samples (dots) and the corresponding
MC (Rabhat, histogram) normalized to the integrated luminosity, triggered
by HiE, where the Bhabha-mask and -veto are applied.
See the caption of Fig.~\ref{fig:rbmcsum_q} for 
each figure.
The horizontal lines at 1.0 in (c,d) show the $Q^2$ region where 
the measurement of the $\pi^0$ TFF is performed using HiE-trigger data.
}
\label{fig:rbmc_q}
\end{figure}

\begin{figure*}
\centering
\includegraphics[width=13cm]{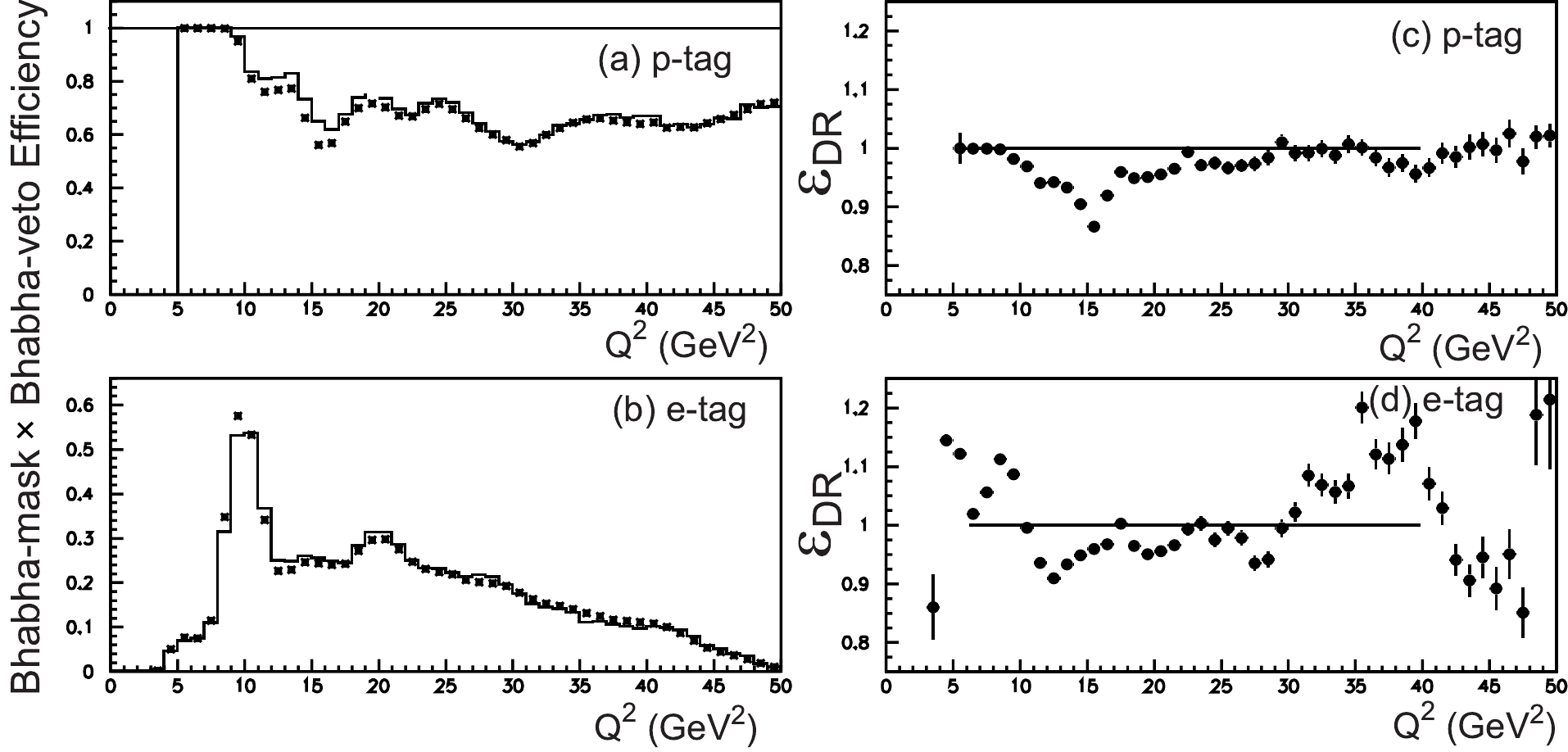}
\centering
\caption{(a,b) Comparison between the efficiency of the ``Bhabha-mask $\times$ 
Bhabha-veto'' for the radiative-Bhabha samples (dots, 
$x_{\rm ratio}({\rm data})$ described in the text)
and the corresponding
MC (Rabhat, histogram, $x_{\rm ratio}({\rm MC})$)
for (a) p-tag  and (b) e-tag events. 
(c,d) Ratio ($\epsilon_{\rm DR}$)
of the efficiencies which are plotted in (a,b)
for the radiative-Bhabha samples and the corresponding
MC for (c) p-tag and (d) e-tag events. 
The vertical error size is derived from the statistical errors
in both experimental and MC samples.
The horizontal lines at 1.0 show the $Q^2$ region where 
the measurement of the $\pi^0$ TFF is performed using the Bhabha 
mask and Bhabha veto. 
}
\label{fig:compeff_rabhati}
\end{figure*}

\subsection{Comparison between the data and MC for radiative Bhabha events}
\label{sub:compar}
In this subsection,
we first compare experimental data for the HiE+50*CsiBB sample with 
MC, where the CsiBB threshold effect is absent, to show that
MC can reproduce various distributions fairly well 
(Figs.~\ref{fig:rbmcsum_a} and \ref{fig:rbmcsum_q}).
Similar plots are presented for the HiE triggered events 
(Figs.~\ref{fig:rbmc_a} and \ref{fig:rbmc_q})
to show that the agreement between data and MC is also adequate.
We use an MC sample where the CsiBB threshold is tuned as described in
Sec.~\ref{sub:radbha}.
We then show the HiE trigger efficiency determined by data using
radiative Bhabha events and compare it to the results from the trigger 
simulation (Fig.~\ref{fig:compeff_rabhati}).

We show the comparison in distributions for
polar angles of the photon and electron, energy-sum
and cell pattern as correlation of the angles
in Fig.~\ref{fig:rbmcsum_a}
for the sample of HiE+50*CsiBB.
The MC events are normalized to the nominal luminosity; 
radiative corrections are included in this normalization.
The horizontal axis for the cell identification
number corresponds to 
each cell in the ($\cos \theta_e$, $\cos \theta_\gamma$)
plane in which the symmetry in interchanging
$\cos \theta_e$ and $\cos \theta_\gamma$ is taken into account
(see Fig.~\ref{fig:cell}).
 
The $Q^2$ distributions for data and MC for the unbiased sample 
(HiE+50*CsiBB) are shown in Figs.~\ref{fig:rbmcsum_q}(a) and (b) 
for the p- and e-tag samples, respectively.
Figures ~\ref{fig:rbmcsum_q}(c) and (d) are the ratio between data and MC for 
the p- and e-tag separately.
In these comparisons, we include the radiative correction of
$\delta=-6\%$ and loosen
the selection criteria related to the $p_t$ balance conditions for
the $e\gamma$ system to $\alpha(e,\gamma)<0.3$~radians and
$|\Sigma \vec{p}_t^*| < 0.5$~GeV/$c$,
for a more meaningful
comparison of the yield, noting that the $p_t$-balance is affected
by the emission of an extra photon; the effect is typically 4\%.

Figures~\ref{fig:rbmc_a} and ~\ref{fig:rbmc_q} present
similar distributions for the samples after applying
TSIM for the HiE trigger including applications of
the Bhabha veto and Bhabha mask. 
The results agree to within the $5-10$\% level (Fig.~\ref{fig:rbmc_q}(c), (d)).
A deficit of events is seen in both the p-tag and e-tag
modes, and also in the  HiE+50*CsiBB sample 
(Fig.~\ref{fig:rbmcsum_q}(c), (d)).
This discrepancy is not due to the Bhabha veto,
because a deficit of similar size is also seen
in the regions where the veto has little or no
inefficiency, {\it e.g.}, for $Q^2 < 10$~GeV$^2$ with the p-tag. 
Moreover, the deficits are of similar size in the unbiased sample
and in the HiE-trigger sample.  
These deficits could be due to the uncertainty of the
radiative correction $\delta$ (4\%), 
an uncertainty in the efficiency due to additional
multi-photon emission, which is not accounted for in the Rabhat MC (4\%),
as well as the experimental measurement itself
($\sim 5\%$, similar to the $Q^2$-independent systematic 
uncertainty in the efficiency for the signal process in
Sec.~\ref{sub:syserr}), which amounts to $\sim 8\%$ in total.

The ratio $x_{\rm ratio}$, defined as
\begin{equation}
x_{\rm ratio} = \frac{N_{\rm HiE}}{N_{\rm HiE} + 50*N_{\rm CsiBB} } ,
\end{equation}
{\it i.e.}, the number of events triggered by HiE
divided by those triggered by ''HiE OR CsiBB'',
provides the fraction of the events that survive the Bhabha veto.
The Bhabha mask is (is not) applied to the HiE (HiE+50*CsiBB) sample.
Thus, $x_{\rm ratio}$ contains
the effect of the Bhabha mask as an inefficiency.
Note that this ratio is not affected by
uncertainties in normalization such as those coming from the
radiative corrections of the Bhabha generators as well as that from
the experimental measurements.
The ratio $x_{\rm ratio}$ is shown as a function of $Q^2$ for the 
data and Rabhat MC events tagged by positrons and electrons in 
Fig.~\ref{fig:compeff_rabhati}(a) and (b), respectively.
The double ratio ($\epsilon_{\rm DR}$) between data and MC
\begin{equation}
\epsilon_{\rm DR} = \frac{x_{\rm ratio}({\rm data})}{x_{\rm ratio}({\rm MC})},
\end{equation}
shown in Fig.~\ref{fig:compeff_rabhati}(c) and (d), demonstrates 
that the Bhabha-veto efficiency
simulated by the MC agrees with data to within $\pm (5\%-10\%)$.



\begin{thebibliography}{99}
\bibitem{babar1}
B. Aubert {\it et al.} (BaBar Collaboration),
Phys. Rev. D \textbf{80}, 052002 (2009).
\bibitem{babar2}
P. del Amo Sanchez {\it et al.} (BaBar Collaboration),
Phys. Rev. D \textbf{84}, 052001 (2011).
\bibitem{cello}
H.J. Behrend {\it et al.} (CELLO Collaboration),
Z. Phys. C. \textbf{49}, 401 (1991).
\bibitem{cleo}
J. Gronberg {\it et al.} (CLEO Collaboration),
Phys. Rev. D \textbf{57}, 33 (1998).
\bibitem{LB}
G.P. Lepage and   S.J.~Brodsky, Phys. Rev. D \textbf{22}, 2157 (1980).
\bibitem{agaev}
S.S. Agaev, V.M. Braun, N. Offen, and F.A. Porkert,
Phys. Rev. D \textbf{83}, 054020 (2011);
we follow the notation in this reference.
\bibitem{rady}
A.V. Radyushkin, Phys. Rev. D \textbf{80}, 094009 (2009).
\bibitem{poly}
M.V. Polyakov, JETP Lett. \textbf{90}, 228 (2009).
\bibitem{pham}
T.N. Pham and X.Y. Pham, 
Int. J. Mod. Phys. A \textbf{26}, 4125 (2011).
\bibitem{MS}
S.V. Mikhailov and N.G. Stefanis, 
Mod. Phys. Lett. A \textbf{24}, 2858 (2009).
\bibitem{roberts}
H.L.L. Roberts, C.D. Roberts, A. Bashir, 
L.X. Guti$\acute{\rm e}$rrez-Guerrero and P.C. Tandy,
Phys. Rev. C \textbf{82}, 065202 (2010).
\bibitem{bcd}
S.J. Brodsky, F.-G. Cao and G.F. de T\'{e}ramond,
Phys. Rev. D \textbf{84}, 033001 (2011);
Phys. Rev. D \textbf{84}, 075012 (2011);
most of earlier theoretical attempts are listed in these articles.
\bibitem{kekb}
S.~Kurokawa and E.~Kikutani, Nucl. Instr. and Meth. A {\bf 499}, 1 (2003),
and other papers included in this volume.
\bibitem{belle} A.~Abashian {\it et al.} (Belle Collaboration),
Nucl. Instr. and Meth. A {\bf 479}, 117 (2002).
\bibitem{ecltrig} 
B.G.~Cheon {\it et al.}, Nucl. Instr. and Meth. A {\bf 494}, 548 (2002).
\bibitem{rabhat}
K.~Tobimatsu and Y.~Shimizu, Comp. Phys. Comm. 
{\bf 55}, 337 (1989). 
\bibitem{bkt}
S.J.~Brodsky, T.~Kinoshita and H.~Terazawa, Phys. Rev. D 
{\bf 4}, 1532 (1971).
\bibitem{treps}
S.~Uehara, KEK Report 96-11 (1996).
\bibitem{shuler} 
G.A.~Schuler, CERN-TH/96-297, arXiv:hep-ph/9610406 (1996) .
\bibitem{expo}
E.~Etim, G.~Pancheri and B.~Touschek, Nuovo Cimento {\bf 51} B,
276 (1967). 
\bibitem{radcor1} 
V.P.~Druzhinin, L.A.~Kardapoltsev and V.A.~Tayursky,
ArXiv:1010.5969[hep-ph] (2010). 
\bibitem{lbbook} 
H.~Genzel {\it et al.}, {\it Landolt-B\"{o}rnstein: Numerical Data and Functional
Relationships in Science and Technology}, New Series, Group {\bf I} Vol. 8 
(Springer, Berlin, 1973), p.314 .
\bibitem{pi0pi0} 
S.~Uehara, Y.~Watanabe, H.~Nakazawa {\it et al.} 
(Belle collaboration), Phys. Rev. D {\bf 79}, 052009 (2009).
\bibitem{sakurai} 
J.J.~Sakurai and D.~Schildknecht, Phys. Lett. 
{\bf 40B}, 121 (1972);
A.M.~Breakstone {\it et al.}, Phys. Rev. Lett {\bf 47}, 1782 (1981).
\bibitem{radcor2} 
S.~Ong and P.~Kessler, Phys.~Rev.~D {\bf 38}, 2280 (1988);
M.~Benayoun, S.I.~Eidelman, V.N.~Ivanchenko and Z.K.~Silagadze, Mod. 
Phys. Lett. A
{\bf 14}, 2605 (1999); F. Jegerlehner, J. Phys. G {\bf 29}, 101 (2003).
\bibitem{footn1}
This contrasts with BaBar where a special trigger was made available
for the signal~\cite{babar1, babar2}.

\bibitem{footn2}
At $25.95^\circ$, $32.27^\circ$, 
$37.81^\circ$, $44.17^\circ$, $51.39^\circ$, $59.42^\circ$,
$68.24^\circ$, $77.73^\circ$, $87.68^\circ$, $97.36^\circ$, 
$107.10^\circ$, $116.27^\circ$ and $124.73^\circ$

\bibitem{radcor3} K.J.~Mork, Phys. Rev. A {\bf 4}, 917 (1971).
\end{thebibliography}
\end{document}